# Speculative Physics: the Ontology of Theory and Experiment in High Energy Particle Physics and Science Fiction

by

Clarissa Ai Ling Lee

Graduate Program in Literature
Duke University

Date: ______________________
Approved:

______________________
N. Katherine Hayles, Co-Supervisor

______________________
Mark C. Kruse, Co-Supervisor

______________________
Mark B. Hansen

______________________
Andrew Janiak

______________________
Timothy Lenoir

Dissertation submitted in partial fulfillment of
the requirements for the degree of Doctor
of Philosophy in the Graduate Program in Literature in the Graduate School
of Duke University

2014

<u>ABSTRACT</u>

Speculative Physics: the Ontology of Theory and Experiment in High Energy Particle Physics and Science Fiction

by

Clarissa Ai Ling Lee

Graduate Program in Literature
Duke University

Date:________________________
Approved:

____________________________
N. Katherine Hayles, Co-Supervisor

____________________________
Mark C. Kruse, Co-Supervisor

____________________________
Mark B. Hansen

____________________________
Andrew Janiak

____________________________
Timothy Lenoir

An abstract of a dissertation submitted in partial
fulfillment of the requirements for the degree
of Doctor of Philosophy in the Graduate Program in Literature in the Graduate School of
Duke University

2014

# Abstract

The dissertation brings together approaches across the fields of physics, critical theory, literary studies, philosophy of physics, sociology of science, and history of science to synthesize a hybrid approach for instigating more rigorous and intense cross-disciplinary interrogations between the sciences and the humanities. I explore the concept of speculation in particle physics and science fiction to examine emergent critical approaches for working in the two areas of literature and physics (the latter through critical science studies), but with the expectation of contributing new insights to media theory, critical code studies, and also the science studies of science fiction.

There are two levels of conversations going on in the dissertation; at the first level, the discussion is centered on a critical historiography and philosophical implications of the discovery Higgs boson in relation to its position at the intersection of old (current) and the potential for new possibilities in quantum physics; I then position my findings on the Higgs boson in connection to the double-slit experiment that represents foundational inquiries into quantum physics, to demonstrate the bridge between fundamental physics and high energy particle physics. The conceptualization of the variants of the double-slit experiment informs the aforementioned critical comparisons. At the second level of the conversation, theories are produced from a close study of the physics objects as speculative engine for new knowledge generation that are then reconceptualized and re-articulated for extrapolation into the speculative ontology of hard science fiction, particularly the hard science fiction written with the double intent of speaking to the science while producing imaginative and socially conscious science through the literary affordances of science fiction. The works of science fiction examined here demonstrate the tension between the internal values of physics in the practice of theory and experiment and questions on ethics, culture, and morality.

Nevertheless, the dissertation hopes to show the beginnings of a possibility, through the contentious but generative space provided by speculative physics, to produce more cross-collaborative thinking between physics as represented by the hard sciences, and science fiction representing the objects of literary enterprise and creative evolution.

.

# Dedication

In loving memory of my father, Lt Commander Han Seng Lee (Rtd) (AMN), who passed away on July 7, 2012 before he see the fruits of this labor that is completed less than two years after his passing in honor of his memory. I also dedicate this work to my mother and sister, as well as the rest of my extended family and friends for their support through tough times.

# Contents

List of Figures

## Acknowledgements

I thank my dissertation directors, Prof N Katherine Hayles and Prof Mark C Kruse, for supporting my dissertation despite my refusal to conform to any disciplinary insistence. I also thank my committee members, Prof Andrew Janiak, Prof Mark Hansen, and Prof Timothy Lenoir, as well as Prof Negar Mottahedeh, Prof Emeritus Barbara Herrnstein-Smith, Prof Arkady Plotnitsky of Purdue University, and Prof Ron Eglash of the Rensselaer Polytechnic Institute, for their support and useful comments.

I thank my friends at the Literature program and beyond (including Femrem and the 'Duke Collective') for their support, regardless of whether they understood my work (I will not name names since this might mean leaving some of you out but you know who are!). I thank Lynn Badia from UNC-Chapel Hill for agreeing to be my respondent during the Literature Colloquium, and for her astute comments to a very early draft of a chapter. For all my six years in the PhD program, I have met many who came and went, and whose conversations had been very useful feedback in the development of this intellectual project.

Finally, I owe the final shape of my work to the wonderful interactions on Twitter that provided me with the intellectual wherewithal and inspiration I needed to go on even when I feel intellectually forlorn. I dedicate this dissertation to all my Twitter followers, the folks I follow, and the folks on #histsci and #phdchat.

# 1. Prolog: the Science Studies of High Energy Particle Physics and Hard Science Fiction

This dissertation extends the discourse of previous generation's engagement coming out of various philosophical, anthropological, sociological, historical, and cultural studies of particle/quantum physics from multi-faceted perspectives, with a majority of the studies centered on epistemic cultures, historical development and representation, laboratory practices, and globally networked collaborations. The story I am interested in presenting takes on a multi-faceted, and theoretical, transdisciplinary engagement with physics that is not merely focused on physics's narrative, but also on the relevance of that story to other disciplines, even if that story is not an easy tale to tell, and one that is filled with pitfalls. From a humanistic angle, I am interested in what the discursive elements of the intellectual projects situated in other disciplines can contribute in terms of the description of certain aspects of quantum physical epistemology, and particle physics more specifically.

Before sociologists and anthropologists became interested in the epistemology of particle physics, especially from the 1980s, philosophers and historians of physics have been considering the implications of developments in the subfield of particle physics because of the latter's close connection to foundational inquiries within quantum physics. In addition, particle physics also expands into the realm of astrophysics, and therefore, cosmology. Cosmology has real philosophical significance that is not only limited to the immediate questions within philosophy of science, but also implicated within the broader perspectives of the philosophy of nature in relation to ethics and morality.

However, to consider the topic from the view of publishing and PhD dissertation history, the publication of Sharon Traweek's seminal work *Beamtimes and Lifetimes: the World of Higher Energy Physics* (originating from her dissertation project), in 1988, has opened up the way for studying the culture of physicists, including the differences between theorists and experimentalists; hence, particle physics is no longer just of interest to philosophers of physics but has also entered the critical embrace of critical and cultural studies of science. It started with a trickle that has turned into a more regular flow, as other publications are produced. Dissertations that consider the sociological, historical, and ethnographical implications of big laboratory sciences that also include high-energy particle physics, or use the methodologies developed out of such studies to consider other related and not-so-related fields of science, have also began to mushroom, as one will discover from the database searches.

Traweek had conducted an ethnographic and observer-participation study into the 'masculine' labor of high-energy physics in the 1970s that compared between the national laboratories in Japan (National Laboratory for High-energy physics or KEK), Stanford Linear Accelerator at Stanford, and Fermilab in Chicago. It so happens that her fieldwork was taking place at about the same time as when exciting 'new' discoveries were made and a revolution in epistemic paradigms was taking place.

However, for all the explosive excitement heralding a generation of new scientific knowledge, Traweek carefully unveils the 'big-boys' attitude prevalent in the social climate surrounding big science and their apparatuses. She also highlights another important feature of her role as a fieldworker: to not be enticed into siding with the physicists and merely reproduce the latter's epistemic priorities and sense of selves even as she takes stock of her position as an 'impartial' observer and outsider. This standpoint must be heeded by researchers (and other interested readers) who are wading through the internal and public discourse of science; with the rise in science communication and the application of scientific rhetoric by the scientists themselves (not just scholars studying their work), perception about the audience and assumptions about their capabilities propel the types of knowledge that get transmitted to the public. Humanistic and social science researchers who are not insider to the knowledge-making enterprise would have to keep in mind, the theory-ladenness that is present in their 'interdisciplinary' interactions with scientists, just as much as the researcher with insider connections and knowledge (such as a trained physicist who later becomes a historian or sociologist of physics) must be wary of the assumptions about the culture he/she might have already imbibed.

Although Traweek is not the first scholar to have written about the social life of experiments and the relationships that physicists have with their apparatuses of work in relation to the construction of scientific knowledge (we can think of earlier relevant works on the subject by scholars such as Michel Serres, Bruno Latour, Harry Collins and Andrew Pickering), she could be considered as the first to address the non-universal values and localized politics that shape the physicists' social life at the laboratories, in relation to their work, while also attending to gendered nuances. Unlike Latour, she makes a distinction between the human subject and the non-human actor by emphasizing how the reproduction of social norms in the scientific world plays a big role in shaping the actual science being produced. In other words, Traweek does not take privilege to be the default even as she remains aware of her privileged position stemming from her personal interactions at the sites of her fieldwork.

Before *Beamtimes*, Andrew Pickering had published a *tour de force* on the world of particle physics, *Constructing Quarks: a Sociological History of Particle Physics*[1].

---

[1] While visiting the CERN library during my short stint of fieldwork back in the summer of 2010, I found this book by Pickering prominently displayed in an area of high visibility to grab the attention of physicists usually too busy just catching up on the immediate literature of their work, let alone read the sociological tracts on their work!

However, the book concentrates mainly on the epistemic turns that separate 'old' standard physics from the new ones, the latter represented by the pivotal year of 1974. 1974, known also as the year of the 'November Revolution' that saw the first in a series of discoveries of "unusual elementary particles" such as the J/psi meson (an example of the hadronic particle constituting two quarks) that marked the establishment of a new physics not perceived as possible under the old model; the 'new' physics at the time were electron-positron collisions, neutrinos, and the quark composite model of hadronic particles. In a sense, it also marks the revival of quantum field theory from its almost decrepit existence, and becomes in itself, the first sign that theory and experiment in particle physics are beginning to reconcile that previously irreconcilable differences that had been part of the crisis in the field.

Pickering situates old physics within the advancement of quantum electrodynamics that ushered the beginnings of a formal system for representing particles in space-time such as was developed by Heisenberg, Schrödinger, Pauli, Dirac, and Born; and the founding of the 'Eightfold-way,' which is a group theoretical pre-Standard Model (to be further explained in chapters two, three, and Appendix A) representation of the strong interaction of quarks. He then differentiates all of the aforementioned entities from the new physics of theories comprising electroweak and strong interactions, and the different categories of subatomic particles. For Pickering, the establishment of 'new' physics stems from discoveries in the areas of mesonic theory (two quarks combined), heavy leptons (electrons, positrons, muons, and neutrinos), and new generations of quarks (three are known at the time of writing).

According to Pickering, the 'new physics' is a new paradigm for explicating phenomena not easily described under older theoretical models. He also traces the dynamics of practices involved in achieving epistemic evolution, gesturing to the still speculative possibility of the grand unification of nature's forces. I consider Pickering's work as providing a more organized, critical, and systematic approach to the various personal scientific essays collected in *The Rise of the Standard Model: Particle Physics in the 1960s and 1970s* that chronicle, though not always in a chronological fashion, the contributions of the experimentalists and theorists in manufacturing the products structuring the Standard Model. However, the essays are written in a way that assumes the reader as having general sophistication and background knowledge on the topic.

While Pickering's work concentrates on the sociological aspects of knowledge transmission and transference that enable some theories to continue while others to dissipate, from reasons ranging from the lack of effective ways for measuring predictions to errors (or misdirections) in calculations, Peter Galison's *Image and Logic: A Material Culture of Microphysics*, published in 1997, is interested in the material culture of particle physics experiments by tracing its prehistory through radiation and nuclear physics. Galison details the development of particle physics in terms of instrumental design and the endgame involved, all the while setting his analyses against the background of real-world politics that can affect the direction of knowledge production in high-energy physics.

His discussion of the politics of big science considers social constructivism (and its limits) and how that could be connected with the epistemic materiality of modern physics: the relationships between scientific valuation/ideologies and instruments of choice, academic structures that promote the advancement of certain research areas, and the contestation of objectivity in knowledge production. He does not shy away from engaging in an interrogation between internal and external values to demonstrate how they converge into a dominant narrative strain in physics. He also considers how experiments and theoretical predictions (or speculations) are able to complement each other, and how these could be communicated within a 'trading zone' where a common goal is aspired to even if the 'language' of communication is fraught with ambivalence, linguistic fractures, and incommensurability.

Possibly taking a leaf out of the work of Traweek, Karin Knorr-Cetina, who had performed comparative ethnography in the sociology of science, wrote *Epistemic Cultures: How the Sciences Make Knowledge* (1999) that considers the differences and similarities between experiments in molecular biology with detection work in high-energy physics at CERN. While these scientific fields work with objects at a microscopic scale that requires many levels of mediation, she highlights the differences in their practices stemming from the different prerogatives of their fields' various scientific inquiries (such as microbiology versus physics) that therefore shape the development of their praxis, signature-representations, and relationship to their experimental instruments. Knorr-Cetina's study takes up the sociological practices in high-energy physics raised by Traweek, Pickering, and Galison by finding common traits between particle physics and microbiology, therefore raising some important questions regarding transdisciplinary interrogations within the sciences that are foundationally different.

However, at a different end of a spectrum, and with influences less direct, are the critical works relating to interpretations in quantum theory, including the history of such interpretations, by scholars such as David Bohm, Karen Barad, Mara Beller, and Arkady Plotnitsky, who had influenced my interest in exploring the possibility of real-time implications and impact of quantum mechanical interpretations on pragmatic questions in physics, especially experimental particle physics. While taking a different epistemic direction from these scholars, I have deployed the discussions they began with regard to developments in ordinary quantum mechanics and quantum field theories to query some of the ontological problems that have arisen in both Standard Model particle physics and in the consideration of as yet undetected, new physics. In fact, their discussions into the workings of interpretations allow me to make the necessary connections to earlier history of physics, which I then relate to the conceptualization and development of the Higgs boson. I went as far back as the early nineteenth century development in classical electrodynamics to better understand the confluence of a multiplicity of scientific and ideological events involved in shaping the course of modern theoretical and experimental physics.

However, what had most directly informed the coalescing of my project around the topic of speculation is the publication of a work that considers speculation as enacted in the history of physics. Helge Kragh's *Higher Speculations: Grand Theories and Failed Revolutions in Physics and Cosmology* (2011) attempts to explicate problems relating to how novel theories are constructed until the emergence of new empirical evidence, or theories that pass the test of scientific rigor, render them obsolete. The book considers the different ways in which the theories are considered as speculative since the nineteenth century such as theories of fields (and vortices) right up to twenty-first century questions on cosmological constants, quantum gravity, and astrobiology.

However, he does not attempt to unpack what speculation means philosophically, nor produce more than a thematic core of engagement for connecting the different cases of physical sciences in the past two centuries. His deployment of speculation positions him as a materialist; he deploys speculation to demonstrate the epistemic nebulousness of scientific theories anchored in metaphysical conceptualizations/mathematical formalisms that do not have counterparts in empirical data. Further, he cautions ebullience over ambitious and theoretically sophisticated ideas that are inadequately confirmed by experiments, or that are resistant to more pragmatic engagements. It is both my contention (and agreement) with his assessment of some of the history of physics cases, as well as my dissatisfaction with how speculative knowledge has been conceptualized in the book, that had spurred me to recuperate speculation towards a more constructive approach for engaging with swathes of knowledge and facts that are never as ontologically stable as they might appear above ground.

In addition, there has been ongoing work relating to the ontology and epistemology of the LHC, and its experimental searches, conducted by a group based at Wuppertal Universität (http://www.lhc-epistemologie.uni-wuppertal.de/) and their external collaborators. They are particularly interested in how physicists (experimentalists, theoreticians, phenomenologists who are experts at creating models at the intersection of theory and experiment) have developed a conceptual understanding and relationship with the different theoretical models in particle physics. One of the collaboration's members recently gave a talk on the result of their work at CERN in Feb 14, 2013 (http://cds.cern.ch/record/1516534). I take cognizant of their work in progress to trace how our inquiries intersect and diverge, while using their access to physicists in Europe to help me track the differences between the discourses emerging out of Europe and North America, though mostly to help me extend from the ideas proposed within this dissertation into future projects.

Beyond the important abovementioned works that help situate the conversation whence my dissertation takes up, there have also been interesting scholarship at the intersection of science and literature, particularly physics and literature, which enriched the literary aspect of my work, even if we consider different epistemic objects, with their different hierarchies and issues. One of the more incisive is Sean Miller's *Strung Together: The Cultural Currency of String Theory as a Scientific Imaginary*. Miller's work considers how string theory is understood in popular culture and characterized in various works of fictions.

Some of these works of fictions he examines take advantage of the rich scientific imagery (and imaginary) underlying the constitution and development of string theory while others are merely content to engage with cartoon representations of the scientific model. He also considers the rhetorical strategies and popular metaphors appropriated by physicists writing to a general audience in an attempt to convey the wonder and beauty of the science, including strategies that could be misleading when the reader does not understand the context behind the rhetoric. Miller's close reading of the works of fiction in juxtaposition with the technical knowledge base of the science behind string theory provides me with a model for thinking how I can best integrate the discussion of science fiction that provides an imaginative prototype for considering the ideas and concepts in physics expansively, with science; both the physics and the works of fiction act as counterpoints to each other while feeding on each side of the epistemic divide to obtain a more aesthetically enriching interpretation of literature and physics. In the process, I have also discovered just how such a contrapuntal interrogation of physics with literature, and vice-versa, has provided a richer philosophical reading than a mere straightforward approach. Nevertheless, the reading itself has to be a case of demonstration rather than pure argument.

In light of the above, my dissertation attempts to explore the concept of speculation in particle physics and science fiction to investigate non-linear approaches to innovative conceptualizations in physics and literature, with the hope of creating new insights that can relate to knowledge systems in these two areas more specifically, and the sciences and humanities more broadly. I hope to demonstrate how speculative physics could potentially be useful as a philosophical and fictionalizing device, especially as it is used to disrupt epistemic assumptions believed to be foundationally strong. In other words, through a confluence of science fiction and the 'edgy' representations of scientific knowledge at the frontier as epitomized by particle physics, I hope to show how literary aesthetics can be enriched even as interpretive scientific insights are extended.

In the preliminary stages of the dissertation research in spring 2010, I began making contact with the particle physics experimental collaboration, ATLAS, at Duke so that I could sit in during their meetings, as well as get to know the physicists and their epistemic priorities. In the summer of 2010, I was able to conduct a field visit of CERN and interviewed a number of leading physicists who were involved in the collaborations, while observing, firsthand, how data-taking takes place (including the problems that arise from data-taking). With my limited security access, I was able to tour the various control rooms and relevant facilities that gave me insight into how everything is connected. Later in Spring 2013, I was able to conduct a more involved interview with an experimental particle physicist based at Duke who is also part of the collaboration of interest to my research. I asked very specific questions about work relating to the search for new physics, including the continuous work that is still being done since the announcement of the discovery of the Higgs boson, including the internal and external challenges faced.

At the same time, I had also conducted a survey with a small group of physicists that include experimentalists and a theorist to gain insight into their attitudes in relation to the physics they work with, specifically in relation to the act of discovery, while also attempting to gauge their estimation of the status of the theoretical, and experimental, models that shape their work. I have also asked them questions about the role of inter-, trans- and cross-disciplinarity in relation to their science, that could, or not, inform their work. Their responses have been useful in allowing me to consider the role of the imagination and creative insight more specifically, and humanities more broadly, for informing epistemic attitudes, including the manner in which the scientific epistemology is formulated. The case studies and observations I performed are supplemented by extensive archival research across multiple institutions such as the Niels Bohr Library, CERN's internal library, the History of Science Special Collections at Oregon State University, and of course, the CERN users database I was able to access with the help of one of my advisors, Professor Mark C Kruse, and by reason of my earlier research visit to CERN. I thank Niels Bohr Library and Oregon State University Library for their generous grants-in-aid that had allowed me to do the necessary archival work.

My intended readers for the chapters are interested physicists (including particle physicists who might appreciate a different than usual spin on the subject of their study), historians of physics, philosophers of science, sociologists of science, literary scholars, media theorists, and other science and technology studies scholars in general. As such, while strenuous attempts have been made to ensure that the chapters can cater to the varying backgrounds of the readership, some of the chapters, especially those pertaining to the theoretical and experimental narratives of the Higgs boson, are primarily aimed at readers with more physics background.

Given the technicality of the scientific subjects I am addressing, and my need to do them proper justice in order to provide sufficiently rigorous demonstrations in my critique, chapters three and five are aimed more at physicists or humanists and social scientists who are used to engaging with the hard sciences, particularly physical sciences. Despite that, other readers should not be prevented from reading these chapters because each chapter has been constructed as such that one can still understand the arguments made without needing full comprehension of all the scientific references.

Chapter seven, while drawing its material from another technical area in scientific computation, is addressed as much to the primary readers of chapters three and five as to literary and media theorists interested in seeing how the philosophy of the science of Monte Carlo simulations can contribute to their own areas of interest, such as in digital theory, theories of the cinema, and even critical code studies. Therefore, even more pains have been taken to make the chapter as accessible to readers with limited or no scientific background as is possible.

As supplementary background reading to bring readers with limited technical background to speed with the science, while providing the more technically savvy reader with a survey of the critical historiography of particle physics, I have included Appendix A as a supplement to both chapters two and three. I cannot emphasize enough the importance of reading Appendix A to be able to be more completely immersed in what I hope are interesting and illuminating critical narratives that demonstrate the capaciousness of the speculative physics I have envisioned.

The second chapter, which is the 'actual' first chapter of the story, targets all of the abovementioned readers as it provides the intellectual motivation to the dissertation while also introducing the concept of speculative physics. After an overview of the philosophy and critical theories of speculative physics, I then venture into more directed examination of the constitution of speculative theory and experiment through the various historical examples representing the development of the Standard Model. At the same time, I also demonstrate how 'experimentation' is performed in fiction through Alan Lightman's *Einstein's Dreams* that entwines the life of the scientist with the scientific theory he (Einstein) proposed. This is in itself, a demonstration to the argument of how theory and experiment are embodied and not merely pure abstractions, even if the illustration of that embodiment is in abstract terms.

The next five chapters are then divided into three sections (measurement and interpretation, observation, and modeling). Chapters three and four take on the theme of measurement and interpretation. Chapter three introduces the Higgs boson through an analytic description of the prediction of the Standard Model by way of the speculative theory explicated in chapter two, beginning from the early days of the predictions on the Higgs in the 1960s to May 2014. The Higgs is a potent entity for working out the demarcation problem of measurement at the experimental and theoretical level, and some of the arguments I intend to make are highlighted through the invocation of the variants of the double-slit experiment, all of which are meant to symbolize the crucial influence of quantum interpretation over how physics is done even though this connection is mostly neglected in the everyday grind of doing physics.

Henceforth, in this chapter, I will also be exploring the relevant questions that emerge out of foundational inquiries into quantum theories, as well as larger scale connections between consciousness and materialist-realist determinism and indeterminism, particularly since such questions will emerge in the novel I will be discussing in the next chapter. Chapter four takes some of the conceptual arguments developed in chapter three, but with minimal technical dressing, and revisits them in relation to a close reading of Greg Egan's *Quarantine*. While this chapter is aimed at literary scholars and philosophers of science who are curious over how interpretations in quantum theory are modeled in and through hard science fiction, it should come as no surprise that the critical intervention made involves the application of not only a continental philosophical understanding of capital and a philosophical understanding of science fiction, but also the deployment of a small area in the philosophy of quantum mechanics to understand the narrative intent of the novel, which I propose to be divided into three epochs for reasons that will become obvious later.

The theme of the following section, covering chapters five and six, is observation. Chapter five continues the discussion started in chapter three but with focus on the meaning of observation in relation to speculative experiment. Therefore, I will be concentrating on the experimental developments that confirmed, within a range of probability, the discovery of the Higgs boson, and the epistemic politics involved. This is also the chapter where I will be entering deeply into the discussion of the role of observation in building the case of the Higgs boson, as well as the development of other candidate theories such as supersymmetry that could supersede the delimitations and constraints of the Standard Model Higgs boson, or prove that our interpretation of the Higgs boson is erroneous. I also explicate on the role of speculative experiment in the different sub-experiments performed to obtain sufficient data for building a composite profile of the Higgs boson.

In chapter six, I will look into two short stories by physicist and science fiction writer, Gregory Benford, and also a three-part serialized novella by Richard and Nancy Carrigan, that were published by *Analog: Fact and Fiction* in the summer of 1976. This chapter continues some of the critical readings I have begun with the thought experiments of *Quarantine* but in terms of how theory bridges into experiment. This chapter is where the historical turns that created the qualities of hard science fiction, as well as the ethics and moral implications of experimental and observational practices, are read with a mind to how critical science studies could speak to the epistemic (and political) problematics generated in works of fiction. Moreover, what is interesting for both chapters four and six, when considered side by side, is the deployment of a range of hard science fiction works to deal with abstract ethical issues and speculative transdisciplinary relationships of the different sciences, and in the process, is able to amplify some of the issues that neither literary scholars, philosophers of science, nor physicists have been able to find satisfactory answers to.

Chapter seven, which is the only chapter on modeling in this dissertation, concentrates on another angle of speculative physics by positioning speculative theory in unison with speculative experiment through the cybernetic constitution of the Monte Carlo method. The Monte Carlo refers to the heuristics of simulating the data produced by experiments to obtain approximations for comparisons to experimental data, as well as to build a fit between theory and experiment. The chapter intends to demonstrate how the coding of information, which extends from the actual to the virtual and back, is integral to producing a composite representation of the Higgs boson. The hermeneutics of the data generated against that of the 'real' data are explicated and discussed throughout the chapter. Simultaneously, I consider how the promulgation of the philosophy of Monte Carlo can also benefit other areas of studies that have no direct physical relevance yet share certain conceptual connections, such as critical code studies and digital media studies.

The concluding chapter eight brings together all the issues that had been raised earlier to explain what have been, or not, achieved thus far, before concluding with an evaluation of speculative physics's contribution to critical science studies. After all, the dissertation aspires to demonstrate, more broadly, how the two-prong approach to knowledge, as represented by concepts in physics and science fiction prototype, parallels the generation and dissemination of knowledge through media-constituted technology, and the political implications of such epistemic mediations. This chapter also considers the potential that speculative physics can make to critical science studies.

Finally, Appendix A provides a critical historical overview of the Standard Model of quantum theory and particle physics that are crucial to helping the reader understand the intellectual and scientific processes that produce the Higgs boson while demonstrating how speculative knowledge-construction is inevitable to the scientific enterprise.

What you will find in the chapters to come, given that three years is an insufficient time to produce a polished product of such an ambitious project that has undergone unexpected twists and turns due to unforeseen connections that decide to emerge beyond the direct control of the author, an experimentation with attempts to bring into conversations, knowledges of disparate disciplines that go beyond 'borrowing from the disciplinary' to an encounter that will bring to head the limits of interdisciplinary intercourse, the assumptions that a discipline could have on others, the place of theoretical interventions in the process of the historicization of knowledge, and the disruptions to the neat logics of flow in any encounters that cannot be reconciled without much difficulties.

Speculative physics is an epistemic driven medium that attempts to provide space for the logics and illogics of such interrogations; however, speculative physics is a methodology that is honest about its own interrogative limitations and the self-reflexivity that is informed by the ineluctable precedence of *a priori* knowledge. What this dissertation offers is not the beginning of a book, but rather, a blueprint, or user guide, for producing more of such cross-disciplinary, inter-learning translations, ruptures, disruptions, and potential collaborations among other seemingly incommensurable disciplines. At the same time, it intends to demonstrate the difficulties encountered by humanistic critical theory when it came in direct contact with physical theories; they might share the same linguistic syntaxes but are operating under very different cognitive maps. In a sense, this dissertation, its failure in achieving thorough synthesis and commensurabilities, also demonstrates that the true model of interdisciplinary work across the sciences and the humanities have not yet been successfully achieved and that changes of attitudes from within the academy must first take place before any of such work can truly be done.

*The partially successful philosophic generalization will, if derived from physics, find applications in fields of experience beyond physics. It will enlighten observation in those remote fields, so that general principles can be discerned as in process of illustration, which in the absence of the imaginative generalization are obscured by their persistent exemplification.*

**(Alfred North Whitehead – *Process and Reality*)**

# 2. What is Speculative Physics: Between Theory and Experiment

On July 4, 2012, the world was abuzzed with the news of an event of a spectacular nature: evidence of the much-vaunted elementary particle, the Higgs boson. In the months leading to its imminent discovery, there were active speculations about how this would unlock the door to the secrets of the universe, therefore filling the blanks with details of how the different elements and forces interact to form our universe; we can finally understand the properties and mechanics that produce certain types of physical behavior, including how we can interpret the ethics of life. Such an understanding can be deployed to our technological and even moral advantage. Presentations and media conferences were given at CERN, the particle and nuclear physics facility at the border of Switzerland and France, and champagnes were popped in exhilaration.

Upon closer examination, one is struck by the careful packaging of the language of speculation (in terms of the measurement, observation, and modeling) that is used to parlay an event with aspirations towards 'recognition' of a discovery that resembles the desirable confirmation of an important prediction. At the time of the announcement, unequivocal confirmation was still about three standard deviations away, placing the discovered particle some distance from its theoretical version. Since then, physicists have managed to decrease the range of deviations as more data are analyzed for theory-to-experiment fitting. The pronouncement on the Higgs[1] discovery involves much behind-the-scenes intensive analysis through a hybrid of related statistical calculus: distributions, Bayesian probabilities of likelihoods and priors, and frequentism. The physicists hope that the calculations and knowledge derived from this Higgs-like boson will set the stage for a range of possibilities that can later be extended, extrapolated, and verified for the examination of other still speculative, and not quite demonstrated, theories in high-energy particle physics.

It is the desire of this dissertation to produce an interdisciplinary framework for tackling the implications of the abovementioned developments that have shaped the framework of speculative physics, which could then be applied towards finessing the ontology of theory and experiment. However, such an attempt will remain a work in progress in this dissertation, to be carried forward and improved upon (or even transformed) through further and other work. Speculative physics is a theoretical framework (and potential methodology) not only for producing a coherent and cohesive epistemic device for the production of the 'new' in physics, including the constitution of what is the new, but also for retheorizing existing philosophical concepts within the disciplinary matrix of scientific understanding: the underdetermination of theory, probabilistic thinking, indeterminism, stochastic processes, physical laws (and rules!), as well as metaphysics and its connections to similar perspectives within philosophy, media theory, and literature.

---

[1] There are also problems raised with regard to the nomenclature of this scalar boson because, as the arguments go, to merely name the boson Higgs is to erase the contributions of others who had made its theoretical prediction, and subsequent discovery, possible. Therefore, it is not uncommon to see variations of the nomenclature in different high-level physics textbooks that make references to quantum field theories. While I will not be delving into this aspect in my dissertation, one can read more about the history of naming the Higgs boson in Sean M Carroll's *Particle at the End of the Universe* and "A Historical Profile of the Higgs Boson" by Ellis et.al.

The 'new,' in this instance, refers to advances in the standard theory of quantum physics that attempt to account for various difficulties within existing theoretical franeworks. The 'new' also represents a more sophisticated interpretation of the role of played by phenomena as our knowledge of quantum mechanical properties and its stratum of action is consolidated against emergent qualities. These qualities are produced from newly 'materialized' entity (an object/subject with corporeal quality and mathematical representation) due to the act of discovery: an observable-made-tangible that either affirms or contradicts our previous speculative predictions, predictions based on best guesses and estimates through extrapolations of the known, with margins of errors that could be rehabilitated for the emergence of the unknown.

But as Stiegler notes, the achieving of a synergistic coherence (and unity) that achieves reciprocal causality is sufficiently unpredictable that one must anticipate object limits while remaining aware that "the technical object is more than the sum of the scientific principles that it implements" (78). Speculative physics, as we will see in the next few chapters, intends to grow a narrative structure for connecting between, and understanding, the meaning of prediction, the role of metaphysics in scientific thinking, and the configurations of entities at micro- and macrophysical scales.

The Standard Model, the object of our discussion in this work, is an outcome of observations in quantum physics phenomena and materiality. The construction of the Model is mostly ad-hoc, with group theoretical mathematics for describing the relationships of subatomic entities; the theories reveal the ontological pitfalls of the model through its overdetermined dependence on epistemologically derived descriptions. At the same time, the Standard Model also draws from theories that are still speculative, however convincing the theory's logic might be due to an ability to apply empirically unconfirmed theory to the production of the most ingenious and elegant resolutions to the puzzles of nature; one such theory of note is the supersymmetry, to be further discussed in chapter five. The Standard Model is best summed up by Andrew Pickering in *The Mangle of Practice* as representing "…the link between existing culture and the future states that are the goals of scientific practice, but the link is not a causal or mechanical one: the choice of any particular model opens up an indefinite space of modeling vectors, of different goals" (56).

In the history of philosophy and history of physics, the reference to speculative physics originated in post-Enlightenment natural philosophy interested in what went into the composition of physical properties. Speculative physics as a concept was first explicitly detailed by Schelling, during the 18th century, in the *Introduction to the Outlines of a System of Natural Philosophy, or Speculative Physics and the Internal Organization of a System of Science*. His exposition on the concept of speculation and nature's organism considers the limits of knowability in epistemic schema; the bounded constraints that determine the degrees of observability of nature's phenomena, the division between empiricism and theory, and the reciprocal (though not always unequivocally determined) differences between the product and productive.

Centuries later, Helge Kragh, in his book *Higher Speculations: Grand Theories and Failed Revolutions in Physics and Cosmology,* sees speculation in physics as representative of that moment when theories were derived of assumptions without empirical legitimacy, or based primarily on beliefs arising from the philosophical and ideological framework that the scientist/theorist subscribes to. Many of the 'speculative' forms of knowledge that Kragh discusses consist of theories proposed for the unification of diremptive and heterogenous epistemic identities. While readily explaining the phenomenal aesthetic of nature, these speculative concepts contain unproven (and often unquantifiable) conjectures.

However, Kragh does not attempt to qualify his usage of 'speculation' beyond the mundane; he merely offers case histories that, for him, illustrate the sort of theories that lack empirical evidence, or physics interpretations that did not stand the test of time, especially when the latter's most ardent supporters fall away.[2] Hence, it appears that speculation, for Kragh, concerns how an object-theory of choice (such as string theory or a cosmological theory) can be subject to the interpretations produced from the epistemological persuasions of the physicist; aesthetics might trump empirical veracity. Moreover, speculation in this sense can include the deployment of theoretically unwieldy and scientifically 'dubious' ideas, the latter of which might only be realized in hindsight as evidence of the latter's scientific 'legitimacy' accumulates.

However, I am taking a more complex approach to the motif of the speculation in speculative physics, one that goes beyond Schelling's conception and Kragh's analyses. My conception of speculative physics overlaps partially with the position of Gabriel Catren's as articulated in his "Outland Empire: Prologomena to Speculative Absolutism." He argues against subsuming theoretical knowledge under the 'ruse' of philosophical idealism and the kind of overt metaphysicalism that displaces theory from the 'real' space of physics. Therefore, in anchoring speculative epistemology on material subjects at the initial stage of exploration, it becomes easier for one to identify the primary and secondary features that arise from the becoming of knowledge, including the approximations of knowability and unknowability, thus allowing us to prioritize, more effectively, the choices for dealing with speculative knowledge in a research program.

---

[2] Some of these physics theories and interpretations have re-emerged in the latter day, but in a form that is a specter of the historical version. An example of such a case is the aesthetical, even if not epistemic, connection that the vortex field has to quantum field theory.

Speculative physics is not merely about the lack of empirical proofs or material evidence for verifying theories and models. Instead, speculative physics can be constituted as a methodology for comparing between more established forms of knowledge that could then be destabilized as new and unexpected information comes to light. In other words, knowledge that is validated through long-term application can still undergo epistemic transformation because of the shifts needed to accommodate new information as they arise.

For example, the raw and unreconstructed data from the Large Hadron Collider (LHC) detectors are rebuilt with the aid of algorithms constructed out of the best epistemic understanding of the scientific knowledge. The raw datum (ontology) is what it is, but our interpretation of it changes as our attitudes with regard to its potential changes (epistemology). One has to understand the difference between ontology (the being or a priori constitution of the most foundational structure) and epistemology (represented here as a theory of knowledge and a way of knowing). The piling of epistemology onto ontology activates speculation because ontology cannot be directly measured and observed. The problem of direct measurability has also afflicted quantum mechanics as the latter depends on apparatuses shaped by classical physics to achieve macro-observability.

Speculative physics takes up the argument that the physical sciences can be ontologically complete (because the latter's ontology persists independently of the knowledge producer) while remaining epistemologically imperfect (because access to epistemology is dependent on the state of relationship between the knowledge and the knowledge producer). Speculative physics recognizes imperfection in the construction of facts because observers cannot attain perfect/complete scientific knowledge (epistemology) with phenomenological certainty (phenomenological in the Husserlian sense of thought-experiences and logical judgment whence pure intuition is derived). The process of knowledge acquisition is always at the point of becoming and striving unto completion, but never attaining an endpoint since the certainty of a finale presupposes omniscience.

Speculative physics can be constituted as a paradigm that cannot be definitely concluded. To be open to speculative physics is to be willing to disown what we thought we knew, and be open to the implausible, including implausibility that is the product of epistemic rupture between theory and experiment. It also means reconstituting the idea of prediction as narrating the delimitations of a boundary whence an event can take place while remaining attentive to potential singularities, singularity as representing poles of exceptions, failures, and the unspeakable. Metaphysics, on the other hand, is read as the modeling of cause-and-effect between the states of the imaginary (also the fictive) and the material real that also includes queries into the role of physical laws involved in inscribing a physical system.

A perfect case study of the unequal relationship between epistemology and ontology is the 'paradox' of wave-particle duality in a double-slit experiment, with the paradox as representative outcome connected to the materiality of the atom. An intuitive way to think about the 'form' constituting subatomic particles would be to imagine tiny ball bearings. In reality, the concept of the particle is way more complex than what a ball-bearing model can represent, and this is amply demonstrated by experiment. Additionally, conceptualizing subatomic particles as ball bearings is merely a method for describing the properties that differentiates between the microscopic and macroscopic. This method attempts to fit the determinacy of Galilean/Newtonian physics into an indeterminate and uncertain quantum world.[3]

---

[3] Quantum mechanics may have an ontological continuum with classical mechanics, mathematically speaking, because of the porting over of mathematical functions (operators, complex numbers, vectors, and scalars) used in the practice of classical mechanics onto the mathematical physics of the quantum as

During the double-slit experiment, a beam of electrons is directed through two narrow slits onto a screen, producing patterns of intensities. At the same time, these patterns also produce variations of ordered intensity, a fringe of light and shadow that could only be explained by considering interferences as containing wave-like properties. Paradoxically, the wave-particle duality problem might not have existed if an observer attempting to formulate a measurement of the experimental system has been able to access all of the variables involved at the same time. However, this is made impossible because of the constraints imposed by the Heisenberg's Uncertainty Principle, which states that the position and momentum of a quantum level object cannot be simultaneously known due to the probabilistic spread in uncertainty. The principle points to an equivalence between the waves and the particles given that they share similar representations in equations and are derived from a single source; additionally, there are also constraints in the linguistic medium used to describe physical indeterminism. The descriptive constraints are the outcome of applying a verbal style of communication that is linearly progressive and determininistic to embody non-linear and non-causal, indeterministic epistemics.

---

a result of their shared ontology in kinematics, dynamics, and statics. An example of such practices would be in the use of the Lagrangian and Hamiltonians, in going from ordinary to relativistic quantum mechanics.

The double-slit experiment is a rich demonstration of the importance of fictive modeling for apprehending the incomprehensible. The narration that embodies the analytics of the experiment parallels the story-building world of fiction because of how the tweaking of initial parameters could change the potential outcomes; the outcomes that emerge are probabilistically represented and open to different strategies of informational, and therefore plot, distribution and sub-narratives. Moreover, the theoretical models of physics and fiction occupy the same state of the imaginary as that of speculative physics: their different imaginaries super-positioned and superimposed upon each other produce an interference of individual events. Fiction occupies the space where physical and cognitive dissonance and estrangement can merge. The fictional space, which can oscillate between high literary value and the poetics of the mundane, is built to be receptive of speculative practices.

Another good example of epistemology's uncertain relation to ontology is gravity. Gravity, despite the existence of a variety of ways for quantifying and calculating its presence, and the best efforts of scientists from Newton to Einstein, remains very much a mystery. While General Relativity is successful in demonstrating all the properties of gravity known in classical physics, what works in classical physics does not fare as well in quantum theory despite the efforts of Matvei Bronstein and other physicists studying the potential in quantum gravity.

Physicists have yet to succeed in making direct correlations between gravity and quantum mechanics, or to conclude with certainty as to whether they would (or should) be able to do so, thereby rendering existing models of gravity on the side of speculative even when there is no question of gravity's technological utility. In other words, the Standard model of quantum physics, as it stands, is unable to provide the framework for integrating gravity.

The story of gravity parallels the story of the Higgs boson up to 2012, when commensuration between theory and experiment was realized for the latter. However, the main difference is that the situation of the Higgs boson is resolvable within the scales of electromagnetic, weak, and strong interactions that are referred to as the gauge interactions of the Standard Model. All of these interactions are further described in Appendix A.

My specification of speculative physics is an attempt at locating the point of speculation through the locus of interpretive practices that go hand in hand with demonstration through a plethora of data; speculative physics is situated at the cutting-edge of theory construction and experiments in quantum theory, and on unstable ground due to the continuous coupling and decoupling of knowledge and concepts, with unions made, unmade, and remade with each new insight, discovery, and more developed mathematical formulation. Speculative physics wends its way around unfamiliar ground with the goal of breaking out from the shackles of epistemic comfort, bringing about revisions to one's scientific presuppositions while maintaining logical rigor.

By being attentive to how knowledge is always at that point of becoming, what appears factual can be dissolved, reversed, and then re-produced as a leap into the ontological, as Peter Gaffney notes in his introduction to *The Force of the Virtual: Deleuze, Science and Philosophy*. This comes about through the making of new judgment as attendant comprehension increases. Gaffney goes on to add, in the same manner of argument I made in the preceding paragraph, that science does not escape the logic of becoming because its object is not a static world, but one always undergoing qualitative change (6-7). The laws in physics are not separable from the epistemology governing the physics, and in fact, enable the categorical schemes that allow the classification of physics objects and interactions.

While there are correlations between speculative physics and the concept of mangle of practice advanced by Pickering, I insist on the importance of emphasizing the epistemic distinction between theory and experiment that I do not think Pickering recognizes, even if I share in his principle of accommodation and resistance as the dialectic of interactions between material and human agency. In fact, this interaction between theory and experiment, which is taken to another level by Karen Barad in *Meeting the Universe Halfway: Quantum Physics and the Entanglement of Matter and Meaning* through her invocation of the inseparable relationship between the real and agential, is intrinsic to the discourse of speculative physics since the interaction involves the narration of fact-building that can be altered and dismantled through the dialectical relationship between the material and their human tinkerers.

A theory, as it stands, is always partial in the perspective it inscribes. This has to do with the fictitiousness of theories, theories that represent an extended narration of models built from a physical phenomenon that may exclude other existing elements not already part of an advancing body of theorems, laws, or principles. According to Collins and Pinch in *The Golem*, there is a need to understand the different strands of thinking and configurations that go into the constructions of theories and experiments for the manifestation of certain properties. The authors were looking to examples across the physical and biological sciences, and discussed what were presumed to be controversial aspects of all the experiments examined (in the case of physics, examples such as relativity, cold fusion, gravitational waves, and solar neutrinos are used as illustrations), the stakes involved, the players and their standing in the scientific community, and the flaws in the results of the experiments.

Nevertheless, given that the demarcation of our knowledge is imperfect, the outcome of the experiment is also imperfect. The experiment may embody a constrained access to ontology, with whatever data present, subject to constant denaturing and reconstruction. The scientific method does not determine what should be the object of an investigation. Instead, the choice is for the investigator to make a compelling case for further scientific exploration.

By the premise of *The Golem*, the theoretical foundation of the science can only be as complete as the observables and measurables that go into the construction of the science, if only because the foundation is built of knowledge that has achieved a measure of recognition, is somewhat established, and experimentally verified. Anything that lies outside the measurable is undetectable, and therefore epistemically void.

On the other hand, in the area where the detection of the material evidence is possible, the testing of various 'strands' of the theories, such as the different forms of experimental confirmation, falsification, and re-testing are what contribute to the "strong rope" of epistemic coherence. The data is there but our interpretation is represented by what we can perceive from the data. Even our experimental controls, such as data produced through parameterized simulations (the Monte Carlo), operate through the extrapolations and interpolations provided by the scientists. This is where I suggest the reconfiguration of the role (and capability) of existing experimental practices through more speculative interrogation.

For instance, let us consider a high-energy physics experiment, ATLAS (A Toroidal LHC Apparatus). A slight variation in the apparatus, such as in its detecting sensitivity, can make a difference in the instrument's ability to measure the different striations and demarcations of a physical phenomenon. Therefore, the sensitivity of the detector plays a role in obtaining or obviating the epistemological (though not the ontological) existence of a physical entity through the collection of measurable and observable evidence.[4] The data could later be used for defining the parameters of a simulation used as comparative control and for generating theoretical predictions that can be reconfigured again after further analysis. The detector, indirectly, becomes the source that would influence the decision to be made with regard to the aspects or segments of the data chosen for further review. The CMS (Compact Muon Solenoid) experiment, a direct counterpart to ATLAS, has a similar setup and goal, but with some differences in the design of the sub-detectors, thereby producing variations in how data selections are procured at the point of collection.

Each of these experiments takes a different strand of the data, with the sub-strands representing the cross-sections being analyzed. From these factors, the different cross-sections of a collision are determined by analyzing the data collected through the different channels enabled by a multitude of sub-detector trackers in each of the experiment's detectors. Cross checking between these two experiments produces a more 'complete' picture of the same events that were initially measured individually. All the different 'live' and simulated data analyzed would then converge to create the metaphorical 'strong' rope referred to by Collins and Pinch.

---

[4] In a quantum system, observables are not necessarily measurable if we intend to constrain the measurement to the parameters of a particular framework that may not be able to articulate the range of properties required for the measurement of a particular event, or to produce quantization of the event. Gravity has been articulated as just such a case in the Standard Model of particle physics.

At the other end of the 'experimentation' spectrum, in specialized fiction genres such as science fiction, are conflicting views of how far one can push experimental boundaries, and if the process of estrangement itself should not deplete the constraints of natural logic, as is found in the opinion pieces/essays of hard science fiction journals such as *Analog Science Fiction and Fact* and the book *Time Machines: Time Travel in Physics, Metaphysics, and Science Fiction* by Paul J Nahin, regardless of the amount of defamiliarization the writer employs. Therefore, if one desires to produce a series of speculations (or forecasts) that approximate most plausibly what can happen in the future as long as certain conditions are met, whatever the level of difficulties one encounters in meeting those parameters, science fiction, in whatever medium or form it takes, is the engine for speculating about science.

I argue for the importance of considering speculative theory as separate from speculative experiment. While speculative theory can be the point of origin for theories, conjectures, and lawful (or law-like)[5] explanations to develop, speculative

---

[5] In *Laws of Nature: Essays on the Philosophical, Scientific, and Historical Dimensions*, edited by Friedel Weinert, he provides a comprehensive introduction to how one would classify laws, such as real laws and 'pseudo' laws, demarcating the 'lawful' as representing true natures of laws and the 'lawlike' as

experiment focuses on the negotiation of a predictive theory, regardless of the formal outcome, while building on experiment-based theorization that brings about its own set of predictions. The Higgs boson of the Standard Model that supposedly provides the verification needed for the validity and justification of the model catalyzes the relationship between speculative theory and speculative experiment so that these critical methodologies can converge to a richer description of speculative physics.[6]

However, to speak of speculative physics in generality without accounting for the differences one can find in theory and experiment would be to ignore the important factors permeating their differences, including the necessity of describing the methodology of speculative physics from the individual frameworks of speculative theory and speculative experiment in order to produce a complete picture of the role of speculative physics for elucidating the subjects discussed, as well as the goals of trans/interdisciplinary thinking, for rigorous interpolations between the humanities and the sciences. The next sections will build on what

---

having the appearance of exemplifying the essence of nature but are merely manifesting certain 'accidental' qualities in a regular fashion, a form of 'pseudo-law.'

[6] For a comprehensive and concise historical, and scientific, overview of the Standard Model, interested readers are encouraged to read Appendix A. While I will reference relevant details concerning the elementary particles of the Standard Model throughout the dissertation, I will not delve too deeply into their backgrounds as I intend to let the ensuing critical arguments illustrate the relevant physics. Appendix A contains all the technical information underlying this work and should provide adequate background information to the reader.

speculative theory and speculative experiment are about. Then, we will consider how the conceptual thinking behind speculative experiment is applicable to speculative science fiction.

Recently, I conducted a survey with a small posse of particle physicists to assess their relationship to the Standard Model through questions on the unification of forces, potentials beyond the Standard Model, and the influence of external push-factors on their work with physics. Most importantly, the purpose of the survey is to venture into their cognitive space and understand better how they think about concepts such as precision and ambiguity, certainty and uncertainty, potentiality and constraints, theory and experiment; as well as the language used to articulate these dichotomous continuum; bringing about investigations into the sociology, history, and philosophy of physics that unveil the nature of knowledge-construction. I include summarized analyses of responses that are relevant to the discussion in this dissertation.

Foremost, language is very important to a physicist as it marks the main differences between new age spiel and scientific rigor. Physicists are particularly insistent that the choice of descriptors used to portray their thinking and work should present a physics that is delimited by the scientific method: a euphemism for taming unruly and capricious Nature. Just as we need rules and routines to give sufficient calm and order to an otherwise frenetic life, physicists need to have the right language to organize the data they collect, and to work out the expanse of uncertainty in relation to what is known. They want precision in the terms used to deal with the malleability and ever-changeability of knowledge at the frontier. Moreover, they do not like to be perceived as ideologues, and are thus ambivalent over the use of words such as "belief." For at least one of the respondents, "belief" has a connotation of "faith" rather than "fact."

When queried about the range of speculative contingencies, for most of the physicists, the speculative aspects of any part of a knowledge merely illustrate the need for more direct evidence to justify the available theories, including corrections that are to be added to the original theories for closer approximation to 'real' data. The accounts of the speculative, and the multiple experimental triggers that are the outcome of significant choices, are all wrapped around the symbolic signification, and delimitations, produced by telling the story of the particles of the Standard Model through abstract algebra and algebraic geometry (group theory included): what mathematical terms get extended and what get dropped whenever physicists consider what it is to move 'beyond' the Standard Model. Even the new physics they envision is shaped by the status of the mathematics deployed for representational purposes, including the implementation of particular mathematical subfields for resolving long-standing paradoxes or unresolved cul-de-sacs.

Additionally, physicists are not always the most united when it comes to believing in what is possible among the not-yet-existing, even if they might possess the same basic knowledge of their fields. This lack of unity can be attributed to differences in how rules and laws are applied to solve a problem (and we know solutions to the same problems are far from homogenous), and how one thinks philosophically about the problem. For example, when asked about the meaning of subjectivity, one gets answers ranging from interpretational preferences, expectation biases, forms of contextualization, to the method of decision-making; all governed by external factors other than the science. Seldom do the physicists consider the data they work with as inherently subjective (even if the constitution of the data itself is not) despite the preference for the logic of one model over that of another due to myriad circumstances that are not usually logically linear.

But, in the case of objectivity, most of the answers appear standard on the surface until one digs deeper; then, one finds responses such as the application of clear and inclusive criteria to the acceptance or rejection of data before applying that same data to the testing of a hypothesis, the repeatability (or as some philosophers would say, the cycle of falsification and justification) and duplication of that experiment, as well as the exclusion of bias in the decision-making process. The attainment of objectivity is a recursive move for the first two instances, even if their starting points can have a range of subjective possibilities. As for the third instance, one finds that there is no clear way for measuring bias other than by tracing the entire process from hypothesis to the final interpretation of analysis. However one might reason, bias does not cease to exist even after an experimental fit to theory, nor after an experimental setup is found to be replicable.

Finally, there are the different backgrounds and trainings of the physicists, boiling down to whether they are experimentalists, theoreticians, or phenomenologists (the latter third being the group bridging the divide between the first two groups). While they emphasize the importance of locating empirical evidence to support whatever predictive theories there are, they do not necessarily have the same level of investment when dealing with more speculative forms of physical theories. While a theorist is more willing to wait until the boundaries of knowledge broaden while continuing to work on theories that have not yet demonstrated experimental promise, experimentalists tend to be more skeptical, even if not necessarily dismissive, because of the need for attention to expediency and practicability in order to succeed in everyday science. However, both sides agree that experimentalists and theoreticians are able to find middle ground to work from.[7]

[7] As the initial survey done is merely a test in preparation for an actual field-survey to be performed as a postdoctoral project, I will not be including the former as a material reference in this dissertation.

## 2.1 Speculative Theory

Before going further, I need to clarify the crucial differences between speculative theory and other theories that appear similar, on the surface. Firstly, speculative theory is more than a hypothesis, with the latter based on the idea that theory-induced predictions are made of axioms that combine established facts and theoretical presuppositions. Even if a speculative theory has begun life as a hypothesis, it will later develop a more extensive narrative life, and a more complex explicatory model, for constructing mechanisms that can produce scientifically causal explanations about the spatio-temporal structure of the model. Moreover, in speculative theory, alternatives to available explanations are considered so that one is not fixated on a single hypothetical problem. The best way to sum up the meaning of hypothesis in relation to speculative theory would to think of the level of convolutedness (or messiness) that each is expected the hold, with speculative theory being more complexly constructed.

A hypothesis may be falsified after different attempts have proven it to be impossible or unverifiable. Or, a hypothesis can become a speculative theory after a series of examples either falsify or justify the former due to either the theory's underdetermination or evident potential for connecting to other established or equally speculative theories. However, the underlying ontology of a speculative theory is not materially reconfigured. In speculative theory, the reconfiguration of epistemology means getting, as close to ontology as possible, and for getting to the bottom of theoretical proofs within existing frameworks while keeping in mind that serendipity can open doors to the unexpected. 'Exotic' predictive models such as string theory, the supersymmetry, and dark matter are examples of actors in speculative theory with attributes that can only be partially explained within the currently knowable while waiting for the emergence of paradigms that can finally unveil what was hidden. The abovementioned exotic candidates have origin stories similar to that of Higgs boson, only that the latter's narrative is more developed despite containing degrees of indeterminacy. Nonetheless, each of them occupies different levels of relationality, and narrative depth, within speculative theory.

Speculative theory can be viewed as a stage ready-made with props and designs, and cast members waiting at the wings to make their way center-stage. The stage is set to perform the script even when the plot does not appear to be progressing linearly. But such is the nature of speculative theory: the narrative of the subjects and objects are reconfigured, with roles switched around, and original cast members and props (differently predicted features of a theory) potentially eliminated or substituted. For a hypothesis, the plot is underdeveloped without identifiable protagonists (or antagonists). There is a greater degree of similarity between speculative theory and underdetermined theory, whereas the difference is related to speculative theory's strong connection to the uncanny and the unfamiliar (with a will towards defamiliarization) that an underdetermined[8] theory does not possess, because the latter mostly represents that point of explicit connection between physical cause and effect.

Speculative theory differs from an underdetermined theory because the former is less about the uncertainty of correlation between cause and effect; rather it is about linking cause-and-effect with epistemically malleable 'placeholders' that can temporarily hold hidden variables as the latter's level of significance to the epistemology is assessed. Nevertheless, an elegant universal theory to explain away the physical paradoxes is not yet attainable given that there is no open access to ontology.

Above and beyond the concept of underdetermination, speculative theory overlaps into speculative model even when the former exceeds the latter; one does not prove definitively the success or failure of a theory until better alternatives embodying the successes while also accounting for the failures are unequivocally demonstrated. The epistemics of the Standard Model is such that it interleaves between underdetermined theory and speculative theory. Regardless of the current state of the model's causal ontology, there is sufficient evidence to assure the physicists that the model is formulating some sort of prediction.

---

[8] In the theory of underdetermination, it is suggested that the evidence available to us with regard to the verification and proof of any scientific theory cannot provide sufficient answers to the scientific beliefs that we should adhere to. Pierre Duhem, a French physicist and philosopher, and W.V.O. Quine, one of the philosophers from the Vienna Circle, had been the pioneers of the work done in this area. For a general overview, check out the entry at the Stanford Encyclopedia of Philosophy <http://plato.stanford.edu/entries/scientific-underdetermination/>.

Speculative theory can be developed to a level that is able to provide contingent explanations to observable phenomena or objects with indeterminate provenance, at least until further evidence emerges to provide more extensive explication of the objects. Even experimental verification can only be partial because the development of a new paradigm to account for the inexplicable takes time to surface, and existing technologies are limited in their capacity for producing conclusive demonstrations. Above all, the resilience of speculative theory is not dependent on a single or a series of 'short-lived' experimental proofs, as resilience is determined by how long a theory remains as the best explanation for a particular event or phenomena.

However, one has to be aware that resilience is not the same as stability; a theory can be resilient while unstable at the same time, which is the underlying premise of speculation. To draw again from the example of the Standard Model, I argue for the importance of resiliency in the model for predicting the existence of particles that can later be verified empirically despite the lack of ontological clarity, and that the recovery of initially undetectable physical clues will destabilize its dominant epistemological structure. In other words, the model should be flexible enough to accommodate major revisions should new discoveries render the current model too messy and complicated. We can think of a model as theory in the making that has not been pushed to the maximum of its descriptive powers.

I have discussed the double-slit experiment and the paradox that is exposed in terms of the ontological framework of theorizations when making the choice between a field and particle for representing quantum events. The preference for either framework can inform the perception on the paradox of wave-particle duality, as well as foreground the phenomenological and hyletic events that implicate the choice.[9] The decision will impact the measurement of objects being theorized; either way, there would still be common factors that connect them even if each of the factors follow different paths depending on how their interpretive preferences are shaped. Each path converges into a composite representation of the factish (an extension to the definition of the 'factual' developed by Stengers that exists autonomously regardless of the changes in its externalities because the factish departs from the concept of facts being never absolute; in a manner of speaking it is

---

[9] Here, I am taking from Husserlian phenomenology that posits structures of consciousness stemming from the sense perception and intentionality of the first-person observer. As Husserl himself puts it most succinctly in *The Crisis of European Sciences and Transcendental Phenomenology* (Evanston: Northwestern University Press) 1970, p. 76 "The lowest stratum of all objective knowledge, the cognitive ground of all hitherto existing sciences, all sciences of 'the' world, is, what we can say, for the first time, called into question in the manner of a 'critique of knowledge.' It is experience in the usual sense which is thus called into question, 'sense' experience – and its correlate, the world itself, as that which has sense and being for us in and through this experience, just as it is constantly valid for us, with unquestioned certainty, as simply there [*vorharden*], having such and such a content of particular real objects [*Realitäten*], and which is occasionally devaluated as doubtful or as invalid illusion only in individual details." Hyletic here refers to that form of raw, pre-processed data that precedes any intentional acts related to it.

Stengers way of laying out experimental and theoretical facts such as I am doing with speculative theory and experiment), where agency and dependence on interpretations occur simultaneously, thereby composing the material profile of the entity, which is the scientific subject of interest.

Another crucial factor to account for is realism. Realism, in the philosophy of physics, contains multiple levels of explication, with its qualities the subject of exhaustive debates among eminent philosophers from Isabelle Stengers to Bernard d'Espagnat. However, there are two points on realism that are vital to the interrogation of theory's materiality, including speculative theory; realism, which is the existence of an object/subject independent of any thought or acknowledgement of its existence that is external to our perception and epistemic frameworks; and realism that is a mathematical simulacrum that influences, within defined epistemic positions (operational, empirical, subjective, or objective), what interactions are useful for providing ontological clues. Moreover, investigation into realism begins at that intersection between macro and microphysics, mainly because the familiar is rendered alien in microphysical reality.

In entering the world of quantum objects, the language of the real becomes anti-real, and one encounters difficulties in forming a clear distinction between representation and reality. Language becomes the epistemic translator between an actual being (scientific subject) and its representation (mathematics and symbolic logic). Therefore, it becomes the physicists' prerogative to decide if they prefer instrumentalism (that is anti-realist in its approach) or realism.

Nevertheless, at the level of the interpretive, there is always a choice: a choice of whether realism should be closely connected to an operation that attributes value and meaning to it, or if it has to be independent of any recognition to its value. This becomes crucial when we take into account the intervention undertaken through acts of measurement and observation. How do we decide on the level of precision for quantifying that observation: how is the measure of precision dependent on the theoretical framework; does the ultimate objective of measurement define the means or are the means more important than the end? For realism to have a definable role for exacting precision during a measurement process, we have to decide whether the former has to be embodied by a particular standard of physicality that is then mathematically approximated, or if there is another more extended narrative of realism that can be enacted through the deployment of alternative modes of comparisons.

Even as the philosophy of physics theorizes realism through means available to quantum mechanics in order to explain the disconnect between the physical states and their observed behavior, as well as the phenomena they exude,[10] the materialism

---

[10] For a more detailed discussion of reality and realism from within the lens of quantum physics, I suggest reading chapters four to ten of Bernard d'Espagnat's *On Physics and Philosophy*. Princeton & Oxford: Princeton University Press, 2006. Most of the discussions on reality and realism, that gesture to,

that embodies quantum-level interpretations requires distancing from the 'original' atomistic materialist tradition instituted by Lucretius and Democritus because the former's relationship to atomist materialism is shaped by different understanding of physical phenomena. Even so, the materialism is directly connected to the physical real, as well as to what is physically applicable and producible.

The subject, a corporeal entity that is the focus of the physical framework of choice, is also temporal; however, the relationship of time to the subject is dependent on whether one investigates the subject from a classically bound or quantum physically defined framework. In the former case, temporality is macrosopically invariant. However, in quantum physics, time is mostly a metaphysical consideration, despite renewed interest in understanding time's role in quantum processes. Such an interest can lead to renewed consideration of the relationship between time and quantum objects, with applications to quantum informatics.

### 2.1.1 Speculative Theory and Mathematical Embodiment

Beyond thought experiments and conceptual developments, speculative theory is primarily configured mathematically. Moreover, the mathematical choices of physicists play a large part in their interpretive vocabulary. For example, a field

---

and acknowledge the issues that were raised in the Bell Theorem, are considered to be representing problems of entanglement, unequal distribution, and incommensurable correlations between observation and measurement.

theorist may decide whether he/she wants to use an algebraic field (through manipulations of complex variables and set theory) or the Lagrangian[11] (an extension of the classical model) field theory, dictated by the level of formalism desired. (Mackinnon, "Generating Ontology" 45-6)

Nonetheless, physicists have been oscillating between different mathematical approaches in working out the foundational issues of quantum physics, especially when confronted with the complications induced by infinities (a problem that arises as we begin calculating the relativistic interactions of particles in quantum mechanics). Renormalization, as a self-consistent algebraic technique used for reconciling long-range with short-range interactions and 'normalizing out' the infinities (divergences) arising from the equations of perturbation theory for waves, is an important tool for ensuring that the descriptions of the Standard Model interactions do not become mathematically unfeasible.[12] At the same time, renormalization requires the quantization of interactions set within the computational framework of quantum mechanics. Gravity's inability to deal with that requirement has rendered it inoperable within the framework of the Standard Model. Even as the Standard Model is already speculative in the epistemic construction involved in the refinement of the model, the paradox of gravity increases the instability of the model. Nevertheless, the resiliency of the model has yet to be challenged to the fullest.

---

[11] The Lagrangian mechanics was derived from classical mechanical application of the variational principle of action whereby the path traversed by macro or micro objects in terms of energy or momenta change, including the conservation of these two quantities, are traced. Lagrangian mechanics has since gone from a Newtonian classical framework to application in quantum mechanics and quantum field theories for the calculation of more than one paths traversed by a single particle (or a multiplicity of possible paths traversed by a multiplicity of particles), so that the amplitude of probability for its final condition can be calculated.

[12] Perturbative mathematical theory arises from the need to 'simplify' an otherwise impossible-to-calculate mathematical 'mess' that constitute the quantum system. However, there are limits to the former's application as it assumes that the particle i.e. electron involved to be free (not bounded). While used in weak interactions, it cannot be used for approximating calculations in strong interactions such as that between the quarks and their respective gluons.

A good example of a type of speculative theory that derives much of its predictive and descriptive function from mathematical formalism is the theory of supersymmetry, which has strong theoretical backing despite the lack of conclusive experimental evidence. However, we will get into a more thorough discussion of this in chapter three.[13] At this point, I merely state that supersymmetry postulates a new symmetry in the universe that is derived from balancing between the fermions (most of the particles in the Standard Model fall into this category) and bosons (Z, W, Higgs), with the intent of alleviating several problems, and observed anomalies, within the Standard Model.

---

[13] Supersymmetry here refers to a partner, or shadow, of the existing symmetry structure of the Standard Model by positing that, for every particle that exists in the Standard model there is a 'superpartner' located half a spin away from the particle (hence every boson has a fermion partner and vice-versa). Supersymmetry particles, when finally experimentally demonstrated (experimental groups in CERN are working hard on that and are noting every possible evidence that will help them construct the picture of the existence of these superpartners), will bring particle physics a step closer to going beyond the existing Standard Model. See Lykken, Joseph D. "Introduction to Supersymmetry." 11 Dec 1996. **arXiv:hep-th/9612114**. One can also look into standard texts in particle physics for more basic instruction.

Abstract algebra undergirds the discourse of the Standard Model (and beyond), including the problems of unification and symmetry. Particular mathematical methods are chosen for their capability (and potential flexibility) in maneuvering (and transforming) very complex computation required to describe any aspect of an embodiment of a microscopic universe; however, the methods can only be the best representation of the elementary entities (representing the different fractions of an atom) making up the Standard Model.

Atoms and subatomic particles are extrapolated from observed phenomena of nature that do not appear to be consistently described by the 'rules' and 'laws' of the macroscopic. The appearance of inconsistency is the outcome of a problem whereby the 'symptoms' of a microscopic state are described in a language developed to deal specifically with the materiality of a macroscopic condition. Even the mathematical functions that are shorthand for producing thorough analyses of physical events that cannot be verbally explicated can only demonstrate what is macrophysically visualizable from the microphysical; visual mathematical diagrams, such as the Feynman diagram used for demonstrating the interactions between the particles in the Standard Model, can contain couplings of highly complex algebraic sets that have been simplified for more efficient computation of subatomic level interactions.

It should be mentioned that the details of the physical states are folded onto mathematical factors that have been coupled together, therefore hiding most of the nuts and bolts when all we see is the final embodiment of a derived equation. Moreover, the interleaving of multiple mathematical factors (of constants and variables), constitutive of equations used for describing the physical states, are not reversible to their original selves if only because, during reorganization, the original elements become completely transformed and considered 'negligible' under the new couplings, with the higher order terms in an expansion series dropped along the way, therefore making a reversion to the original impossible.

These eliminations are calculated in order to simplify an otherwise tangled mess of equations. The transformation requires an epistemologically informed decision-making process with regard to what can be included and excluded. While some of these eliminations are trivial, they can open up questions pertaining to whether the elimination process means a permanent exclusion of certain theories for good. Therefore, to understand what is at stake, and what may already have been excluded in the process of transformation, one must consider the precise point of that transformation in order to estimate, better, the areas of interest in relation to the areas excluded. This is important for providing an informed, and coherent, explication on why certain interactional processes behave in the way they do.

In addition, mathematical formalism images theories of the quantum in juxtaposition to the actualities of the world, and is therefore a useful medium for interrogating established facts; for destabilizing dominant epistemic assumptions by having us consider propositions that are physically counterintuitive (yet mathematically logical) at a glance, but which would make sense down the road as our knowledge of what is out there grows.

However, we have to remember that there are as many elements of irregularity in nature as regularity, with not-so-easily decipherable logic other than that they all converge into a pre-determined physical state. For instance, some physical states (momentum, position, energy) may or may not commute (commute here refers to when the ordering of two operations, i.e. measurements of a particle's properties, does not matter), depending on the qualities measured, or on whether they can be measured simultaneously (without neglecting Heisenberg's Uncertainty Principle).

While there is a pattern of behavior and organization among physical observables that mathematics tries to illustrate through the symmetries of the Standard Model, there are also micro-entities that are beyond the reach of observation because of the technological limitations of the measuring instruments. The incompatibility between the ontological and the observable is the reason why one cannot have an absolute standard for deciding, arbitrarily, as to what stands for the regularities and irregularities in a system. What the symmetrical features of the Standard Model (and the breaking of the symmetries) provide are the mathematical and visual tools for approximating our conception of nature's particles and fields.[14]

## 2.1.2 The Activism of Speculative Theory

Speculative theory exists at the highpoint of activist philosophy whence speculative experiment can later emerge. Activist philosophy is informed by the notion of *radical empiricism* whereby all that is experienced is real in some way and everything that is real is also experienced (Massumi 6). In the real of the experience, there is also the problem of cost, and the cost is not only in the material (as well as financial) cost, but also the ontological and epistemological costs that are the outcome of making choices to activate specific nodes: the nodes consisting of theoretically-led experimental design, observational priority in the process of data collection, base level calibration of instruments, affective engagement of the experimenter with their team of collaborators as well as accelerators and detectors, the organization of authority and decision-making concerning the release of data, and the naming process (including the 'gender of the name') of recently discovered particle/physical phenomena or constructed instrument.

---

[14] It is important to consider these particles as 'imaginary' and an abstract conception of how we imagine the building blocks of the universe to be, rather than to take them as literal corpuscular entities.

When activist philosophy provides the schematics for comprehending scientific knowledge production in terms of epistemic (and ontological) costs, the activated nodes can represent the intensity that stems from navigating dense concepts located in the concrete and the abstract, as well as the different epistemic responses that might arise from the different approaches for getting at the ontological structure of nature. An example of such an activated node is the point where two (or more) events representing two (or more) arrows of time can merge into one, as and when it becomes convenient and possible to do so.

Another more routine example can be found in introductory particle physics textbooks on relativistic kinematics discussing the center of mass of the particles and the laboratory framework of reference, where single to multiple interactions are framed at a specific instant or focal point. Such contextualizations represent a shift between the perspective/'viewpoint' of the particle and that of the laboratory's human or non-human observers. Agency is situated at that intersection between the two, the duration of both inextricably bounded by the interacting particles that remain invariant. This agency is also at the point where "practice becomes perception" (Massumi 11). The relationship between the objects of theory and experiment is diagrammed through the different self-reflexive and agential practices of theory and experiment.

Activist philosophy brings speculative theory a step closer to speculative experiment by accentuating the philosophical phenomenology of relationality between humans, machines, and microscopic entities at the intersection of theory and experiment. More importantly, activist philosophy foregrounds the experiential points that lay hidden in the more formalistic features of speculative theory. The activism of theory and experiment enables a more holistic presentation of realism (through materialist practices) to be constructed from the differential approaches in these two areas of speculative physics.

## *2.2 Speculative Experiment*

Before proceeding into the discussion on speculative experiment, I would like to point to the differences in the preoccupation between a theorist and an experimentalist: how the different interpretive practices from the two different camps, even if they converge at certain points, can lead to different emphases on theoretical goals and choices of frameworks. For instance, quantum field theories are unequivocally important in high-energy physics, both experimentally and theoretically, for formulating predictive models in particle physics, whether within or beyond the parameters of the Standard Model.

For a theorist, the important question is which of the quantum field theories would enable the greatest ontological access and success in prescribing microscopic phenomena. An example is the ongoing debate about which quantum field theoretical approach is the best for dealing with the complexities of renormalization: there are those who feel that an axiomatic algebraic approach has a better track record than the conventional version, while others consider both approaches as merely rival theoretical frameworks without either being better than the other.[15] Nevertheless, the choice, in the end, is about obtaining the best solution to a problem. The decision over which mathematical approaches are the most useful for interpretive purposes does not necessarily influence how experiments in high-energy physics are set-up and performed. Experimentalists, while remaining aware of the role of theories in influencing the different types of experimental design, are more concerned with the material applicability of the theory to data analytics than in sweating over which theory is more mathematically elegant/comprehensive (unless the elegance/comprehensiveness allows the experimentalists more extensive material reach).

Hence, speculative experiment can be focused on experimentally conceived choices to search for a still undetectable entity within the effective range of its existence. In a more mundane sense, speculative experiment conforms to the fields of action in a selected model (that can be based on the best likelihoods established through formal theory and statistical calculations). Repetitions may never be acquired outside of the habit, and there might be no going beyond the cycle of falsification and justification. However, I intend to extend the notion of the speculative experiment to embody the dialectical interface between micro-objects and macro-events. The millions of repetitive collisions are measured through the scattering effect; a specific choice of data 'cut' can lead to the unveiling of observables that are hitherto hidden by the presence of other material effects such as dark matter.

---

[15] See the debate between Wallace and Fraser in *History and Philosophy of Science Part B: Studies in History and Philosophy of Modern Physics*. 42.2 (2011): 116-35 on the axiomatic versus conventional algebraic method.

Firstly, it is important to highlight the difference between speculative experiment and the product of pure uncertainty in an experiment. As Ian Hacking and Allan Franklin have both pointed out, experiments persist beyond the shadow of theory and can exist independently as a solid body of scientific research even without *a priori* predictive theories to initialize the experiment. Moreover, experiments can form and produce their own theorizations strictly through the analyses of known cause-and-effects, which would then feed-back into revising, altering, or modifying existing experimental designs. The aspects of experimentation leading to the formation of theories to explain and describe the physical effects observed had always been part of early modern and pre-twentieth century scientific work in physics.

It is under such conditions of speculation that one observes the tension between measuring and observing, modeling and measuring, or modeling and observing. The speculative method in experiments allows us to think about what it means to perform measurements in experiments, the same decision-making that goes into deciding the scale and range of the measurement, and what observation comes out of a measurement that is conducted almost entirely with digital instruments today.

As the ontology of the physical structure that the data is supposed to explicate approximates ideal conditions, Monte Carlo simulation, or model building with parameters under 'perfect' conditions that preclude instrumental artifacts and other experimental errors, is produced as control data for comparison with real data. This in itself is another level of speculation as the experimentalist moves between virtual data and actual data, while attempting to detect what is contained in the gap between both types of data. In speculative experiment, there is no determinate end to an experiment: there might be a physical end but the theoretical part of the experiment does not end because the next stage of becoming is already set in motion for the next phase of experiment to occur.

Also, speculative experiment recognizes that measurements are bounded within the confines of selected parameters and is aware that experimental biases are as much informed by predictive circumstances as by the models (both theoretical and computationally simulated ones) they choose to adopt for the design of their experiments. However, the story of its empirical discovery is another interesting tale of speculative experiment that does not always coincide with speculative theory's choices, if only because experiment is more constrained by external circumstances that frame the possibility of the experiment.

In particle physics, speculative experiment diagrams and activates specific nodes of matter by unmasking an increasing array of emergent particles and sub-particles, all playing to the tune of different fields of interactions, from Newton to Coulomb to Feynman, their identities and appearances manifesting, and remanifesting, through every new theorem and quantizable phenomena. An example that best represents the activation of nodes in speculative experiment that also involves speculative theory is Fermi's theory of beta-decay, known as the spectral emission of electrons from which the original discoveries of neutrinos (particles without mass produced during the process of decay) were made.

Fermi had wanted to maintain an experimental analogy with electromagnetic theory when he found that his theoretical calculations agreed with experiment. Therefore, he maintained only the use of one out of five allowable interactions (mathematically signified to represent directed and un-directed actions) that could take place, so that the wavefunctions of the electrons and neutrinos can be considered. These are mathematically detailed through the use of transformation operators involving leptons (electrons and neutrinos) and hadrons (protons and neutrons). In fact, what differed among the five interactions would be the particular positioning of the transformative operators in relation to each other and in relation to the dimensional algebraic sets meant for denoting three-dimensional spatial elements.

The work on the theorization and confirmation of the beta decay, first by Fermi, then by others such as Konopinski and Uhlenbeck, as well as Beck and Sitte, was meant to synthesize the experimental data that had been obtained from the positron emitter and from bismuth, another form of radioactive element. Much of the theoretical fine-tuning had to do with organizing how one intends to balance the energy produced during the experiment. There was also the work by others such as Gamow and Teller, who went back to the fundamental basis of Fermi's work to demonstrate how Fermi had neglected to include certain effects. This means revisiting the earlier arrangements of the mathematical operators and then finding another interaction that works. Knowledge not made manifest in the first run of theory became available when considered from other angles, and as further experimental data were obtained. This led to a better fit with the model envisioned by Gamow and Teller than with the original Fermi theory. The process would repeat and new difficulties arise during the repetition of the experiment, so that new curves from newer data plots were produced and the theory adjusted. In a sense, the abovementioned examples represent the correlation between the real and the imaginary, with neither quite exceeding the other. But it is also representative of a speculative theory that fueled the speculative experiment, which in turn, channels the next round of experiments, and so forth.

Therefore, speculative experiment does not fear the deconstruction and re-construction of original theories and experiments to produce increasingly solid, yet still speculative, theories[16] that are demarcated by experimental choices built off the under-developed, inchoate, empirical possibilities of mathematically-driven predictions. This may mean decisions have to be made with regard to how cuts are made at points where the entities are supposed to exist while leaving a spectrum of possibilities as flexible as practicable so as to have greater flexibility for managing expectations in potential outcomes.

There are some examples connected to speculative experiment in the year of the discovery of the Higgs boson. One involves the detection of a supposedly heavy subatomic particle at Fermilab in Batavia, Illinois; the physicists had hoped that the discovery of the unknown entity could be the break they need with regard to exceeding the delimitations of the Standard Model. Among the conjectures connected to what the unusual bump detected in the plot line was that the bump could be nothing more than an artifact produced by background noise and a detector that had not been properly calibrated. On the other hand, it might be an actual lead into new physics. It was soon found that the perceived bump was an accident.

---

[16] See chapter one of Franklin, Allan. *Experiment, Right or Wrong*. Cambridge & New York: Cambridge University Press.

A similar situation happened in 2011 involving the OPERA detector in Italy and the supposed detection of superluminal (faster-than-speed-of-light) particles that would have falsified Einstein's theory of special relativity, were they proven true. But as the story went, there was a mistake within the setup of the experiment. In this specific example, one crosses that grey line between speculation and human errors. However, by allowing for any such discoveries to be located in the space of the speculative, one engages in every form of the possible, in every mangle of practice, and every question that will converge to the physical object of interest with newer insights. Speculative experiment analyzes the facts and errors of material entities immersed within its bounds, therefore allowing each fact or error to inform the interpretation of the other.[17]

An experiment, even a non-speculative one, is about finding a language for describing the observable, including the observation of unknown entities. Many of the works in the history and sociology of experiments in physics focus on the culture of epistemic production, as well as the various philosophical interpretations accorded them, depending on whether one takes on a constructivist, realist, positivist/antipositivist, or relativist view of the experiment's material practice. The historiography of the accounts constitutes an examination of the correlation between the institutional and disciplinary discourses that shape the choices made on acts of social-epistemic agency (à la Andrew Pickering) such as deciding upon points of measurements; tracing the emergence of intentionality in the interaction between the subject and object in science; analyzing the material traces left behind due to the interactions between human 'experts,' instrument/machine, as well as the data produced; and finally, dividing-up between the concepts, facts, and representations that are discussed extensively in the social life of the natural sciences.[18]

---

[17] These are the two most accessible articles that address the problems denoted by the example I have mentioned here. For more on the Fermilab incident, check out Maugh, Thomas H. "Research Points to a Fundamental Change in Physics -- or Else a Fluke." *LA Times* 7 Apr. 2011. [Accessed 20 August 2013] < http://articles.latimes.com/2011/apr/07/science/la-sci-anomalous-physics-results-20110407> and for more on the OPERA story, check out Strassler, Matt. "OPERA: What Went Wrong." *Of Particular Significance: Conversations About Science with Theoretical Physicist Matt Strassler* 2 April 2012. [Accessed 20 August 2013]. < http://profmattstrassler.com/articles-and-posts/particle-physics-basics/neutrinos/neutrinos-faster-than-light/opera-what-went-wrong/>.

Among the speculative aspects of experimental practice (especially when one thinks of particle physics) that differ most significantly from speculative theory is the design of experimental decision-making for conducting speculative searches: these searches could be theory-driven as well as independently experimental in that they draw upon the analyses of data from earlier experiments to produce theories with origins in the experimental.

---

[18] Among the works that come to mind are those by Peter Galison, Andrew Pickering, Trevor Pinch, Harry Collins, Karin Knorr-Cetina, Bruno Latour, and Imre Lakatos, among others. Many of these works address the contours of material and relational agency (that could imply autonomy, epistemic determination, intentionality and intra-reaction), transmissions and dispersions, the politics of hierarchy ranging from organizational layout to experimental choices (whether working towards demonstrating a specific theory or using a particular instrument), specific constraints and parameters of an experimental research program (that includes comparing between the results of predictions and experiments while trying to understand the deviations of the latter from the former), collaboration and integration (or intercalation, to borrow the term from Galison's *Image and Logic*). There were also attempts at delineating some of the problematics relating to how experiments are demarcated and situated within the current experimental framework of practice.

In experimental particle physics, professional physicists and graduate students are working on counting, scanning, and detecting traces of still invisible entities with the hope of providing further refinement to confirmed theories, and to 'discover' evidential traces that may have been lurking beyond the theory. With the spectrum of possibilities waiting to be uncovered at each energy band, experimentalists work in concert with accelerator physicists to define best practices for increasing the possibility of obtaining 'quality' data within the parameters of the targeted searches.

## *2.3 Speculative Experiment of Fiction: Science Fiction and the Politics of Speculation*

Having discussed the role of speculative experiment, I argue that the critical theories that emerge can also be extended to imagining the role of science fiction as a medium for prototyping highly abstract and philosophical ideas. The idea of the science fiction prototype is not new, for it has been used by futurists, scientists, engineers, information technologists, artists, and even market forecasters as a way of imagining a world that could-be as well as could-have-been.

In his book, *Science Fiction Prototyping: Designing the Future with Science Fiction*, Brian Johnson David discusses how science fiction, in popular culture and fiction, has stirred the imagination of those who then decided to make it their career to be involved in the creation of new possibilities for the future. He provides examples of technological forecasts, such as in robotics, that are constituted within a chain of causal what-ifs for extending the spectrum of possible occurrences through the introduction of seemingly innocuous (or sometimes more radical) variables. Science fiction, for all its tension in dealing with science, and thus, the facticity of legitimized knowledge, is a comprehensive medium for social, political, and technological challenges.

In considering science fiction as part of a "datum in that act of experience," I am partial to Whitehead's critique of the doctrine of subjectivist principle where he insists on the distinction between universals and particulars, and how the act of experience can only be adequately parlayed through the universals. Even so, this does not mean that particulars have no place; I argue that the particulars are important for localizing and situating the experience while the universals allow a grander extrapolation of the experience to produce a global fit that is made of a series of particulars standing for series of experiences. In addition, Whitehead's discussion of the relationship between causal efficacy and presentational immediacy in relation to symbolic reference is important for comparison between perceptions and sensations in relation to representation – where the comparative relationships are simultaneously antipodal and inter-connected. Such relationships are part of the building blocks informing the universals.

All of the relationality discussed is part of a phenomenology that travels the routes of the past and the present to gather causality and immediate sensations into a state of affectiveness surpassing the parameters of linear logic. They are part of Whitehead's attempt at developing a philosophy of organism that can surpass the limits in philosophy of science. If we think about science fiction as the creation of an "inorganic world," we can then consider science fiction as that world where "causation never for a moment seems to lose its grip." Whitehead argues that what gets lost is the "originativeness" and evidence of "absorption in the present" which leave no traces behind.

Whitehead's injunction parallels Lem Stanislaw's gruff observation. In his essay "On the Structural Analysis of Science Fiction," Stanislaw claims that science fiction is a chance mistake that defamiliarizes through the play of inversion. For him, science is a knowledge that can be realized and set in distinction from logic, even if such a distinction may not always be possible. Further, science fiction is ontologically different from the rest of the world. In the *Dimensions of Science*, Bainbridge states that science fiction is an imaginative interpretation of science and technology, communicating to a wide audience, ideas guiding the future of our civilization (4). It is my intention to consider how science fiction can become a platform for thinking through the fundamental theories of physics as well as for working out other scientific possibilities. As literature, science fiction produces a critique of politics as they come into play in science; the politics of science colliding with the social aspects of the real world, and the ensuing results are viewed in relation to scientific epistemology. In other words, science fiction is a mixture of speculative physics and speculative politics, and there is no better way to introduce a concrete illustration of this than by Alan Lightman's *Einstein's Dreams*.

*Einstein's Dreams*, based entirely on thought experiments represented as dreams, is focused on the inhabitants of Berne, Switzerland. Lightman's novella, with the exception of the occasional appearance of Einstein and his friend Besso, is centered around the destinies of various, seemingly random, individuals whose lives were intricately connected to the flow and movement of time; forward and in reverse, as well as across infinity or a lifetime truncated into a single day, with the looping of time and different configurations between time and aging/rejuvenation. The multiple plotlines, connected through the larger theme of space and time, configure interactions of space-time in the parallel universes of our world and another world with amplified traits that are not visible to our world. Sociological and scientific practices become entangled and inseparable, with memories bound to the actions that are to come.

In a dream sequence that continues daily, the lives of the individuals take on the shape of particles interacting with each other within a closed environment, an 'enworlded' space.[19] Hence, the mostly 'invisible' observer, the active dreamer who is Einstein, becomes the experimenter who tries out different permutations of time within the framework of time-dilation, of relative time between two inertial frameworks and the curvature of space and time, as well as the problematization of 'lifetime' from both humanistic and scientific contexts. As Barad argues in *Meeting the Universe Halfway*:

> …It's not that the experimenter changes a past that had already been present or that atoms fall in line with a new future simply by erasing information. The point is that the past was never simply there to begin with and the future is not simply what will unfold; the 'past' and the 'future' are iteratively reworked and enfolded through the iterative practices of spacetime mattering – including the which-slit detection and the subsequent erasure of which-slit information – all are one phenomenon. (315)

The novella begins by establishing the reader in the primary space where all these 'wild' ideas are produced: the patent office where Einstein worked. We are given a glimpse into his disheveled personal life; it appears that the Einstein of Lightman's construction was a man so completely obsessed with his dreams that normal life became a distraction and irritation. The story tracks through the entanglement of metaphysical suppositions with physical realism. The narrative began in April 14 1905 and ended on June 28 1905, flanked by a prolog and epilog.

---

[19] As is mentioned at various instances in Vivian Sobchack's *Carnal Thoughts: Embodiment and Moving Image Culture* (Los Angeles: University of California Press, 2004).

Each day, except for undated interludes when the reader is brought back to the reality of Einstein's physical presence, marks a succession of dreams where the anonymous characters populating these dreams interacted with each other in a highly mathematical fashion, even as the impact of time-imposed (or time-less-ly) imposed fates on their lives remained significant to each of these individual actants. Psychoanalysis is put in conversation with physics in the construction of parallel worlds, where three different decisions were made by a young man to, or not to, pursue a particular woman, and how that pursuit would affect his life.

The novella also contains the sequence of a man who watched himself played-out in difference and repetition, as if the events took place in a house of mirrors, except that he was multiplied across different spatial dimensions while retaining a common ontology of time through nested concatenations. In the repetition is the infinitesimal; the man found that the duration played out in his life, repeated by others, was slowly reduced to the most important fundamental entity, which is music in this case. In the novella, Lightman takes creative license with Einstein's work on electrical-technical time and foregrounds phenomenological time in the process (of the Husserlian/Heideggerian orbit) while interrogating the very notion of time and its symbolic function. He illuminates the argument by narrating the tale of two groups of people living in the same world; one who is always going after time and the other with no clue about the meaning of time in a linear, progressive sense. While we may think this to be the fictionalizing of fact in order to set up a stochastic condition, it has strong relevance to how different cultures themselves view time, and how time therefore defines the way by which civilizations rule over their lives.

Lightman turns the Einsteinian models of special and general relativity, theories that grew out of scientific thought experiments, on their head, by laying out the confluence of the mathematics with the subjective-experiential. The former's fictionalizing of scientific theories has enabled us to venture into an interdisciplinary interchange I refer to as psychoanalytic physics. In this simulated laboratory, human subjects are studied within the entanglement of physics and psychoanalysis where the maneuvering of physical realities in these different universes have repercussions on the cognitive and psychological worlds of humans living through different cosmic episodes but possessing similar biology. Lightman has limited the setting of the story to Switzerland so we could not place the story within the larger context of the greater European continent, or some other geographical location.

A brief discussion of the novella demonstrates, preliminarily, the inter-relationality between the speculative physics developed by Einstein and the social reality foregrounding interesting ideas not previously noticed. Science fiction can deploy the empirical data of the micro-world in conjunction with the analogous features observable in the macro-worlds. It can also become the site where the discrepant and incommensurable qualities of physics and literature can stand in contestation, with the potential for producing hitherto unimagined theories and new explanations to existing quandaries or paradoxes.

## *2.4 Theory and Experiment: the Dialectics*

While one cannot posit a divorce between theory and experiment, it is possible to isolate specific speculative practices that are particular to experiments. Some of these practices involve the material cuts enacted through the data collected, which include deciding which portion of the data from the trackers and scanners should be the focus, on the design of instruments to isolate and amplify certain unique characteristics of the object's profile (magnetic resonance, direction of spin, thermodynamic qualities, etc.), and the distribution of data-checks across a network of physicists among different collaborative groups who analyze the data cuts.

These data cuts include an in-depth study into the top and beauty quarks (two of the heavier quarks of the Standard Model), as well as interactions between the gamma rays, bosons, and supersymmetry particles (known also as superpartners), all of which could show up the possibilities and limitations of the Standard Model. At the pre-experimental level, one has to deal with the calibration of the sensitivity of the detectors against background activity, and the specific practices of engineering in connection with experimental physics that stem from working within conditions that are neither as ideal nor as 'free' as what theoretical physics can model.

At the same time, one must also be cognizant of the politics that choose to validate particular instrumental techniques and analytics over that of the others.[20] Such politics will be more extensively discussed in the fifth chapter, which is where I explicate on the particular experimental channels aimed at studying the interactions of the fermions and bosons that are coupled to the Higgs (as part of the process of revealing the properties of the Higgs empirically), then relating these searches to the determination of experimental triggers and data cuts.

When drawing on the work of the theorists, experimentalists appeal to the phenomenologists as the mediator. The phenomenologist would then take the abstract theories that are generalizable from the chains of connected but individual events, and use numerical methods to produce an algorithm that allows one to rescale analysis out of the generated, but with an eye cocked towards further data to-be-generated. Such speculative practices pay close attention to refining techniques that will enable them to detect predicted entities, and these entities are the populace that inhabits the space of the microphysical.

---

[20] This is a definite preoccupation with historians and sociologists of science such as Galison, Pinch, Collins, and Knorr-Cetina. Pinch discusses this extensively in *The Golem* in the explication of experiments ranging from the chemical transfer of memory in the study of earth worms to experiments that 'prove' (indirectly through observed phenomena predicted of theory) the validity of the theory of relativity to the experimental narrative of the 'unobservable' and 'theoretically constructed' entity of the neutrino. These examples represent that direct and indirect chain of relationality between speculative theory and experiments (some of which are considered speculative at the time when they were first performed) especially when one takes a closer look at the subjective relationship of the theorists and experimenters to their objects of examinations, observing that point when different elements constituting the experiments are deconstructed and put into dialog with theory.

Despite the existing dialectics between theory and experiments, there still exists a gap between the practices of experiments and theory. Primarily, there are differences in formalistic priorities: most experiments are interested in interpreting empirical results through the existing models (until they encounter a phenomenon that could not be reasonably explicated within established frameworks), or to simulate results using experimentally realizable models with hypothetical constraints (that may or may not be realizable within current technology) enacted through computational algorithms. On the other hand, theories are interested in working out interpretive models with multiple epistemic probabilities, observables, and symbolically represented ontology that can only be partially approximated and might not be experimentally realized.

Another difficulty is how one prioritizes the physical states in the explication of any phenomenon. Connecting back to an earlier discussion on speculative theory, there have been criticisms on how theory privileges the abstract illustration of physical phenomenon rather than consider a more intuitively 'physical' way for producing interpretations of physical observables without overdependence on mathematical abstraction. An illumination to this question has been provided by Helge Kragh, by way of an example in the history of quantum mechanics where the utility of theory has little to do with comprehension at the theoretical and ontological level. According to Kragh, "Max Born believed in 1923 that 'the whole system of concepts of physics must be reconstructed from the ground up' (Forman 1968,159). It was also Born who coined the term 'quantum mechanics' in a 1924 paper, whereby he dealt with the problematic translation of classical formulas in their quantum-theoretic analogs by means of the correspondence principle" (159). The inadequacy of the Bohr-Sommerfeld theory (known as the classical quantum theory) was recognized and the name of its successor – quantum mechanics – was coined at a time before anybody knew how to provide a coherent explanation. Quantum mechanics might be foundational to the formation of the standard theory of particle physics but the ontological crutch provided by classical quantum physics is still in use in the standard theory because of the prescriptive delimitations of quantum mechanics.

## 2.5 Conclusion

As this chapter establishes the theoretical motivations whence my work departs, it grounds the multiple facets of theories, at a cultural and scientific level, that will become relevant and fundamental to the discussions on measurement/interpretation, observation, and modeling in the next chapters. In these chapters, I hope to demonstrate the important connections between experimental particle physics and its theories, foundational questions in quantum theory, and science. Then, we will see how the investigations into these areas produce critical readings that can be applied to other relevant subjects.

By the time I get to the concluding chapter, I hope to have demonstrated how speculative physics can be a productive methodology for formulating novel scientific and philosophical relationships and constructs; providing new insights into the collaboration between physics and literature, by way of science fiction, for inscribing a transdisciplinarily enriched and staunchly rigorous theoretical interface constructed of physical and literary theories; producing a politically yet scientifically informed critical philosophy of the cybernetics by way of understanding its relation to quantum states and the Monte Carlo; and finally, the potential for re-reading a different form of media-technical history through the frame of a new theoretical space through the new theories that emerged out of the Monte-Carlo-cybernetic dyadic relationship. I hope to demonstrate how understanding the historical and philosophical implication and narrative of the Higgs boson can provide useful insight, albeit indirectly, to the reading of speculative epistemology in science fiction.

*…not many days later, someone came to me and expressed his bewilderment with the fact that we make a rather narrow selection when choosing the data on which we test our theories. "How do we know that, if we made a theory which focuses its attention on phenomena we disregard and disregards some of the phenomena now commanding our attention, that we could not build another theory which has little in common with the present one but which, nevertheless, explains just as many phenomena as the present theory?" It has to be admitted that we have no definite evidence that there is no such theory.* ***(Wigner, The Unreasonable Effectiveness of Mathematics in the Natural Sciences, 1960).***

# 3. Interpreting and Measuring: the Fiction of the Higgs and the Fable of Fundamental Interpretation

In his recent essay "The Garden of Live Flowers,"[1] noted theoretical physicist and 1969 Nobel Laureate, Murray Gell-Mann, reminisced about the important formative influence he had as a graduate student:

> I suffered, at least as much as other students, from an infatuation with beautiful formalism. Working with Viki Weisskopf was a most effective remedy against the excesses of such an infatuation. He never ceased to harp on the importance of 'pedestrian' work in theoretical physics and on understanding, by means of simple arguments, the physical meaning of a theory and its implications. He also stressed the need to learn about the relevant experimental evidence and to use theory to interpret that evidence and predict the results of new experimental work (111).

Were we to extrapolate the meaning of 'experiment' beyond Gell-Mann's original intention, we can represent 'experiment' as a an emerging praxis that puts physics theories into interactions with fictive models of science. I aspire to complicate the denotation of the "pedestrian" beyond a mere monolithic reference to routine scientific work that would consider metaphysical interventions into scientific questions that complicate the current debates in physical realism and materialism.

[1] Murray Gell-Mann: Selected Papers. World Scientific Series in 20th Century Physics Vol 40. Ed. Harald Fritzsch. Singapore: World Scientific Publishing Co. Pte. Ltd, 2009.

Therefore, through the deployment of the critical framework of speculative theory that I aspire to re-read the foundations of the Higgs boson prediction and discovery against the ongoing elucidation and interpretive enigma that inform the work within the foundations of quantum physics. As for what the historical and philosophical narrative of the Higgs boson can do to contribute to reconfiguring the epistemic foundations of quantum physics, or to the reconstitution of physical/scientific laws in our approach to new physics, one might consider the significance of the Higgs boson's epistemic position in relation to larger metaphysical questions in quantum physics, the questions that underlie the constitution and problematics of the Standard Model.

In a sense, the manifestation of the Higgs boson is a consequence of how the quantum physical interactions have been interpreted and identified, therefore pushing forward the development of one particular form of analysis over that of another. This goes as far as how the quantum measurement can then be reconstituted in terms of how the quantum subsystems are perceived to behave within spatial conditions, especially in terms of how they communicate among themselves and with the physically positioned observer. Therefore, how we think about measurement and observation at the fundamental level, and what we choose to ignore, is what also gives shape to theoretical methods in the less interpretation-directed, and more prediction-concerned, subfields of physics.

In light of that, I have assembled three versions of the double-slit experiment to provide the most fundamental demonstration of a compromise between the rules of classical physics with that of quantum physics at varying degrees, a compromise that is also symbolized by the Standard Model. However, the double-slit experiment I am depicting in this chapter is but a fictive simplification of the more complex question of entanglement, and environment-system communications, involved in the discernment of local-non-local causality. The fictive models push the logical boundary of metaphysics and physics as are embodied by the manifold representations of the double-slit experiment that also involves the foregrounding of logical gaps that are the cause of the physical paradoxes.

As had been discussed through the presentation of the double-slit experiment in the previous chapter, speculative physics, by way of speculative theory, foregrounds, and fictionalizes, in the sense of artificially suppressing certain effects to highlight the what-ifs of other more subdued and less observable effects, in terms of the privileging of a particular interpretation of quantum theory for reasons having to do with mathematical formalisms that could then produce the narrative structures for interpretation to take place. It is this narrative interpretation that sets up the chain of events involving the different interactions of nature embodied by the Standard Model. The stage for that interpretation, with the two main actors being the Higgs boson and the double slit experiment, is set when we see what it means to talk about the classical to quantum journey of the double slit experiment that produce effects that could later be the Higgs boson, with the latter representing the next level of development in the sage of particles versus fields.

Therefore, I am proposing that the realization of the Higgs boson within the Standard Model, and the information produced from the interference and decoherence of the double-slit experiment, represent two spatially separate subsystems forming a composite system of epistemic micro-actions. The two subsystems are connected through a complex network of interactions involving microphysical entities that 'perform' to classically situated detectors and channels transmitting their properties and structures to the physicist-observer. Moreover, ongoing research into these two subsystems, in high-energy particle physics and quantum foundations, means the production of experiments that could change our understanding of classical-quantum interactions, and reconstitute current computational tools.

The seemingly tenuous connection between the prediction and discovery of the Higgs boson, with that of the double-slit experiment, is merely deceptive; in actuality, both processes are sub-parts of an ontological entailment of the field-particle model of symmetry. Both are components that play important roles in the unfolding drama of speculative theory, being as they are, part of a chain of events that had come a long way from the early years of quantum interpretations even as their origins preceded the quantum revolution. Further, the physics foundations underlying the phenomenon produced in the double-slit experiment are of the same kind that underlies the Higgs boson. In a sense, the double-slit experiment interprets the structure, and epistemic path, surrounding the Higgs prediction, one of which involves the uncovering of certain properties of non-relativistic quantum mechanics that would have implication on quantum field theories. The decades of fretting over the (in)completeness and informational instability of quantum mechanics has come full circle with the development of quantum computational technology that can have reflexive impact on how quantum measurement will be performed and comprehended in classically-operated communications. This will be discussed more of in chapter seven.

First and foremost, the double-slit experiment can be used to demonstrate the qualities of the electromagnetic force by illustrating both the field-like and particle-like qualities of the photons. That the electrons behave in the same manner as the photons shows how the quantizability of the most fundamental boson and fermion would allow unification to take place between the electromagnetic and weak interactions to form the electroweak coupling necessary as we approach an increasing possibility of unification between the different dimensions of interactions. Further, the double-slit experiment, by way of the Feynman experiment with the electron to be discussed later in this chapter (and briefly mentioned in Appendix A), is the simplest representation of the properties of electrodynamics.

In the second instance, the double-slit experiment can also be used to demonstrate quantum entanglement, and the role of the Bell theorem, in understanding the epistemic formation of the Higgs boson and the potential for the future simulation of the latter's property through quantum computation. In fact, the finessing of the Bell Inequality theorem is necessary because of the conceptual difficulties of (in)determinism, (im)precision, and (un)certainty that have demonized quantum causality in the eyes of physicists while also finding greater commensurability between technological applicability and ontological prescriptivism.

In the third instance, the multiple variants of the double-slit experiment, evolving since its earliest demonstration in the 18$^{th}$ century, are deploying increasingly massive particles to produce a more sophisticated approach to measurement needed for excavating an increasing amount of information pertaining to the Higgs. The rationale for producing the three versions of the double-slit experiments, the actual and fictive revisions from the real, will be explained later in the chapter.

Nevertheless, I would like the reader to understand that the connections between quantum informatics and particle physics made in this chapter via the double-slit experiment and the Higgs boson is only a basic demonstration that could be developed in a separate work. The purpose for presenting the connections as such will become clearer by the penultimate chapter, as I proceed through alternate readings of speculative theory and experiment as they apply to the analysis of the narrative of the Higgs boson and selected works of science fiction centered on the physics. Moreover, in my readings of the works of fiction that have been selected for this dissertation, one will be able to see the re-enactment of the aesthetics and philosophies through the variants of the double-slit experiments, not to mention the scientific imaginary that can be conjured up as a result of the synthesis between the conceptual embodiment of the experiments and science fictional works.

This is the first time, as far as I am aware, that a form of critical longue durée of the history of the Higgs boson has been put together, in this chapter and Appendix A,[2] with the objective of making explicit, the inseparability of the

---

[2] The most recent attempt at a historical narrative of the Higgs boson was produced by Ellis et al. "A Historical Profile of the Higgs boson" (2012), that provides a good recent technical history that explains the current status of the discovery of the Higgs boson in relation to the incommensurability during the early days of its theoretical formation and experimental interest (or lack, thereof). However, the paper is

fundamentals of quantum theories from the preoccupations of high-energy particle physics. In retrospect, certain conceptual and ontological puzzles in the history of quantum physics can either be clarified or amplified when viewed in relation to the historical development of the Standard Model Higgs boson, and beyond the Standard Model.

The discovery of the Higgs boson can be read as a recovery of a specific epistemological pursuit of quantum theory, and therefore, as part of a contiguous narrative (though not continuous because of a multiplicity of disruptions in the historiography of its narrative) marking the early developments of modern (quantum) physics for configuring the different scales of interactions in fundamental physics. On the other hand, there is also the element of surprise in the sense of the unexpectedness of discovery despite a setup premised on particular expectations of discovery; the unexpectedness that justifies the existence of the Higgs boson is a consequence of the serendipitous coming together of the right data in the right places, where the act of discovery is not coincidental.

With regard to how we can situate the acts of discovery and justification of the Standard Model Higgs boson in relation to extant physical laws, I problematize

---

nowhere as extensive as what I am doing here since I am building on more materials that have emerged since that time.

the conventional wisdom of the laws governing the physical real that insist on thorough independence from being implicated by all of the other life forms, without considering that one cannot, with certainty, know the difference between laws that are anthropically dependent and laws that are not, given our ability for measuring such differences is weak and constituted by our position as an observer. In other words, even if we can imagine a distinction between the universals and the particulars, we could not do so physically without being implicated in the outcome of that observation, as demonstrated by the paradox of the wave-particle duality and Heisenberg's Uncertainty Principle.

The problem of such ontological distinctions has also been the Achilles heels of the Standard Model. However, as Norman Swartz notes in his essay "A Neo-Humean Perspective: Laws as Regularities," in the Regularity Theory, contingency and the conditional are properties of the physical laws, but both properties, in his opinion, have strictly physical (non-organic) characteristics (71). Perhaps, in this case, we need to differentiate between a purely metaphysical artifact and a scientific datum before considering how that distinction can then be applied to the redefinition of the Standard Model Higgs within the framework of new physics.

Measurement conceived at a metaphysical level acts as counterfactual to the physical measurement. However, a metaphysical measurement lays no claim on being able to provide a quantifiable prescription of a phenomenon, as the purpose is not the production of a definite empirical value. Instead, metaphysical measurement supplements the subjective domain of a physical measurement by becoming a platform for modeling the what-ifs derived of varying parameters that are neither homogenous nor linearly computable. Those outputs are not confined to arbitrary values, but rather, point to a domain of potential possibilities produced through the transformation of initial values to obtain a range of non-trivial possibilities.

Then, there is an important linear differential equation function (where the problems of a geometrical and algebraic nature converge) known as the Green's function that shapes the path integral computations in quantum field theories; this function is an example of a negotiation between the physical and metaphysical because of its position as a mathematical operator that serves the specific non-trivial needs of second-order differential equations, whereby particular homogeneous boundary conditions of the inhomogeneous linear equations could lead, or not, to the construction of the Green's function for obtaining unique solutions. The Green's function also allows discontinuities to be calculated. Therefore, the metaphysical quality of the Green's function supplements the fulfillment of a physical need by generalizing from particular solutions to more general principles, counterfactuals, and physical laws. One can interpret the role of the Green's function as that of an 'observer' that has to deal with a probabilistic range of possible solutions in second-order equations when a road-block emerges in the form of a discontinuity.

As Norton asserts in *A Primer on Determinism*, "observation combined with inductive reasoning may recommend one hypothesis over all others, but rarely, if ever, does the combination yield a confidence that approaches the certainty with which Laplace's demon went about its prediction tasks" (66). In other words, the recommendation of a solution can at best only be probabilistically determined by a measurable datum, for as Whitehead argues, "the data upon which the subject passes judgment are themselves components conditioning the character of the judging subject" (203).

Moreover, the Laplacian ideal of deterministic finitude, where the instantaneously conceived physical magnitudes and time derivatives of all mathematical orders are contained, can only work when the quantum system utilizes semi-classical physics approaches grounded on the macrostate. Even so, one has to make sure that there are no contradictions and violations of any laws in how the evaluative analytics are done in moving back and forth between the classical and quantum worlds. Nevertheless, given the relativity of the metric used for the measurement of any observation, can one decide as to what constitutes the most reliable method for dealing with truth-claims without including the epistemic context? Could preparation at the pre-measurement stage that sharply foregrounds the entity of observational interest only provide as much information as the deployed epistemic context would allow, therefore shaping the ethics of metaphysical interpretation?

As John Bell suggests, in a different context, one's 'moral' relation to a physical representation is a cumulative effect of the interaction between a quantum system and its measuring instrument, where a normalization factor is needed to ensure that the probability of the outcome of a measurement made of two eigenstates (or states of definite values), either simultaneously or one after the other, remains the same. This is what Bell refers to as Wigner's "reduction of the wave packet," which was, ironically, a method for expanding the initial conditions of observable quantities by 'cutting' out the unobservable before renormalizing the remainder. The metaphysical ethics (morality) in quantum mechanics, as proposed by Bell, advocates for a measurement system that does the least damage, due to the production of erroneous readings, to the initial state of the system that is prepped for measurement, so that the outcome is "consistent with the requirement that an immediate repetition of the measurement gives the same result" (23).[3]

When Bell wrote on the aforementioned question, he was thinking about the process of scattering and interaction between particles such as the proton and neutron, and his suggestion that one might want to measure the process rather than the initial and final state, considers process as a way of circumventing the physical ambiguity that arises from certain mathematical derivations of quantum mechanics, even if there are problems with measurement reproducibility due to the unpredictability of the outcome. Hence, one is faced with a morality tale that defines the analytical and interpretive choices made in the correlations between the prediction of the Higgs boson and its potential empirical representation. Further, one might have to consider the ontological decoupling of what we understand about the Higgs boson from the constraints of the Standard Model to get to new physics.

---

[3] See "The Moral Aspect of Quantum Mechanics" in J.S. Bell's Speakable and Unspeakable in Quantum Mechanics. Second Ed. Cambridge: Cambridge University Press, 2004.

The winding narrative of the Higgs boson provides rich material for illustrating how speculative theory works. In the five decades between its prediction in 1964 and materialization as a result of progress made in the discoveries of other elementary particles, the Higgs boson holds the promise of reaffirming an almost century old history of particle physics while straddling the increase in the permeability of the theory into higher degrees of freedom, and therefore, extending the reach of the Standard Model beyond its current realm. In other words, the Higgs boson can be both a counterfactual to the Standard Model as well as a bridge to beyond the Standard Model. However, that will be dependent on what the current data that has accumulated from experimental triggers of the Higgs search can provide; such as what can be extrapolated for further analysis when seen a framework located beyond the Standard Model, with epistemic meanings foreshadowed by seeming anomalies that no current theories could adequately explain away.

The historiography of the Higgs boson contains multiple points of experimental and theoretical convergences. The narrative of the Higgs boson also highlights, albeit indirectly, the metaphysical conceptualization of energy that follows from developments in eighteenth and nineteenth century history of science of microphysical matter that constitutes physical quantities with both traceable and less traceable causalities. The ontology of the bosonic field is mediated by different energy states, with their eigenfunctions encompassing a range of predictability that could be reshaped, depending on how we analyze the interactions between the micro-entities constituting the energy and its environment.

Since we know that the quantum field theories underlying the Standard Model are constructed because of a need to project the quantization of energy onto space-time, the most basic description of the QFTs are the oscillating/vibrating fundamental particles mediating the various interactions that can be experienced by matter and the interactions that perturb the bosons at their ground, and most stable, state. The formalism that grounds quantum field theories is subject to multiple re-theorizations by philosophers and physicists, each working towards the best theory for an experimental production across different scales of quantizations, such as that of energy.[4] Further, quantum field theories provide the platform for developing metaphysical conceptualization of the unitarily symmetrical arguments pertaining to the relationships between the different elementary particles necessary for understanding their interactions.

Philosophically, I am interested in expanding the discussion of ontology, particularly with regard to how existing theorems can be reconstituted for producing more ontologically causal interpretations. David Bohm and Basil Hiley's *The Undivided Universe: An Ontological Interpretation of Quantum Theory* revisits the original equations of quantum mechanics to reconsider previously neglected parameters for producing solutions that can amplify outcomes not already made clear in previous iterations of solutions. Such considerations may lead to the convergence of more deterministic outcomes by concentrating on what is observable while minimizing the non-observables (paralleling Bell's own interpretive suggestion). For the most part, the Bohm-Hiley interpretation is considered a maverick and not given sufficient attention by physicists who are not among those already interested in the emergent studies of quantum theory. Nevertheless, physicists and philosophers have similar vested interests in ascertaining how much of the classical approaches to quantum theory can be retained and what had to be relinquished.

---

[4] On the theoretical side of things, there have been no shortage of claims for having successfully found methods to achieve that quantization and that not all means for doing so are purely mathematical in nature. One of the latest examples is Michael Epperson and Elias Zafiris's *Foundations of Relational Realism: A Topological Approach to Quantum Mechanics and the Philosophy of Nature*. Lanham, MD: Lexington Books, 2013 that deploys Whiteheadian realism and topology towards that end. Epperson and Zaffiris's return to ordinary quantum mechanics and propose sheaf theory; a theory that grew out of the abstract algebra of topology and set theory; as a solution to the stubborn paradoxes found in quantization attempts. They then read the theory's interpretive value against Whitehead's category scheme found in Process and Reality.

Hence, while measurement is the bread and butter of the foundations of quantum interpretation, in the recent decade, attention has shifted to the experimental implications of realizing the different 'realities' of what would have been formalistically derived through experiments. This is highlighted by developments in the applications of quantum theory, where the earlier thought experiments that Richard P Feynman discusses in his *QED: the Strange Theory of Light and Matter* are materialized by way of the detectors and instruments built to collect and contain the data. In fact, members of the quantum foundations community have been discussing the experiments produced as variations of the classical-quantum interactions of light and matter, as represented by a zero-mass boson (photon) and a fermion (electron). The division of these 'particles,' each representing the most fundamental particle of each category, instantiates the formal theories of measurement that would describe the interactions of the Standard Model gauge forces whence prediction for the Higgs boson, and its accompanying mechanism, is derived.[5]

In their capacity for highlighting the ruptures and contiguities in the Standard Model, I posit photons and electrons as fundamental entities amply suited to act as measurement controls in the quest for getting as near as possible to the ontology of quantum physics, which are then magnified in mixed-state multiplets (subatomic groupings of particles) of the Standard Model. At the same time, the discovery and confirmation of the properties of these two particles represent the turning point whereby microphysical entities can be fully considered within a quantum model without the crutch of classical theory. Each has an important role within the hierarchical network of the Standard Model. While both exude similar material phenomenon in the double-slit experiment, the fact that the electron produces wave-like interferences was not confirmed until fairly recently.[6]

If the photon is located at the intersection of the space-time symmetries of relativity with quantum energetic transmission, it is the electron, with its small mass unit charge, and point-like countenance, that is used for demarcating interactive capabilities of a single and many-body particle system (even if most examples do not name the electron specifically). The energetic exchange processes between electrons (spin ½) that produce photons (spin 1) as the energy absorbed and emitted represents the electromagnetic force. Moreover, the photons and electrons that are everywhere indistinguishable in their behavior, when considered individually, exhibit more distinguishable markers in collective interactions with other fermions and bosons.

---

[5] There has not been a consensus in the physics community over the naming of this mechanism, usually referred to as the Higgs mechanism, because of the perception that the actual work done on it had been due to the creative contribution of multiple individuals, some of whom may not be completely aware of the game-changing importance of their respective contribution at the time.
[6] See "Feynman's Double-Slit Experiment Gets a Makeover"" in the Physics World. May 14, 2013. < http://physicsworld.com/cws/article/news/2013/mar/14/feynmans-double-slit-experiment-gets-a-makeover>.

Therefore, what I would like to do is to show how the concept of measurement can unite the seemingly unrelated foundational theories in non-relativistic quantum mechanics with the quantum field-like developments of the Standard Model Higgs boson. Further, interpretation has a role to play in rationalizing the observability of physical properties (both the hidden and not-so-hidden qualities). The virtual particles, because of the role they play in mediating the interactions of real particles (best defined by the Feynman diagram), perform interpretation through those interactions. Additionally, interpretation is critical for differentiating between measurements in classical, semi-classical, and properly quantum mechanical systems, given how that would modify our interpretations of measurement.

Energy, in its philosophical and material invocation, is also a protagonist in the story of the Standard Model Higgs boson. One cannot properly explain why the Higgs boson appears to be the unsurprising outcome of an ontologically impotent Standard Model if we do not account for the phenomenon of energy, the same kind of energy that formulated the aforementioned Bose-Einstein and Fermi-Dirac statistical rules. Without the concept of energy, and a systematic way for demarcating its existence, one might not have witnessed the mechanical revolution of physics, the boom of big science, Einstein's theories of relativity, both physics and science fiction's conception of time travel, and the discovery of the Higgs boson that earned the 2013 Nobel Prize in physics.

There is also a sense of precarity in energy at multiple levels: a precarity that is bred by the instability of our epistemic presentation of empirical evidence gleaned through a manipulation of physical techniques to access the energy. While attempts are made to render equivalent the different formal representations of energy for calculational efficiency; as were developed in the works of Lagrange, Hamilton, Bohr, Einstein, Heisenberg, and Schrödinger, among others; computational equivalence does not translate to philosophical equivalence, if only because of the metaphysical features of the theories.

We might ask: how does understanding energy, in all its abstract and material manifestations, help edge us closer to the ontological question of life? The crossing between different scales that energy performs with impunity, from cosmology/astrophysics to particle physics, and a mesoscopic spectrum of physical sciences between the aforementioned two, are scaled up or down, and are plugged into different types of models used for interpreting the long-standing fictions of both cosmological and subatomic scales, such as dark energy and the magnetic monopole. In the next section, I consider why the prediction of the Higgs Boson is inevitable by examining the confluence of events in the development of quantum field theories.

## 3.1 The Higgs boson and its Speculative Denouement for the Standard Model

The development of the Standard Model Higgs boson (or the Higgs boson that resides beyond the Standard Model) is a composite for approximating the relational ontology between different elementary building blocks to form a unified ontological understanding of the universe. The bosons have a pivotal role in the quantization of SU(n) interactions such as the electroweak (W/Z) and quantum chromodynamics (the theory of strong interactions). However, there has not been a clear explanation of how supersymmetry connects to the prediction of the Higgs and experimental explication. The theoretically developed supersymmetry might not be a physical entity even if it is a legitimate computational method that supplements the inadequacies of the Standard Model; one might want to consider it, instead, as a mathematical stopgap with a metaphysical rationale. Therefore, supersymmetry has a phenomenological role for bridging the goals of theory building with that of experiment.

Fermions, via the Fermi-Dirac weak interactions, are important as the first step towards achieving unitarity and symmetry. With the unification of the electromagnetic and weak force, physicists began working on experiments to search for the existence of very specific bosonic particles predicted to mediate the electroweak force (specifically massive W and Z bosons). The process of the mediation of the electroweak force, and other quantizable forces of the Standard Model, is supported by gauge theory as the latter is important for resolving the problem of infinities and mathematical divergences surrounding self-energy interactions that have been epistemically problematic since the 1930s, coinciding with the early development of quantum electrodynamics bearing vestiges of its classical origins. At that time, quantum electrodynamics had been coming to terms with the problem of so-called radiative corrections, which are quantum effects stemming from the differences between classical and quantum-level results. Field theory was supplied for the consideration of the behavior of elementary particles not well understood ontologically.[7]

---

[7]Freeman Dyson wrote an interesting article about the problems of Field Theory and what it still could not do for the 1953 issue of Scientific America. It can be found as an offprint in volume 28 of the magazine's collected papers.

In the 1940s, the physics communities were trying to find ways to reconcile the field theoretic method of Schwinger with the particle-centric approach of Feynman and the others. One of the solutions was to focus on a single spatial component of the particle and assume the particle to be invariant. At the same time, there was also a need to reconcile quantized states with relativity; therefore, the Dirac equations came to the rescue by enabling a compromise between gravitational and electromagnetic fields. Feynman had been working on simplifying the method for performing highly convoluted and error-prone calculations. He crafted a theoretical framework enabling an equivalence treatment of the relativistic equations of Dirac and the non-relativistic wave mechanics of Schrödinger so as to meet the objective of creating a particle-centric view of the different interactions in nuclear physics as an antidote to the highly complex field-theoretic methods proposed by Julian Schwinger.[8] In fact, that was how Feynman developed his path integral method.

What is more, Feynman was interested in the renormalization process for overcoming the problem of divergences, this problem was only partially resolved and would not be completely dealt with until the introduction of gauge fields by Martinus Veltman and Gerard 't Hooft to renormalization. Feynman had been working mostly with perturbative (approximative) field theory even though he was interested in how his path integrals could be applied to the development of non-perturbative field calculations.

[8] See Adrian Wüthrich's book *The Genesis of Feynman Diagrams*. Dodrecht: Spring Science+Business Media B.V., 2010, especially chapters one, four and five.

Renormalization, which works well in quantum electrodynamics, would prove to be an integral part of the Higgs mechanism for providing the gauge bosons with finite mass.[9] At the same time, one might ask if renormalization as a process of solving a physical improbability might be limiting to the conceptualization of these microscopic entities and the manner in which the properties of mass and energy are distributed among them, since it would push for a constrained conception of gauge field theory. The issue of probabilistic constraints was tackled when Freeman Dyson took the matrix formulation of the field theoretic approach, and correlated that with Feynman's diagrammatic approach for the purpose of simplifying and rendering more pragmatic, a calculation process, thereby producing a consolidated approach to a more rounded method for tackling highly complex scales of interactions in particle physics and quantum theory. Dyson's discussion of the calculations in his paper "The Radiation Theories of Tomonaga, Feynman, and Schwinger," drove the work of Murray Gell-Mann as he wrestled with the so-called 'S-matrix' constitution of particles in place of a field theory unable to describe strong interactions.

---

[9] See Sidney Bludman's "The First Gauge Theory of the Weak Interactions" in The Rise of the Standard Model.

However, the insistence on a flat hierarchy of elementary particles in the S-Matrix model, stemming from a desire for a democratic ontology, landed the metaphysical inspiration behind the model in metaphorical hot water, especially when further experimental excavations proved the model untenable. Would one consider this to be a bad deployment of metaphysics in action? Not necessarily, especially since the model has been recuperated for considering the unstable particles in quantum mechanics in relation to high-mass Higgs boson.[10] While the model is useful for navigating the role of the observed in relation to the observing system,[11] the micro-entities are resolved to macro-measurement apparatus because it is impossible to perform an objective measurement as an embedded observer.[12]

Nonetheless, it was Gell-Mann's work on the strong interactions that produces the partonic model of hadrons. Much of the early developments in QED concentrated on free particles such as the electrons and muons. The bound state of the nucleus became the new site for consideration, with the development of quark theory and sustenance of short-range interactions of the partons. At the time when the quark model was proposed, the model was greeted with derision by physicists, as was explained by the 1990 Nobel Laureate Jerome I Friedman during his talk at the Rutherford Centennial Colloquium at CERN in 2011.[13] Incidentally, Friedman shared the Nobel Prize with Henry W. Kendall and Richard E. Taylor for their work on the deep inelastic experiment that finally validated the quark-parton model that helped to transition high-energy particle physics into a new era.

---

[10] Kunszt, Zoltan. "Unstable Particles in Quantum Mechanics, Analytic S-matrix Theory and Quantum Field Theory." Presentation at CERN on May 14, 2012. PDF. <http://indico.cern.ch/getFile.py/access?contribId=0&sessionId=0&resId=0&materialId=slides&confId=174430>.

[11] Stapp, Henry P."S-Matrix Interpretation of Quantum Theory." Physical Review D. 3.6.March (1971):1303-20.

[12] Incidentally, Henry Stapp of the quantum theory and consciousness fame had also been working on the S-matrix early in his career.

However, that experiment did not take place until 1968, four years after the initial prediction. For years after it was first proposed in 1964, the quark model was ridiculed for daring to propose that the atom is a composite model, six decades after the 'plum pudding' model of Thomson had been disproved, and at a time when much of the particle physics experiments were targeted at the nuclei of the atoms. In an article published in the February 1967 issue of the popular hard science fiction magazine, *Analog Science Fiction and Fact*, Margaret L Silbar satirizes the sentiments of the day with:

> This is a scientific-fact article in that scientists seriously considered the quarks to be fact. My personal opinion is that they must be getting desperate!"

Silbar's sarcasm points to how the partonic model, and its accompanying quarks, were seen as metaphysical constructs making extensive claims about changing how nature would be understood; claims that require a shift in experimental perspectives. Nevertheless, she was right in predicting, were one to read the rest of the article, that the discovery of quarks did not translate to greater ontological oversight on the substance of matter. Instead, the discovery of quarks opened up even more difficult questions![14]

---

[13]See http://indico.cern.ch/materialDisplay.py?contribId=3&materialId=slides&confId=147420. Refer to slide number 12.

The top quark, as we will see in chapter five, is experimentally important for probing at the interaction vertex of the Higgs boson, even if not much data could be uncovered about the boson from the beauty quark at the time of writing. But the fact that a considerable percentage of the Higgs bosons were discovered through decay in the top quark loop demonstrates the compatibility between the conceptual structure of the quarks and that of the Higgs boson. Moreover, the prediction and discovery of the pion in the photographic emulsions (a technique developed out of studying the cosmic rays) was a precursor to the discovery of the quarks, as well as to symmetry breaking. It was in trying to work around some of the issues that arose from the relationship between massive/massless bosons and symmetry breaking that the predictive theory of the scalar (Higgs) boson was first derived. In addition, the Glashow-Weinberg-Salam (GWS) description of the weak interaction is also needed for explaining certain electromagnetic features of the quarks. In chapter five, we will see how the strong and weak theories work together to produce a solid understanding of the Higgs boson.

[14] A reader unfamiliar with the history of the quarks may do well to check out a brief summary of it in Appendix A.

Nevertheless, there is less symmetrical stability in a multi-dimensional geometrical entity than a two-dimensional surface because the former faces greater difficulty in maintaining symmetrical balance due to the higher number of variables involved, including indeterminate variables. The same sort of complexity is involved for quantum symmetries due to the need in balancing the different scales of interactions (not to mention quantum states) compared to the classical symmetries operating within the same macrophysical continuum. An example of a quantum symmetry that has much relevance for the constitution of the Higgs boson due to the act of spontaneous electroweak symmetry breaking is the chiral theory of symmetry: this is where complex numbers enter the narrative for explicating the phase of a quantum wavefunction during the rotation of handedness and helicity in the course of rotation, important for working out the balance between matter and anti-matter and predicting new generations of particles.

Prior to the development of path integrals for resolving problems of physical symmetry, there were also attempts made to reconcile the problem of field theory with the particle framework by working out the internal invariance of a system under observation in space-time. The development of such a system is concentrated on a 'free' particle since a particle that is involved in interaction with another particle or its environment would increase the number of variables that have to be accounted for. The early interest in transformative invariance for maintaining the symmetry of a system coincided with developments in Einstein's theory of relativity, such as through the Lorentzian transformation. Of course, the transformation would later be expanded to include velocity and momentum, heralding the observations made about the internal symmetry of a particle system integral to describing the conservation of energy. These symmetrical structures would become important in dealing with the Higgs boson's seeming ability to 'give' mass to other particles.[15]

The prediction of the Higgs boson started with attempts at working out the problems surrounding the massless Goldstone bosons and the requirements for producing a massive scalar boson. The Higgs mechanism gives mass to an otherwise massless scalar boson of zero spin, by producing field equations that can interact with other bosonic and fermionic field equations in the aftermath of spontaneous and explicit symmetry breaking. But as we know, quantum symmetry breaking is not the main claim of particle physics, for as David Gross pointed out in 1997[16], and a physics professor, Shivaji Sondhi[17] stressed in an opinion editorial in November 2013, the study of superconductivity by Philip Anderson had elucidated the principles of broken supersymmetry long before it was understood by particle physicists. In addition, there was in existence, then, the Yang-Mills theory that would become important for reconciling strong interactions with field theory, but was unfortunately not well understood in the 1950s. Ironically, it was also the development in superconductor technology that enabled the design and building of the Large Hadron Collider to perform experiments at unprecedented energy levels that could then culminate into new physics potential.

---

[15] This blog post by Flip Tanedo of US LHC at a physics communication blog, Quantum Diaries <http://www.quantumdiaries.org/2011/06/19/helicity-chirality-mass-and-the-higgs/>.

[16] "Asymptotic Freedom and the Emergence of QCD" in The Rise of the Standard Model. Pp 207-8

[17] Sondhi, Shivaji. "The Tao of Modern Physics - Indian Express Mobile." N. p., n.d. Web. 18 Nov. 2013. < http://m.indianexpress.com/news/the-tao-of-modern-physics/1195124/>.

Nevertheless, the confirmation of the Higgs boson brought together the theoretical and experimental particle physics communities to consider theoretical candidates for extending the possibility of the Standard Model; the most promising candidate in this regard is the supersymmetry. Further, supersymmetry has provided the leading order quantum corrections for solving the hierarchy problem of the Standard Model, and for reconciling data with theory by adding appropriate fermionic and bosonic superpartners in order to equalize the different interactive scales of the Standard Model. Physicists have been working with the minimally standard model supersymmetry as a way of attenuating anomalies not properly explained through the Standard Model symmetrical features.[18]

---

[18] See Christoper P Hays's "Doubly.Charged Higgs Bosons," Search for the Higgs Boson. Ed. John V.Lee. New York: Nova Science Publishers, Inc, 2006: 7-9 and Gordon Kanes's Supersymmetry: Unveiling the Ultimate Laws of Nature. Cambridge, MA: Helix Books, 2000. Chapter four.

While the confirmation of the properties of the super-heavy top quarks appears to give hope that one might uncover more direct evidence of supersymmetry, there has been much back-chatter in conferences, meetings, and online spaces with regard to letting nature takes its course; then regroup to decide on what to do next should further experiments draw no further material evidence of the supersymmetry model. Additionally, there are also the large extra dimensions and the hidden valleys, all of which are extensions of the exotic theoretical speculation for extending beyond the Standard Model and bridging into other experimentally uncertain worlds such as string theory.

Even as these supersymmetry particles are more complex due to their relativistic properties, they draw a parallel to the ordinary quantum mechanical debates of (non)locality because of derivations from similar quantum states, and the fact that measurement enacted in both cases are drawn from similar basic principles in quantum theoretical measurements that would be represented in myriad ways in the chapters to come. In addition, the aforementioned example resembles the parallels of symmetry breaking in condensed matter and particle physics due to a shared, even if non-visible ontology. As mentioned in the previous section, the common ontology that each of these physics share could potentially be investigated as a representation of mixed-state measurement. Further exploration might consolidate, and produce more persuasive evidence, on the arguments regarding how the choice and method of measurement, whether as a thought or an actual experiment, could influence the epistemic contexts of physical events under dissection.

One might ask whether the status of physical and scientific laws would remain unperturbed regardless of the choice of theory, and whether experiments' attempt at coaxing nature's entities into manifesting within a physically observable range (or the range of instrumental sensitivity) could change the state of the observable in relation to the actual (non-interacting) state. What are the new physics that one can aspire to in this century, and would experiments be the ones to shape new physics, given that there are still theories waiting in the background? If so, how would we like to reconstitute the logic that drives the experimental design, and could a metaphysical intervention decrease the gap between epistemology and ontology? A summary of such aspirations are outlined in the New Particles Working Group report published in the summer of 2013 that details the not-so-new (in terms of their theoretical 'existence') 'new' particles that might be uncovered during the second run at the Large Hadron Collider, to begin in 2015 after the upgrades are done, as well as the next steps to be taken should there be no new particles discovered.[19]

---

[19] See Y. Gershtein et.al. "New Particles Working Group Report of the Snowmass 2013 Community Summer Study." <http://arxiv.org/abs/1311.0299> .

Even before the ‘discovery’ of the Higgs boson, there have been many speculations on the next step for elucidating the way forward, as physicists have been frustrated with the epistemological shortcomings of the Standard Model in explicating the behavior of matter and prescribing the ontology underlying the Model’s interactions. There are attempts to dig deeper into potentially suppressed physics stemming from the choice of data cuts made that provide the elusive solution to some of the difficulties faced by the Standard Model. The discovery of the Higgs boson is both a success and a failure of the standard quantum theory: a success because it shows the Standard Model as not being completely misguided and a failure because the discovery of the Higgs boson appears to have raised more ontological questions than it has solved.

Perhaps, what is needed for this century is the rethinking of the foundations of quantum theory in relation to phenomenologically engaged causality; to find an access to ontology that transcends ‘object-gazing’ by penetrating those objects. One will have to re-examine the apparatus available for more direct investigation into the structure of the universe.

For particle physics, it is an aspiration that the confirmation of the Higgs boson will now take physicists away from the comforting security of the Standard

Model, towards less standard measurements. The next section attempts to demonstrate how quantum theory, as we know it, is shaped by certain interpretive practices and epistemic choices. The initial and final state of a measurement process contains as much important information as the interval between them.

## 3.2 The Foundation of Measurement and (Non)-Standard Interpretation

Now that we have finally laid out the (composite) profile of the Higgs, we can proceed with the promised discussion on the double-slit experiment in relation to the former. As was explained in the previous section, the conceptual rationale underlying the experiment that sits at the intersection of realist, idealist, and relativist principles in the philosophy of science has crucial similarities to ontological issues arising in quantum field theories. Therefore, the bridge built of the two outwardly dissimilar physical events can then be used to account for bi-directional developments as their syntheses contribute to a comprehensive characterization of the properties underlying the ontology (and multiple epistemologies) of the quantum micro-universe, since the interpretive method used for making sense of computationally processed data is derived from choices made out of expediency, pragmatism, reliability, and consistency across all states.

The classical experiment on the wave interference of light was performed at the turn of the nineteenth century by Young to explain the color spectrum that is the outcome of light diffraction by a prism once observed by Newton.[20] An Italian mathematician, Francesco Maria Grimaldi, discovered that diffraction could take place when light passes through a barrier, either around the edges, or through a small slit, but his findings were only published, posthumously, in 1665. The non-consensual views about the characteristics of light, as constitutive of either particles or waves, have been at play since classical antiquity. However, experiments with optics in the early modern period unleashed physical phenomena that were inexplicable until the late nineteenth and twentieth century.[21]

Having discussed the conventional setup of a modernized double-slit experiment involving photons in the previous chapter, I will proceed directly into an explication of the 'real' and fictive variants of the experiment, before explaining my rationale for doing so. One can consider this a form of metaphysical fictionalization that also anticipates modifications in outcomes stemming from modifications to the input. The modification in the setup is a thought-experimental manipulation of John Bell's concept of the 'beable' that symbolizes a theoretical construct with a materially tangible counterpart of more than a single physical potential.

Imagine having a highly sophisticated two-photon beam, double barreled, ray gun that can control for pairs of photonic beams at the point of release, including the behavior of the beams. This gun can also time the release of each pair of the beams, so that a photon from the first beam entering one slit is followed closely by a photon from the second beam, in accordance to a pre-set timing. In this fictive version of the double-slit experiment, I imagine that one can manipulate the timing and also which slit the photons can enter by (a fictive re-enactment of the which-way experiment). Consider also, that the ray gun has a special function that could be set to deploy the process of entanglement at will; in that the gun could determine which path each photon emitted would take, with the intention of increasing the localization behavior of the input object. Therefore, the uncertainty of the position and time are narrowed, while the probabilistic range of the momenta of photons is increased to correspond to any increase in energy.

---

[20] See Young. "The Bakerian Lecture: Experiments and Calculations Relative to Physical Optics." 1-16.

[21] Holton. *Foundations of Modern Physical Science* 543-63.

The ray gun produces three different pairs of beams with different phase and interactional properties. The first type of interaction involves two beams that pass through the double-slits as usual (in the actual experiment, it would be one beam at a time; but in my adjusted version, it can be one beam at a time, two beams at once, or two beams almost simultaneously although one will not know which beam goes where until after detection has taken place); the second type of interaction consists of two beams that are intentionally suppressed after they passed through the slits (they do as the first type except that the informational path gets suppressed immediately after detection so that one does not know whence each beam originates, an reenactment of the quantum eraser); and the third type represents a state of semi-decoherence, whereby a selection of photonic waves from the second beam are superposed with a selection of the photonic waves from the first beam, leaving some photonic waves unsuperposed (this means that part of the waves become quantum entanglements while the other part still behave by semi-classical rules). These superposed waves then hit the screen, after going through the slits, at the same time as the other 'regular' unsuperposed photons. To throw another wrench into the wheel, so to speak, the ray gun has been set in auto to emit the beams at random so that one cannot pre-determine the type of interaction occurring at each time.

The first example is an enhanced version of the double-slit experiment where the infamous (or famous) wave-particle duality can be observed[22]. The second example represents points of decoherence when the interaction of the photons in their different states (differently localized states, or, if in terms of a classical description, 'spookily connected' states) are intentionally suppressed, while the final example represents a state in semi-collapsed conditions where the effects may be observed to be partially decohering at the classical level. However, in the third example, what one sees is the ensemble in action of a mixed density state with the micro entities in their initial and final states.

---

22. While the classical version (Young's version) of the double-slit experiment had been in existence before anyone has even heard about quantum mechanics or wave-particle duality, the phenomena of the wave-particle became more manifestly expressed because of the discovery of the photoelectric effect, which changed the way physicists understood the properties of light. The photoelectric effect is an observable phenomena of photo (light) emission caused by light-waves pushing through the surface of a solid to free the electrons at the surface. This has been observed since 1887 by Heinrich Hertz, and was further studied by J.J Thomson. While it is not hard to explain why light of particular frequency that is shined onto the surface of a solid with loosely bounded electrons could lead to the freeing of the latter, what is less easily explicable is the level of threshold frequencies required of the light-waves for the electrons to be tipped over and out of their bounded conditions, or how the threshold is different for different sets of light waves even when the energy-frequency for all the light waves across the spectrum remain the same. In other words, in a curved graph, they all have the same slope, showing how the energy-frequency relation is constant, and this constant is called the Planck constant. One might speculate that this change of understanding transformed the manner in which observation and measurement are constituted for the physicist.

The double-slit experiment on particle-wave duality produces quantum interpretations that were applied to the development of quantum electrodynamics (as was inferred in the beginning of this chapter), especially since the experiment itself demonstrates the limits of the classical electrodynamics. Therefore, the tale of the double-split interference as a narrative of decoherence between classical and quantum theories converges in an electromagnetic radiation that is the first stop in symmetrical relations that would later be part of the Standard Model. Then, there is the Copenhagen interpretation of complementarity that posits how two observables cannot be measured simultaneously because the classically located measuring apparatus and the quantum object of measure each abides by different rules pertaining to determinism and precision.[23]

---

[23] In more than one instances, scholars have pointed out that the term 'uncertainty' in the famous Heisenberg principle was a result of a disagreement and final compromise (or sorts) between Bohr and Heisenberg, stemming from the discussion about how one can talk about measurement and the finitude of quantum theory. Barad (*Meeting the Universe Halfway*), Beller (*The Genesis of Interpretation of Quantum Physics 1925-1927*), and Jaeger (*Entanglement, Information, and the Interpretation of Quantum Mechanics)* are among those who had extensive accounts of this issue, in relation to indeterminism and precision, and they have cited some of the pioneering quantum physicists who were also concerned about this issue during the difficult teething years of quantum theory.

However, what are the qualities of measurement that change if both the measuring apparatus and the entity being measured obey only the laws of the quantum? What about the possibility of making the interactions sufficiently deterministic so that certain outcomes can be controlled without necessarily going against the Heisenberg Uncertainty Principle for quantum interactions, such as is the case with weak measurements?[24] There have been a number of theoretical speculations raised on this issue but we will have to await further progress in quantum informational experiment to see how we can change the current epistemic language of precision, indeterminism, and technological expediency when it comes to quantum theoretical interpretations. But for certain, a deeper exploration of the double-slit experiment can lead to development of instruments that can change how we model the theories of new physics, or even how we simulate these theories.

Moreover, the double-slit experiment tells us about some of the self-reflexive fundamental problems of measurement, involving the simplest of entities, before we can attempt to measure even more complex interactions. When measurements are performed, there is a desire to regularize the measured entities to fit a system, to produce the best theory for explaining the different locales of the observables. But in the process of making that connection, entities with characteristics that appear to be behaving differently could be set aside for separate considerations. Even if the intention is not to ignore the signals these 'odd balls' bring to the table, over time, the system that excludes these signals, even if only temporarily, while focusing attention on experimental searches that can 'induce' empirical evidence of other more 'harmoniously' coordinated entities, could end up never dealing with the paradox until such a time when the occurrence of an unpredicted event brings about epistemic crises. But more importantly for the discussion here, the Copenhagen interpretation of quantum mechanics, and therefore the conceptual underpinnings of the quantum physical double-slit experiment, fosters the context enabling the theoretical development, and experimental searches, for the Standard Model Higgs boson. One might say that the philosophies underlying the double-slit experiment and the Standard Model Higgs boson are nomologically constituted through a nomological network, with the point of origin being the Copenhagen interpretation.[25]

---

[24] Svensson, Bengt E. Y. "Pedagogical Review of Quantum Measurement Theory with an Emphasis on Weak Measurements" Quanta 2.1: 18–49. doi:10.12743/quanta.v2i1.12. Retrieved 7 April 2014.

Further, a parallel problem exists in semi-classical physics with the hypothetical graviton because of lack of fit with the spin-environment models that are core to quantization. It does not appear that the current physical system of the

---

[25] See http://www.socialresearchmethods.net/kb/nomonet.htm. Even though this concept was originated for conducting psychological measurements of relationships in psycho-social settings, I find it useful as an analogy for depicting the not always explicit relationship between foundational physics (the double-slit experiment) and particle physics (the Higgs boson), as well as in explaining that speculative physics, and speculative theory and experiment more particularly, shares a philosophy with a nomological network because of shared interest in the process involved in theory to experiment relationship, including how one can construct a valid relationality between both.

Standard Model could deal with this problem until we extend the scale of its dimensional operations. There have been multiple mathematically-induced thought experiments that ask one to reconsider the ontological correlations that might be found between gravity and the other interaction fields. At the same time, there are systemic attempts at working out a semi-classical framework that can supplement what is lacking in quantum field theory.[26]

Thought experiments are all very well but effecting a practical application requires the capacity to transform these thought experiments into engineering and computational parallels, which would also require the experimenter to engage in other forms of speculative practices through the construction of instruments and the mechanics of data-collection that include the excavation of impossible-to-get-at data due to technological limitations. If thought experiment's objective of effecting sufficient determinacy succeeds so that more information can be experimentally obtained, how would that radically alter our perception of what constitutes as measurable and scalable at the level of apparatus-observer relationship, therefore producing a new perception of quantum field theory?

---

[26] See Gorelik, Gennady."Matvei Bronstein and quantum gravity: 70th anniversary of the unsolved problem." Physics-Uspekhi 2005, vol 48, no 10, pp. 1039-1053. The paper addresses a rather interesting history on the development of quantum gravity in the 1930s which connected with the very early stage of quantum field theory's emergence out of work in electrodynamics but before the development of Feynman's path integrals. The English and Russian version of the paper can be found <http://people.bu.edu/gorelik/cGh_Bronstein_UFN-200510_Engl.htm>.

I argue that the performance of measurement on a quantum system represents a speculative mode of ontological reading that excavates the relationship between micro-entities and their environment within a bounded system; the sum of the relationship can then be extrapolated for analyzing causal effects such relationships demonstrate in an overall state. It also means reading beyond what can be easily embodied mathematically, to account for greater subjectivity and epistemic complexity. Current models of quantum theories contain mathematical non-equivalence between the classically derived metric of gravitation and the theory of quanta. In performing an ontological reading, one does not merely reconsider how the physics of gravity can be scaled in relation to other known interactions, but also asks which regions of the interactions would be most affected, in the sense of physical phenomenology, when new orders of quantizations are performed; how would the outcome, should they come into effect, fit with the logic embodying the different interactive scales?

Further, while it is possible for ontology and epistemology to converge at the initial and final states of the physical space one measures, it is also possible for epistemology to take its own direction independent of what the ontology might be, especially when paths to ontology is largely underdetermined. An example of such thinking would be if one chooses between whether to recognize non-locality, a condition conventionally represented as 'hidden' variables due to their non-observable states. Therefore, we could decide whether we would prefer a quantum interpretation that privileges greater levels of determinism, or not (to go with Bohr or Bohm?). At the same time, one might posit a one-on-one correlation between the properties of quantum systems and their projectors, so that one might ask if an individualized and 'atomic' representation will bring about a different outcome than an approach that is premised on statistical ensemble.

Nevertheless, the need for experimental relevance requires one to reconsider the mathematical formalism embodying the particles/mattresses/fields, and what is most suitable for the consideration of entities as are represented by the gravitons/gravitational fields and the Higgs bosons/fields. Physicists delineate the operative framework of physical assumptions, hypotheses, and states while inculcating them with a semblance of mathematical rigor. If the foundational question defining the preferred approach for dealing with quantum theory tends to preoccupy foundational theorists, experimentalists are more interested in figuring the best informal way for modeling their experimental results. Metaphysics can supplement phenomenology in providing the platform by which theorists and experimentalists locate conceptual parallels and overlaps, where fictional models (and computational algorithms) can then be developed side-by-side, such as what I had attempted to do through the fictional reconstruction of the double-slit experiments.

However, given that experiments have only been able to operate within constrained degrees of freedom represented by the dimensions of space in which the detecting apparatus is able to reach, the question would be what should be done to expand their reach into dimensions that are counter-intuitive to our material quotidian. We need to break out of a recursive and nested formalistic approach, one where the intervention to knotty problems involved patching up an already problematic framework. The formalistic patchwork approach has been consistently a part of the epistemic narrative of the Higgs boson, an approach premised on the need to retain the role of the Higgs boson for maintaining a symmetrical apprehension of our space-time. However, its origin story was a demonstration of a determination by groups of physicists coming together to work out what can be done with regard to its masslessness, and unseemliness, by finding loopholes around the problem. The Higgs boson, because of its position as one of the force carriers, is inevitably embedded within the ontological makeup of the Standard Model.

At the end of the day, the development of the Higgs boson has not always developed in congruent with that of its experimental searches. While the discovery of the Higgs boson points to seeming convergence of the theoretical and the experimental, in reality, that convergence is still not a perfect fit, not until it becomes possible to resolve the ontological crisis of the Standard Model in a manner where the theoretical and the experimental can speak. Maybe revisiting the historical evolution of quantum theories can help but much work still has to be done to make the current elucidation of the theories of the Higgs boson, and its exotic accompaniment, more explicitly concerned with what is going on in quantum foundations, as a step, even if still inconclusive, to getting to the root of the problem.

### 3.3 Conclusion

To recapitulate, were one to appropriate the language of quantum field theories, it would come as no surprise that the various approaches involved in the consideration of the Higgs boson revolve around acts of creations and annihilations as represented by the vertices and lines of particle interactions, both symbolic and actual. The narrative of the Higgs boson forms, grows, and coalesces into materiality through the explicatory affordances of speculative theory.

In a sense, the Higgs boson is the interpreting subject of speculative theory as the former could only have descended from combinations of underdetermination, causal choices, formal developments, and ontological anticipations of the possible, while not quite a sum of all these features. At the same time, metaphysical approaches appear not to have impacted the theory-to-experiment relations in the search, and the discovery of the Higgs, and this indicates how the approach to the Higgs's conceptualization and eventual empirical confirmation follows a very particular epistemic strategy that skirts more problematic ontological issues, such as the question of what the Higgs mechanism can do. Yet, one could surmise that the problems of the Higgs boson are both reduced and amplified within the classical approach to matter: the quantum double-slit experiment amplifies the unresolved issues that dog quantum theory at the foundational level even as the current epistemic structure of the Standard Model is only able to manifest a partial, and greatly reduced, representation of the Higgs boson that does not present new physics.

The signification of the Higgs boson can be read against the context of constructivism, realism, idealism, and instrumentalism. It is a metaphysical object with a physical counterpart, of sorts, that seems to fit well into the parameters of the Standard Model while pointing to the disquieting precarity of the Model. But most importantly, it reveals how we need to rethink measurement at the intersection of the known and the unknown, the uncertain and apparently determinate. It also provides the subject orientation for discussing the fiction of indeterminacy (which compels indeterminism) and control in a science fictional micro-universe, which I will begin to address from the next chapter.

*Indeed, "touching," the lexicon of touch, strikes a grammatical pose and heads off on quite diverse rhetorical side paths. It carries a semantic tenor whose specter seems to obey a subtle and ironic play, both discrete and virtuoso* ***(Derrida, On Touching – Jean-Luc Nancy, 2005).***

## 4. Rapprochement of the micro and macro in Egan's *Quarantine*

Fans and writers of hard science fiction, as well as literary scholars, do not always agree on the contents of hard science fiction, if only because of the presupposition that hard science fiction demands technical expertise and scientific competency from its writers. The strict hardcore enthusiasts of the genre can be classified as 'disciplinarians' who enforce a particular "field of claim, counter-claim, and struggle with one another rather than through a relationship to a transcendental object or investigative goal defined anterior to discourse [the object of science fiction] itself" (Lenoir 52). To appropriate Lenoir's argument (made in relation to the institutional practices of science) for staking my argument about science fiction that takes a science studies approach in this chapter: the scientific ideas are experimented with the aid of literary techniques, representing a particular "…flavor of contemporary technoscience, which is simultaneously political and technical" (52).

Hence, the disciplining of the content and rules of this genre is used as a method for legitimating fiction's interpretation of the social in the eyes of science, as well as in justifying the objective of this genre. Hard science fiction writers, for the most part, have a strong preference for scientific accuracy and would not shy away from indulging in highly technical discussions on the progress of science in the fiction they write, including subordinating the plot to the science. In turning hard science fiction as a creative platform of elucidation of the techniques and sociality of physics, I hope to take a literary step in the direction that has been set by last chapter's exposition on the critical-historiography of the Higgs boson and fictive re-narration of the double-slit experiment. In the process, I hope to illustrate how fiction holds the key for a less linear demonstration of theoretical speculation that takes place at the intersection of the sciences and the humanities.

That said, there are a number of supposed hard science fiction that do not always meet the criteria of scientific integrity and rigor, as critics would be quick to point out. The previously mentioned Paul J Nahin did a critical survey of such works in the introductory chapter of *Time Machines: Time Travel in Physics, Metaphysics, and Science Fiction*, whereby he pointed out the foibles and wrong-minded deployment of attempts at time travel by a rather substantial number of science fiction authors. Time travel is not just a fantastical plot device but with real physical and metaphysical ramifications, and has been of interest to physicists dealing with the idea of the 'arrow of time' from different angles of both fiction and natural philosophy, culminating in the special theory of relativity of Einstein, to be further discussed in chapter six. This topic, in itself, is worthy of greater exposition but is not the main point of the dissertation.

At the other end of the spectrum, critics who prefer to read the project of science fiction through the lens of literary criticism, or merely as an aesthetic choice, may find the direction taken by hard science fiction to be discomfiting. For instance, Frederic Jameson considers the 'paradigmatic' shift from physics to the life sciences to be problematic for the conventional science fiction narrative and representation because the increasing abstractness of the science makes the science fiction text incomprehensible to him (and potentially other literary-minded readers); he argues that "…the complexities of biology and the genetic, indeed bio-power itself, offer a content and a raw material far more recalcitrant to plot formation than even Einsteinian cosmology and the undecideability of atomic sub-particles" (67). Indeed, he finds the latest hard science fiction based on informatics (or informational science), such as that by Greg Egan, to be inscrutably unreadable even if the former had no issues with breaking down the densest of critical theories!

Following in almost the same critical vein but attending more to the notion of scientific accuracy than its inscrutability, is literary critic and science fiction writer Adam Roberts. He asks incredulously, in *The History of Science Fiction*, on the need for a demarcation between 'soft' and 'hard' science fiction, before going on pejoratively about what he considers to be an insistence on unloading rigid scientific orthodoxy (drawing on his earlier discussion of philosophers such as Karl Popper and Bertrand Russell with their 'rigid' notion of the scientific method) onto science fiction and the inconsistent deployment of the rules of scientific facticity by science fiction critics and enthusiasts. The latter half of the critique unwittingly reveals Adams's ignorance of the nuanced differences between axioms, physical laws, theories, hypothesis, and facts; and how speculation operates at various levels of acceptability in science based on the aforementioned categories and the choice of an interpretive mode.

> Application of conventional scientific orthodoxy as a criterion of judgment for an aesthetic object is fundamentally foolish even when applied with absolute consistency; and when applied inconsistently, as is often is (swallowing the camel of faster-than-light travel but straining at the gnat of, for instance, S-shaped ballistic trajectories inside spinning environments) it combines deadness with muddle. Our choice between a textual universe running along the oppressive lines of Russell's scientific world government, or a science fiction that plays anarchically with 'science' along the lines Feyerabend suggests. This seems to me no choice at all. (16)

His prejudice to rigid scientific accuracy (or at least a pretense to that accuracy) aside, he addresses a number of important issues that tend to get swept aside, or are less foregrounded, in the more conventional discussions of science fiction, such as the genre's genealogy, the parallel developments in history of science with science fiction (which demonstrates succinctly how the 'science' is generated within the fiction over time in light of new understanding and changes in attitudes), and scientific ethics in relation to society (the latter third question will come up again in chapter six). I am in agreement with Adams over how science fiction, even hard science fiction, should not be so be held so tightly to a particular scientific reading as to dismiss unpopular and minority positions, given how some of these positions are read as 'crank' science by scientists and some science fiction fans alike.

Nevertheless, there is a critical reading of science fiction, including that by Jameson, which privileges the scaling up of micro psychophysical interactions (and characteristics) to fit the macro-level sensibilities of mundane politics, even if such a transition would mean leaving out the same qualities found in the former interactions that made for such fascinating study in the first place. Such critiques are attempts at disciplining the genre of science fiction into a specific utilitarian mold. However, it is possible that one's attitude to science fiction is determined by the choice of reading strategy (Roberts 2), one that is "shaped by one's imagination to synthesize the information given him, and so his perception is simultaneously richer and more private" (Iser 283).

The 'private language' of literary criticism and the disciplinary practices of science, however well understood they are by others sharing the same linguistic trading zone, to invoke Galison, could still culminate into a particular and exceptionalist understanding. While the private relationship between the text and reader can increase the richness of the interpretation by not having to acquiesce to particular rules or laws beyond that set by the reader, such an experience would never be replicable in the same manner as that of a scientific experiment. However, even in the case of a scientific experiment, there is much more going on than merely the repetition of an experiment to replicate an outcome. Therefore, the end product for both experiences is one that is ideologically infused and theory-laden, rather than universal.

Most conventional approaches to practices in literary criticism do not fall far from the macro-perspectives of classical physics (including our comprehension of the biological *Lebenswelt*), informed, as they are, by Galilean/Newtonian 'thought-styles,' so much so that even attempts at reading literature against the non-intuitive world of quantum operations would fall quite short. Therefore, in a number of instances, we end up with critical readings of texts that did not fully make use of the physics subject deployed for the performance of that critique, as is the case with the various attempts at using quantum theory as social theory in literary analysis.[1]

However, the non-intuitive, but classical physics of Einsteinian relativity contains attractive meta-narratives that could spice up the what-if imaginary of an otherwise mundane world because the aesthetics of relativity is premised on a familiar world (even if the theory defamiliarizes how space and time might appear to us), and can therefore provide a germane backdrop for sociopolitical critique. The fascination that literary scholars have towards Einsteinian relativity appears fixed on the twilight and 'crises' laden years of the 1920s, a moment induced by the "so-called 'failure' of mechanics" (Seth 27-29) due to ongoing ontological difficulties experienced by quantum mechanics, whose importance is increasingly validated but whose hidden relations need to be further finessed.

---

[1] See Samuel Chase Coale's *Quirks of the Quantum: Postmodernism and Contemporary American Fiction* for such an example.

Primarily, this chapter is interested in considering science fiction as a critical historiography of techno-scientific interrogations (critical in the sense of being sensitive to ideological and cultural valuations), while also considering the not-yet practically viable interfacing between certain sciences that are, theoretically, not improbable. Secondarily, the chapter is interested in considering the distinction between science and technology, one that is not always clearly demarcated because of the conflation of techno-to-science. *Quarantine* is discussed as an example of that porous dichotomy between science and technology. It is also the aim of this chapter to explore the fictive re-interpretation of the real-time underdetermined quantum theories at its most foundationally enigmatic, in a manner that would best illustrate my more theoretical arguments in both chapters two and three.

The techno-scientific interpretations of physics that is present in *Quarantine*, and in the scientific issues raised throughout this dissertation, can be read against Suman Seth's characterization of the development of postwar German theoretical physics: 'the physics of principles' and the 'physics of problems.'

> The physics of principles, which had its most prominent proponents Poincaré, Planck, Einstein and Bohr, can be seen as the most significant

> continuation of and response to *fin de siècle* debates about the foundations of physics, offering in place of any particular materialist ontology a physics based on generalized principles. The physics of problems both was newer, beginning esssentially with Sommerfeld's move to Munich in 1906, and largely avoided the questions of foundations. Sommerfeld once quipped to Einstein that 'I can only further the engineering of the quantum [die Technik der Quanten]. You have to make it philosophy'. (41)

Fiction like *Quarantine*, their shortcomings included, are not evaluated in the context of a micro-politics produced of laws governing the physical realities of the quantum worlds with their own intense interactions; such interactions cannot be scaled-up and measured against the laws of the macro-world without acknowledging problems of physical and behavioral incompatibilities between micro and macro entities. In other words, when one conducts social experiments that involve 'crossing the borders' between differently 'acculturated' physical realms, one would need to negotiate a space for mediating between the realms to maintain the ontological consistency of the physical laws regardless of the interpretive modes of choice.

In addition, for Jameson, the 'hard' science is "…a miniature sociology of the scientists, a history of their funding, and an account of the role of experimentation and of scientific publication as well" (108). Experiments become the center whereby funding flows in and scientific publication flows out, with clusters of scientists and institutional administrators adjudicating the boundaries and parameters that will determine how the experiments should be performed. Experimentation also enables, depending on the apparatus and ontological attitudes of the subject that is the target of experimentation, the re-interpretation of multitudes of theories that might converge, or not, into successful predictions. Experiments can amplify uncertainty even as they desire to achieve descriptive clarity. But most importantly, experiments premised on creative science (which is not the same as imaginary or fictive science) take the risk of transcending the law-like restrictions of a scientific discipline to imagine an outcome involving the 'unrealistic' mixing of temporal scales in order to complicate the dichotomy between the historical and the present, the macro and the micro. In a sense, to develop a philosophy of experiment is also to develop a philosophy of an always-transformative identity that finds commonality in difference.

The interpretive processes forming the core of problem solving in evidence-based science are also rigorously pursued in hard science fiction, except for having not to deal with hindrances that are results of mathematical constraints. Thought experiments, considered as theoretical physics' closest fictional kin and analytic philosophy's favorite method for conceptual analysis and logical narrations, could only highlight an unknown entity, beyond the accounted for parameters, through the appearance of ruptures and breakdowns during the process of clarification, with the emergence of new paradoxes. This is because most conventional thought experiments have never been constructed to deal with the logical capriciousness of knowledge, derived from the performance of speculation, particularly when the knowledge does not have stable corollaries.

Nevertheless, thought experiments, for all the controversies regarding its usefulness, and its problematic relationship to public experiments[2], could be the hypothesis out of which a science fictional prototype can be built. At this juncture, I will not engage in a circular argument on whether thought experiments can validate the 'real' experiments, given how unproductive such attempts are. Instead, I would allow the rest of the chapter to demonstrate as a thought experiment, in and of itself, on what science fiction can do to enrich further, and extend the interpretive possibility, of science as knowledge in circulation above and beyond its own immediate milieu.

Some hard science fiction authors believe in adhering as closely to the logics of established science as possible, while keeping an eye out for informed speculation. Sometimes, pondering over how to reconcile the demands of fictional narrative with the physical constraints posed by science could bring about an unexpected outcome for the science. In his essay, "When Science Writes Fiction," Robert L Forward provides an example of how, while trying to come to terms with a knotty problem concerning where to put a neutron star in relation to a spaceship that had embarked on a mission to study the frontiers beyond the solar system, he came upon an idea involving a thought experiment and mathematical calculation that would allow the starship to stay near to the star without being torn apart by the star's massive gravity: this is done by the placing of six ultra-dense masses as counterforce.

---

[2] In this book, *Thought Experiments*, philosopher Roy Sorensen notes that "…Those who believe that thought experiments justify and test hypotheses face a dilemma that we can formulate in terms of their connection with public experiments:

1. If a thought experiment can be checked through public experimentation, then not actually checking leaves the results unverified, and an actual check would render the thought experiment redundant or misleading.
2. If a thought experiment cannot be checked through public experimentation, then its results are unverifiable.
3. Any experiment having results that must be either unverified, redundant, misleading, or unverifiable is without scientific value.
4. No thought experiment has a scientific value." (48)

More remarkably, Forward was able to turn this idea into a publishable scientific paper even if his earlier intention had merely been about writing a credible science fiction story![3] However, I am not sure if I agree with his contention that science can write the fiction if one follows its lead, as this goes against the point of what fiction is meant to do, as was earlier argued. But, should one be able to stay true to the science while allowing aesthetics and social purpose to flourish, one could demonstrate more conclusively that the internal values of science cannot be dissociated from the social conditions enabling the former's production. After all, why should the actual world be constricted by realist interpretations that are just our way of straining at objectivity in the course of our knowledge production, itself a questionable virtue? As Whitehead puts it astutely:

> That actual world, in so far as it is a community of entities which are settled, actual, and already become, conditions and limits the potentiality for

---

[3] See *Hard Science Fiction*.Eds. George E. Slusser and Eric S. Rabkin. Carbondale & Edwardsville, IL: Southern Illinois University Press.5.

> creativeness beyond itself. The 'given' world provides determinate data in the form of those objectifications of themselves which the characters of its actual entities can provide. This is a limitation laid upon the general potentiality provided by eternal objects, considered merely in respect to the generality of their natures. Thus, relatively to any actual entity, there is a 'given' world of settled actual entities and a 'real' potentiality, which is the datum for creativeness beyond that standpoint. This datum, which is the primary phase in the process constituting an actual entity, is nothing else than the actual world itself in its character of a possibility for the process of being felt. This exemplifies the metaphysical principle that every 'being' is a potential for a 'becoming.' The actual world is the 'objective content' of each new creation (65).

While many hard science fiction writers have successful careers in the sciences prior to writing fiction; some either become fulltime writers or continue with their scientific practice while writing on the side; an extensive background in the sciences is not prerequisite to one's ability to produce believable science in fiction (though, in some occasions, one might have to be prepared to engage with scientists, who as science fiction readers, may volubly disagree with the writer's fictional representation of a science). Nor should one be forced to conform rigidly to scientific facts in order to produce good hard science fiction, as good science does not always translate to readable fiction (as Jameson duly notes).

Science fiction is more than a platform for prototyping scientific ideas; within the former is the seed for foregrounding the form as well as the practice of science. Science fiction showcases a literary form that combines scientific praxis with socio-scientific ideology. For instance, there are elements of the fictional model at work in theoretical physics because the latter is about imagining a multitude of ways in which a problem can be approached that is dependent on experiments to materialize the solution. Until materialization can happen, the imagined narratives are uninstantiated potentialities.

Therefore, the closest physics counterpart to the hard science fiction writer is the theoretical physicist, if only because the latter shares the same passion for envisioning unrealized potentialities and for pushing the boundaries of what is possible while maintaining an allegiance to the laws of nature. When theoretical physicists decide on making specific arguments for the physics, they emphasize the important descriptive, and explanatory potential, of the theory; the theory could be deployed for producing tangible measurements of a phenomenon whence practical applications can be later derived. However, unlike the science fiction writer, the bread and butter of the theoretical physicists, particularly in this day and age of funding constraints, involves making predictions, and forecasts, that are not technologically impossible and physically insurmountable.

Even if the theoretical physicist is willing to speculate within the range of available horizons, his/her speculation is very much delimited by the mathematical and computational tools he/she works with. For the hard science fiction writer, on the other hand, there is less pressure to adhere to physical plausibility or to ground the characters in the fiction within our mundane everyday. Instead, there is more room for a science fiction writer to negotiate conceptual questions that are more metaphysical than physical, even while speculating on future scientific possibilities by 'inventing' a non- existent technology as a catalyst or mediator of that futurity.

The abovementioned contention parallels that between philosophers of physics and the physicists in general: the former is interested in recuperating older formalism that have fallen out of use with most working physicists, in case there might be an overlooked potential while the latter might not consider the revisiting of older theories to be productive (although some theoretical physicists might differ on this point, especially those who work on the foundations of quantum theory). Indeed, physicists, by and large, are more concerned with whether the shortcomings of the theories advanced would impede the development of a research program grounded on those theories.

Nonetheless, in the same vein that a science fiction writer might be interested in furthering a choice in the design of their subject-narrative, a theoretical physicist might be rooting for a theory they believe to be robust, with unrealized potentialities, whatever the theory's empirical status. Further, the preference for a particular theoretical method could also be the outcome of the philosophical beliefs of physicists who might perceive certain political imaginaries as arising from certain preferred theories; these theories not only inform our views of nature but also our attitudes towards life, given the interest in seeking a grand unified theory of the universe since the early twentieth-century. An example of such philosophical, or almost ideological, convictions can be found in Erwin Schrödinger's series of lectures that are compiled in *What is Life with Mind and Matter and Autobiographical Sketches*.

During the lectures, Schrödinger plays the devil's advocate by taking the role of a "naïve physicist" so as to be able to ask the big question: how can all of the physics produced contribute towards enlightening one on the constitution of life? This answer might be delimited by laws of physics that only knows how to locate life within spatiotemporal equilibrium without accounting for the immateriality of the preternatural. However, science fiction is able to provide different degrees of constrained speculations about observed and unobserved phenomena, with outcomes ranging from the next logical step all the way to the unexpected and even terrifying!

Such outcomes are not unimaginable were we to consider the consequence of a massless Higgs on all the known elementary particles of the Standard Model, and how that would change our view of the physical universe in relation to our quotidian world, particularly if physical laws can be imagined to behave differently. The effects of particular microphysical interactions on the socio-politics of our universe, and in both biological and cosmological space, are very real in *Quarantine*. The novel bridges the internal epistemic concerns located within quantum physical interactions with socio-political factors external to the hard sciences.

Darko Suvin, in the chapter “SF and the Novum,” discusses how science is an all encompassing “horizon” of science fiction, though not in the vulgar sense of “gadgetry-cum-utopia/dystopia;” and that a refusal to countenance the cause and effect of science in one’s fiction is to reject science fiction (67). But Suvin hastens to add that the credibility of the science fiction is independent of the rationale of the science (where the science may or may not be accurately depicted). Rather, we have to consider the point of that rationale in terms of what is displaced and what gets interpreted. Suvin expects the science fiction he advances to be explicable by the scientific method, however astonishing the phenomena portrayed, or else it would be nothing more than a fabulous tale.

However, this does not mean that the appearance of anything fantastical or ‘paranormal’ should be immediately dismissed from science fiction because the former’s existence merely hints at the limits of our epistemic situatedness and limits as observers. For instance, those, whose daily relation to physics is informed by the logics of a Newtonian/Galilean framework, might view the popular theme of time travel as fantastical. While real-world physics is interested in the potentiality of time travel, such an interest also comes with an understanding of the tremendous amount of physical (and metaphysical) barriers that one would have to overcome and how one conceptualizes the spatiality of time, as had been explicated by the principle of no discontinuity in Einstein's field theoretical equations of gravitation.[4] Even so, not everyone agrees on how to portray the essence of time.[5]

The common chord shared by science fiction and speculative theory is; eschew a definite end in favor of a stochastic approach; the novum that contains hidden variables could eventually be foregrounded as we dig deeper into the mesh of interactions within scientific fields of interest to isolate the faint signals being emanated to produce increasing evidence, and a composite profile, of a potentially novel entity or organism that had been far hidden. By integrating science fiction to the hard sciences, a nucleus of thought, fomenting in a critical methodology, is produced and then used to model, heuristically, more interesting, and epistemically varied, thought experiments. Given how measurements and interpretations are thought about through the strict parameters of quantum and classical praxis, the introduction of subjectivity brings with it new judgments, so that one may break out of what Whitehead refers to as the Cartesian "substance-philosophy."

New anticipations arise as new environments come into being, with the possibility of a new reordering or flattening of the currently dominant hierarchies stemming from a shift in original logic. Unfortunately, resistance to extra-normal shifting of logic will happen because of the differences in agreement over how one should conceptualize a scientific effect over a longer term, especially if the differences stem as much from ontological incommensurability as from the different physical scales whence the objects are interpreted. Moreover, we have to heed how the transposing across different scales can produce an illusion that hides potential paradoxes and the different manifestations of the same qualities because of the different epistemic rules governing the interpretation of these qualities. They all form the driving goals of speculative theory.

---

4. De Sitter, Willem. "On the Curvature of Space." Proceedings Royal Academy Amsterdam/ Koninklijke Akademie van Wetenschappen te Amsterdam 20 (1918): 1309–1312. Print.

5 See Louis de Broglie's "The Concepts of Contemporary Physics and Bergson's Ideas on Time and Motion." In *Bergson and the Evolution of Physics*. Ed. and Trans. P.A.Y. Gunter, where he discusses Bergson's resistance to considering time in spatial terms (the over-geometrization of time) while remaining aware of the delimitations of how time is thought about in classical physics.

Greg Egan, a professional computer programmer,[6] wrote *Quarantine* as a ludic experimentation with scientific interpretation. He models his fictional what-ifs against a futuristic dystopia that has little to do with actual quantum physics in practice, but rather, what he envisions as scientific concepts containing promising aesthetics for fictive deployment. In his dystopia, human characters have the unprecedented capability for performing the quantum mechanical act of 'smearing,' where they primed themselves in readiness for potential entanglements with one another and the environment they are a part of.

Most interventions into the narratives of quantum theory have been formalistic, if only because the narratives are grounded on formalistic logic derived of mathematical elements. However the origin story of the Higgs boson, while itself a narrative that began in mathematical formalism, has also become one about exploiting epistemic loopholes in order to tell a more exciting story about the properties of the interactive forces of nature. Had groups of physicists shied away from playing around with mathematical inelegance in the initial stage of its theoretical development, the narrative of the Higgs boson might have been different from what it is today. But in terms of the science fiction portrayed here, I am advancing a still rather preliminary view that was articulated in chapter two about speculative physics, without further argumentation, on the deployment of the narrative structures within science fiction to explore the ontological shortcomings currently encountered within the philosophical elucidation of the Higgs boson. However, that could only be accomplished beyond the dissertation, for this chapter aims first at establishing a critical dissection of the construction of such method by examining how a work of hard science fiction can be turned into work for prototyping philosophical ideas within physics, and vice-versa, beyond the constriction of mathematical formalism, before the latter is thrown into the mix (which is another trajectory worthy of further contemplation beyond this dissertation).

[6] See his website http://gregegan.customer.netspace.net.au/.

In *Quarantine*, Egan does the same thing with quantum mechanics, exploiting its philosophical and conceptual loopholes towards a creative end. While some might accuse him of discarding even the most flexible parameters of higher formalisms in his version of 'quantum interpretation,' I would argue, instead, that Egan is finding ways to navigate topics that have failed to make headway through scientific actualization, by textually imagining the what-ifs of a direct interface between the microstates and their environment, while 'compactifying' the interactions between events across different scales to magnify their effect. One might accuse Egan of focusing too much of his narrative energy on the *character* of quantum mechanics in relation to technologically-driven sociology to the exclusion of subjectivity in his human characters, therefore leading to an appearance of sterility in the characters' portrayal. However, Egan's characters are like nesting dolls; in order to penetrate into their core, one has to break through their layers, therefore delineating all that is observable about their behavior through their interactions with the science that shape their decisions and actions.

## 4.1 Reading Quarantine as Speculative Theory in the Epochs of Capital

*Quarantine* is set in a future not too distant from the time of writing, but which would have felt rather distant when the novel was published in the final decade of the twentieth century. Egan artfully combines the near-randomness of a semi-classical quantum theory (semi-classical in the sense that the measurement quantum theory is still dependent upon classical physical structures) with the mesoscopic world of neural networks in brain science.

I read the novel as divided into three epochs of individuated and collective consciousness. Each of the epochs can be read as variations on the original theme of the double-slit experiment, or what Karen Barad refers to as the which-path experiments that are alterable in accordance to quantum entanglements and the quantum eraser (where a chosen direction can leave behind traces or have the traces removed away from the path of detection). Such a reading parallels the creative license that Lightman took with special and general relativity theory in *Einstein's Dreams*, that is simultaneously surrealistic and impressionistic, yet succeeds in immersing the reader within the structural workings of the physics theories that inspired the narrative. There is little doubt that Egan is serious about the science(s) he chose to portray.

The players in the epochs are coupled in an ensemble to one another, not unlike the fictional Ensemble that represents both an emergent view of quantum states and an assemblage of collectivities (that are assemblages that will inform the three epochs): Nick Stavrianos and the corporations (the Hilgemann Institute and the Ensemble); Stavrianos and the 'quantum-like' women, Po-Kwai and Laura Andrews; Stavrianos and his dead/virtual wife Karen; Stavrianos and the Bubble. Temporality in the form of dates and numbers also take on significance as they mark points of transitions in the novel's plot, even if the transitions may not be overt, such as the background story involving the terrorist Children and the effects of the Bubble that shroud parts of Egan's alternate world in "darkness" by "shielding" the sun from the other galaxies. For the Bubble is "…an immaterial surface which behaves, in many ways, like a concave version of a black hole's event horizon. It absorbs sunlight perfectly, and emits nothing but a featureless trickle of thermal radition…(19)."[7]

Stavrianos is the consistent subject for all of the three epochs, with the temporality of the epochs being at once cinematic yet marked by extra-ordinary events that point to ruptures in expectations, and the clashing of different ontologies. Stavrianos is the center of power in the novel. Just as there are three epochs, there are three zones to power such as are explicated by Deleuze and Guattari: the zone of power, zone of indiscernibility, and zone of impotence. These zones are set in terms of flows and segments, of "financing-money, or credit money [that] involves the mass of economic transactions" (*A Thousand Plateaus* 249). Risks and credits flow through Stavrianos from various behemoths as he switched between his allegiances; from the flow came information that are cycled through his mods, which could be turned into Capital (the capitalization of capital represents a desire for totality) so that more trade could take place. All of these activities culminate into an abstract machine that churns out figures representing near-arbitrary values.

---

[7] The description provided by Egan is a fictive extension of Stephen Hawking's own explication of relativity and the event Horizon in *The Brief History of Time* (1988).

At the same time, we might ask if it is really possible to de-anthropomorphize human conceptions, given that the apparatus for performing the measurement, and the body that observes the measurement, require the measurement to be reconstituted into a sense-making that is always-already anthropomorphic. We get an explicit sense of embodied interpretation at work from the circular conversations Stavrianos had with the other characters in the book. Every character, however minor, plays a part in putting obstacles (constraints) in the path of the protagonist who had been sent to recover a severely mentally disabled woman who appeared to have escaped from a secured facility. Each of these characters, even the most improbable ones (such as the "aliens" who created The Bubble), are necessary as pawns (and needed for human interest) in an environment that offers an array of physical states, where the structuring of the chain of events have greater import than their contents.

On the surface, the quantum mechanics out of which Egan draws his inspiration looks like the standard Copenhagen interpretation that emphasizes the correlations between classical and quantum descriptions of the physical states. The description of the effects of physical interference and suppression of certain physical behavior, as represented by the concept of decoherence, points more plausibly to Egan stringing together the different readings of quantum mechanics in existence; from the famous Everett's many worlds and the de Broglie-Bohm deterministic version, to the lesser known objective collapse theory, and the Von Neumann/Wigner's interpretation that includes consciousness - the latter of interest to physicists and philosophers intrigued by how one can get from metaphysical causation to the production of physical phenomena.

One can accuse Egan of cherry-picking through the more intriguing aspects of the interpretations to formulate his own interpretation that employs fiction to make more literal, the speculations taking place in the micro-narratives of quantum mechanics, sometimes to a comic end, through the psychedelic invocation of smeared humans in a smeared world in the final pages of the novel. Even as Egan admits to certain idealistic, and inaccurate, understanding of quantum mechanics that have driven his ideas[8], the 'mistakes' he made are less important than the possibility he provides for other ways of thinking about science no less intellectually provocative and stimulating. In a sense, the idea of science, and less its content, shapes the narrative.

The first epoch of the novel constitutes a state of individuated being that relates to the force of Capital, yet is capable of holding the latter at bay; the ground of this epoch is represented by strict binaries, determinacy, and control. One catches an antagonistic tendency, for Deleuze and Guattari explain, "...Politics operates by macro decisions and binary choices, binarized interests; but the realm of the decidable remains very slim. Political decision-making necessarily descends into a world of microdeterminations, attractions, and desires, which it must sound out or evaluate in a different fashion" (244).

---

[8] See Egan's post-novel manifesto at gregegan.customer.netspace.net.au/QUARANTINE/QM/QM.html

The first epoch begins when Stavrianos, a former cop turned private investigator/consultant/missing-persons retriever, aligned himself with Capital's interest in 'covering-up' a potential liability in the form of a missing brain-damaged patient, Laura Andrews. In the early stages of the novel, Stavarianos was in his own state, though a state sliced through by a mod that enabled access to a virtual reconstitution of his wife. In fact, the mod-produced incarnation of his wife provides a rich illustration of the delimitations and extensivity of the next generation cybernetic construct that may become a possibility in the future, even as one ventures into the deconstruction of the gendered cyborg embodied by the Nick-Karen dryad, but I will not go into that.

Stavrianos speculated on the connection between Laura's disappearance and the activities of an apocalyptic group called the Children. It is here that Egan first projects onto a collective that is always in the background but never quite took center-stage, yet is the glue that link together events that are apparently non-causal or non-deterministic. In addition, the tragedy of his past had contributed to Stavrianos leaving the employ of the police force (thereby extricating himself from one form of collective) and becoming a mercenary (a return to the individual).

Despite his appearance of individuation, Stavrianos had spent much of his adult life looking for a 'fix' that would allow him to extricate himself from a chaotic (emotionally-impacted) existence so as to be able to live in a world where the cause and effect of actions are purely deductive/inductive, and therefore, more predictable. In fact, many of the events of the novel such as the availability, for purchase, of devices that could modify one's biological programming, particularly the section that controls for conscious behavior that include voluntary and involuntary neural functions. He has startling voluntary control over every aspect of his physiology

> The woman moves in front of me, holstering a gun. From a pouch on her belt beside the holster, she produces a small hypodermic capsule. Stepping over Laura, she takes hold of my jaw with one hand – I *lower my hear rate* – slides the needle into a vein in the side of my neck – I *constrict blood flow to the area* – then squeezes the capsule…Reduced circulation will buy me a few seconds, at best, but that should be long enough for P1 to make an assessment…(75).

However, such over-reliance on body and mind 'apps' can also be read as a critique of societies enamored by, and addicted to, artificial modifiers and contraptions, as means of control and power. After all, it is possible for these modifiers to produce unintended effects, as is demonstrated multiple times in the novel such as in the not-always-predictable behavior of the mod-induced Karen.

With this mods, Stavrianos becomes a cyborg with enhanced mind control options for managing the natural needs of his mind and body. However, unlike Haraway's cyborg that is aware of, and struggles with, the ideologies that keep the body shackled, Stavrianos, at least in his initial state, embodies indifference to the ideology of his paymaster; his risk-taking comes not out of political conviction, but rather, of an informational clout that could be applied as leverage. Stavrianos could aptly be read as the Foucauldian liberal biopolitical subject, a subject that is governed by the multiplicity of mods that enable self-regulation while assuring some level of juridical oversight. He is antithetical to conservatives, such as his own late wife, who frowned on such liberal applications of the invasive mods, until circumstances force compliance:

> Karen had no professional mods; doctors, the eternal conservatives, still frowned upon technology – but differential malpractice insurance premiums, amongst other things, were gradually eroding their resistance. (59)

Yet, Stavrianos represents a form of resistance at the level of the ontological, even if not epistemological; ontological in the sense of the awareness he has of the mod-induced code-switching he engages in that shapes the peculiarities of his individuality down to the level of the molecular. However, he is considered to be epistemically invariant because he remained reliant on the mods and the programming of the mods shaped his engagement with the world. Yet, in the Simondonian sense, he moves beyond a single individuation to begin incorporating these individuations into an enworlded system, to produce states where the individuation of a collective can emerge.[9]

---

[9] See Gilbert Simondon's *The Position of the Problem of Ontogenesis*. Trans. Gregory Flanders. *Parrhesia* 7 (2009): 4-16.

Egan uses his own knowledge of computing technology to draw parallels between neuro-genomic potential and the neural networks performed by artificial intelligence (and other computational models), including the nanobots, as is demonstrated in a scene where Stavrianos obtained a nasal spray containing a heavily modified bacteria that can conduct "nanomachines" to his brain and inhibit the act of 'collapsing' states. Ray Kurzweil discusses many of the same ideas in his book *The Singularity is Near: When Humans Transcend Biology*, by positioning the evolution of technology against biological evolution, inspite of their different timescales.

What Egan began writing about in the 1990s (although such ideas were already circulating in niche speculative science fiction publications such as the *Analog*), he predicted many of the same ideas that Kurzweil did, writing in the twenty-first century. The 'imagined' mods that allowed Stavrianos to mediate the individual psychic with the collective are not unlike the quantum-theoretical Fock space that 'builds' into a single particle greater 'flexibility' for dealing with other similar particles. Therefore, the first epoch shows how that even if Stavrianos appears to hold onto his agency, that agency is entangled with the affective memory of his late wife as the voice of conscience that he tried to quash, though not always successfully, since her 'superposition' onto him can invoke unforeseen somatic and cognitive interference that destabilizes his certainty and conception of the self.

However, for the most part, he was able to maintain a certain control, up to when their increasing divergence from each other gave shape to her increased autonomy.

In the second epoch, suppression is intermingled with resistance; desire for knowledge is coupled to the deployment of the same individual agency for strict control in the distribution of that knowledge, and the zone of power becomes both centralized yet de-centered by an internal rebellion. The process of 'smearing' out of a single concentrated zone of power, often complicated by entanglements that are not all positive to those involved, takes center-stage in this epoch even as the ethical relationships between the characters, and between the characters and their environment, are ambiguous. While Capital wants to appropriate the individual's agency and indoctrinate the latter with its ideology, thereby ensuring compliance from its subjects, subterfuge could still take place even among Capital's most loyal adherents, if only because they do not trust the former's implicit ability to maintain the purity of the ideology. But what is circular about the argument is that the insistence on the purity of ethics (or purpose), in a sense that suggests an arbitrary distinction between ideological purity and pretense, does not account for how that distinction can only happen after a choice has already been made and the implication of that choice revealed.

The circular argument of ideological purity is demonstrated when Stavrianos goes from 'free' agent to a supposed 'loyal' employee of an organization (or an assemblage of organizations) standing in for the Ensemble, where the notion of free-will, survival, and intrigue are not so clearly demarcated. As we have seen in the first epoch, Stavrianos's decision-making process is amplified to a state of unwavering belief in the strict order of logic due to his mods. But his capture by a member organization of the Ensemble led to their commandeering of his mods, and the installing of a loyalty mod. Over the course of his biologically mediated indoctrination, Stavrianos was convinced, or was made sufficiently suggestible to be convinced, that his role was to serve the Ensemble; he was to guard Po-Kwai, a newly 'recruited' college graduate who was to become the observer-participant in an experiment involving quantum states that was reminiscent of the original Stern-Gerlach experiment on particle spin.[10]

Nevertheless, Stavrianos's newfound role was quickly destabilized when it was challenged by one of his superiors, Lui, who insisted that they had been tricked into serving a "sham" Ensemble. The former was then introduced to the "Canon," a secret society of sorts who saw themselves as the upholder of the true Ensemble, though their strict insistence, as Stavrianos observed, did not provide greater clarity on the meaning of that claim. However, relative to the vision, the one goal shared by the rebels, as represented by the "Canon," was to retrieve the eigenstate mod that had been programmed from the thought experiments performed with Po-kwai.

---

[10] In this experiment, a beam of silver atoms is shot through an inhomogeneous magnetic field and the direction of its deflection is noted. Since only neutral atoms are used, the deflection is completely dominated by the spin effect of the atoms. The inhomogeneous field is needed to ensure that no classical effects will cancel out the deflection of the particle. More of the fascinating story of the theory and experiment that do not always match up can be found in the December 2003 copy of Physics Today, in the article "Stern and Gerlach: How a bad cigar helped reorient atomic physics" by Bretislav Friedrich and Dudley Hershbach, both scientists. Karen Barad also discusses the quantum foundations of the experiment in the chapter "Entanglements and Re(con)figurations" in *Meeting the Universe Halfway: Quantum Physics and the Entanglement of Matter and Meaning*, pp 258-65.

In the conversations Stavrianos had with Po-kwai and Lui, they discussed how the superpositions of particular eigenstates (which refer to unique microstates whence a fixed value of a physical property is assigned to each individual state), with an initial indeterminate condition, could bring about a determinate final outcome. At one point in the narrative, Stavrianos had a close encounter with Laura, or one of many versions of Laura, who, despite her mental incapacity, had successfully smeared (or tunneled) herself out of physical barriers, and was able to find her way to him in a manner that both confused and intrigued him.[11] In their conversation, she warned him that the attempt at collapsing an infinity of states was what brought about the Bubble in the first place; the latter was caused by an unknown intelligent life that had the sophistication to manipulate the cause-and-effects of quantum mechanics in a way that could affect the how other intelligent lives would end up perceiving tangible outcome of that seeming entanglement, and collision, between events in the microphysical and macrophysical world.

[11] The creation of a mentally challenged character that appears to have mastered, physically, the manipulation of quantum mechanics, adds a notch to Egan's intentionality, and to what might have been his theories of the mind in relation to the neurological function, none of which are given much space in the novel. However, his deployment of the neuronal modes as a way for manipulating and controlling the most subjective elements of the mind, such as emotion, appears to suggest that certain behaviors of the mind are attached to one's somatic condition, and can be altered at will once we have access to the right code for doing so (a subtle allusion to Egan's background in computer science). However, it is clear that Stavrianos is intrigued by this seeming capability that Laura possesses even though he could not explain how she had done what she did so successfully. He had to choose between accepting, or not, the rationale provided him concerning how Laura had been able to smear her way out of the corporeal trappings of her disability. Her ability to smear, and therefore create instances of herself in the process, denotes personal agency, even if the outcome seems to be involuntary.

At the same time, she revealed the mystery behind her supposed escape from the Hilgemann Institute. In the dialog between Stavrianos and the not-Laura (but one of the smeared states of Laura), Egan attempts to demonstrate the collapse between perception/consciousness and physical reality, a view that resonated with some earlier views of Bohm and Eccles. Here, we can see a pseudo existential crisis of morality that resonates with the point about quantum states and morality by Bell that I had discussed in the previous chapter. However, the point of existence is not in dispute; what is in dispute is the veracity of the claim to existence:

> Who are you? Are you Laura? *Are you real*?
> She laughs. "No. But your perceptions of me will be. I speak for Laura – or Laura-and-the-smeared-Nick-and-Po-kwai, and others. But mostly Laura."
> "I don't understand. You 'speak for Laura?' Are you Laura, or not?"
> "Laura is smeared; she can't talk to you herself. She's talking with the smeared-Nick-and-Po-kwai, but she's created me to talk to you."
> "I -"
> "Her complexity is spread across eigenstates; the two of you would never interact directly. But she's concentrated enough information into a single-state mode to communicate the essentials. She's made contact with the smeared-Nick-and-Po-kwai – but they're childlike, unreliable. Which is why I'm talking to you." (230)

Is the process of 'smearing' in the quotation above merely a neurological process with physical consequence, and if so, how does crossing between the different states take place? According to Bohm in *The Undivided Universe: An Ontological Interpretation of Quantum Theory* that was published in 1993 (Egan's book was published in 1992), neuroscientists perceive the brain as derived from classical concepts of physics, with humans perceiving the classical world, with all its implications, happening simultaneously in the quantum world. Bohm adds:

> However, some neuroscientists, notably Eccles [12], have suggested that quantum processes may be important in understanding the more subtle activities of the brain. For example, as has already been pointed out, we know that retinal cells respond to a few quanta at a time and that this response leads to a multiplication of their effects to a classical level of intensity. But the retina is just an extension of the brain. There could evidently be other parts of the brain in which such a sensitivity may exist, e.g. in certain kinds of synapses. If this were the case, then the brain would be a system that could, like a measuring apparatus, manifest and reveal aspects of the quantum world in the overall processes. Such quantum sensitivity would imply that in more subtle possibilities of the behaviour of the brain, a classical analysis would break down…All this means that as the processes of perception unfolds into the brain, it may as it were connected to the subtle quantum domain which latter [sic] may in turn reconnect to the classical domain, as outgoing action is determined through amplification of quantum effects (179).

One could not be accused of suspecting that such ideas in circulation at the time would have influenced Egan's own views of how the brain performs quantum collapse, even if he had not read Bohm. Nevertheless, at the fictional level where preferred realities are amplified while others ignored, Stavrianos and his neural mods embody these ideas; the confluence of the man and his mods were sufficiently intense that Stavrianos was able to erupt out of classical determinism and connect with the quantum world.

Moreover, there has been a time, in the 1980s and 1990s, when interest in connecting the mind with quantum mechanics became popular, particularly as the shrinking scale of the electronic instruments coincided with developments in supercomputing and artificial intelligence (it so happens that a matchbox size supercomputer also made an appearance towards the latter third of the novel). The consciousness and quantum theoretical connections had especial resonance with the work of David Bohm and Henry Stapp; both men went back to the most fundamental questions asked in the early days of quantum theory to steer quantum mechanical inquiries into the zone of ontological questions, questions that had been set aside for reasons of expediency.

When the information retrieval processes of the brain is exponentially increased as a consequence of quantum manipulation, Egan predicts the perfect quantum supercomputer, the kind that Kurzweil also predicted in connection with the human intelligence architecture. While Kurzweil stays mostly within the boundaries of the biological, Egan is interested in correlating the quantum with the biological through his demarcation of the role of observation in 'smeared' quantum mechanical states (or states that act simultaneously across different spatio-temporal dimensions), leading to the collapse of one preferred state out of a range of possible states, where the preferred state is not always a process of intentional preference. Egan then correlates these states with the mesoscopically-scaled neural circuitry.

With Stavrianos, Egan is able to demonstrate how a neuronally connected quantum computer can break through the infinity of states to pick out a slice for further exploration, and immersion.

Po-kwai, who played the role of a 'neutral' observer, led a highly regimented-lifestyle to maintain her 'purity,' one that is as much literal as it is metaphorical. She is representative of a 'pure' physical state prior to entanglement, and her subsequent entanglements would edge her towards a state of such metaphysical finesse that she was able to weave out of her macro-corporeal being at will. While in the initial state of physical 'purity,' she is as much an observer as she is part of the measuring apparatus, in her 'smeared' state of 'unlimited' potential, with almost no, or merely negligible, external interference prior to her state of 'collapse.' As a macro-classical being, she was also a detector, who in the process of engaging with a quantum state, became sufficiently entangled so that she was able to influence the eigenvalues of the states, and therefore, reached a point where she could manipulate her way into a 'preferred' state.

Soon enough, her entanglement with the quantum state made her a part of its continuous probabilistic structure (known as density matrix), and she was able to affect certain properties of a fundamental particle that 'tunnels' through the walls that stand for energetic potential barriers. A realist might insist that Egan's portrayal of his character's ability to walk through solid walls and firmly shut doors, while in the state of hyper-excitation, as flights of fancy, a misapplication of quantum interpretations, and utter disregard for mathematical and physical constraints.

However, Egan is less interested in performing just any ordinary thought experiment in quantum physics than in breaking down the tendency towards normalization by utilizing the cognitive estrangement science fiction affords. Such an acts of estrangement increase the level of bizarreness in the already non-intuitive interactions of a quantum world to create an information rich environment, with unlimited degrees of freedom (or freedom as constrained by the text). In other words, quantum theory, and its epistemic concerns, is supervened by a fictional counterpart for extending measurements of interpretive entities that are even more complex than the multi-particle interactions.

The connection Egan draws between neuroscience and quantum mechanics may be tenuous, but not entirely imaginary. Tractions in quantum computing and quantum artificial intelligence have opened up new vistas to decade-old questions on the connections between mind, matter, and, consciousness. Moreover, merely operating from the confines delineated by the logic of physics have not necessarily brought about a breakthrough for explaining the more 'exotic' speculations whereas the injection of a 'fantastical' imaginative process could push one into defamiliarized territory, and in the process, enable new 'flavors' of thought experiments to burst forth.

Quantum mechanics and neuroscience are subjects of sexy popularizations but remain shrouded in mystification. No physicist can own to understanding the ontology of quantum theories. As more experiments for demonstrating the interpretation of quantum theories are realized, a greater variety of experiments become available for the testing of an array of interpretations, particularly of quantum information theories. The availability of these experiments allows one to probe the materiality of what has been imagined by rendering more transparent, the 'blackbox' of processes that enabled the inexplicable phenomena in the first place.

It is in the third epoch that we find a quantum-to-mind interfacing that releases Stavrianos from the ideological stranglehold of the corporation's loyalty mode and giving in to the demands of the 'Canon.' He went back to becoming an individuated being, but with a difference: he has now become the master of deception and dis/misinformation as he no longer served anyone but what he considered his principles. However, it was not only Stavrianos who was changed, but also Po-kwai, as their smeared selves gained increasing strength so as to become independent of the true Stavrianos and true Po-kwai for longer periods of time, until the possibility of permanent smearing is achieved. Nevertheless, readers (at least the author) are left with unequivocal certainty that it is the 'true' Stavrianos and 'true' Po-kwai who are the ones speaking to them because nowhere in the narrative, nor in the actions following the conversations between the characters, is there a challenge to that assumption. One can argue in this case that Egan had failed to marshal the affordances of indeterminacy to challenge readerly comfort by being less predictable in his narrative technique.

No longer bowing to capitalist demands, and knowing that a hidden motive existed whichever way he turned, he applied whatever assets he had left to prevent 'reality' from being completely wrested from the control of the true selves by attempting to prevent the entire world from smearing, which would have happened were Lui to let loose the pathogenic vectors. Were universal smearing to take place, the veil created by the Bubble would be torn, and chaos could reign.

> So I wait like a human: sick with pointless unproductive fears. Trying to imagine the unimaginable. If the whole planet smeared, permanently…exactly would people experience? Nothing – because there is no collapse to make anything real? Or *everything* – because there is no collapse to make anything less than real? *Everything, separately* – one isolated consciousness per eigenstate, like the many-worlds model brought to life? Or *everything, simultaneously* – a cacophony of superimposed possibilities? What I've been through myself – or at least *those memories which have survived the collapse* – might bear no resemblance to the nature of things when there'll be *no* collapse at *any* future time (253).

In this epoch, we reach a state where the properties hitherto suppressed in epoch two, and which were nascent in epoch one, rose above ground and bloomed. However, unlike the variations in the Stern-Gerlach experiment and the double-slit experiments (and even thought experiments of the Schrödinger cat), Egan's finale does away with all physical barriers by making the dispersion of information load, through the metaphor of public infection, the beginning of the end of certainty. The time of the capitalist, a time considered as analytic by Antonio Negri, is broken, and a new productivity system that is not a mere reproduction of the original command line is now possible.

If the climax of the novel signals the end of bourgeois science, it also highlights a new form of measurement and interpretation, where there is no longer a distinction between the quantum rules regulating the entities measured, and the semi-classical to classical embodiment of the apparatus performing the measurement. Reproduction is still taking place in the closing chapter, as signified by the vectorial multipliers, but it became less clear, ethically, if a new bio-terrorism had just broken forth or a new physical (and therefore mental) liberty of a new order had been achieved.

> The streets seethe with transformation. Some people's features are shifting, flowing smoothly or jumping between alternatives; walking in a daze, they seem oblivious, and I touch my own face, wondering if the same thing is happening to me. Vegetation is sprouting everywhere – patches of wheat, sugarcane, bamboo; stretches of wild-looking tropical undergrowth. Some stalls are simply crumbling into fine dust; others are mutating into exotic architectural pastische – and the walls of one have turned to flesh, blood visibly pulsing through veins as thick as my arm. I stare up at the skyscrapers, most of them surreally intact – but even as I wonder at this, the fractal cladding on one tower starts drifting down like confetti (273).

The view one gets is almost cinematic as the reterritorialization of the metaphysical and the temporal as a new punctual system with new creations is elaborated, and the becoming is made into a state of purity unconstrained by the imposition of mundane laws or reductive formalisms. Were we to view the last segment of the novel from a media archaeological standpoint, one can treat the events as representative of rapid evolution into new technical bodies as differentiations converge into unifying processes parameterized by newly developed haecceities.

The third epoch of the novel feels uncompleted; the veil has been broken but the reader is left wondering as to what might have been. We could speculate on the philosophical conclusions to be drawn about measurement had the narrative not ended just as new predictions enter. However, the intention in the ending of the novel can be read through the limits by which one imagines a new order that does not have a counterpart in actuality, especially when imagining the steps of problem-solving involving fictional modeling that can later be actualized into a physical representative mode. One can imagine how the current known methods of measurement, and the ways for demarcating them, might change once we enter into new realities, new physics, and new scientific hybrids. But the fact that we have premised our understanding of the new as counterfactuals of the old makes it difficult for imagining a new that is dissociated from current epistemics, until something that shatters our epistemic comfort comes along, to demand completely new interpretations.

## 4.2 Conclusion

My objective of reading *Quarantine* as three epochs is to demonstrate its physical (and metaphysical) instantiations that are the products of Egan's own iconoclastic reading of quantum theory in combination with developments in the power of computing and neuroscience. But whatever the original intention of the author, the layers, segments, and meta-narratives contained in the novel provide both literal and less direct illustration of the philosophical difficulties of conducting measurements across different scales (and standards of comparisons), and what could have taken place should the borders of delimitations (and natural constraints) be lifted from the interactions of entities that do not usually cross paths.

Each of the epoch is replete with speculative extensions of a science while subtly critiquing the institutional Capital that enables the production of the science, as well as the emphasis on the dominance of a subfield of a science over that of the other, or even to privilege one form of interpretation over another. What I had not been able to further explore, but which *Quarantine* provides an allegorical insight into even if only in a highly mediated sense, is the transmission and dissemination of an interpretive mode through institutionalized politics and epistemic practices. Such interpretive modes are highlighted by the foundations underlying the creation of the mods constituting the cyborg human race of Egan's future, as well as the 'secret' experiment conducted by the Ensemble with their strenuous efforts at confidentiality. Further, Lui's almost fanatical desire to spread, or 'infect,' the public with the organic vectors that 'carry' Egan's interpretation of quantum mechanics represent the risks, utopian thinking, and self-interests of the supposedly politically impenetrable scientific knowledge.

Just as the novel is a fictional reconstitution of actual physics and neuroscience current at the time of their depictions, it also foregrounds the goals of speculative theory through the former' demonstration of the fictional modeling of re-interpreted facts while highlighting the problem of epistemic choice and how to circumvent a potential epistemic disaster by taking advantage of the loophole provided in a theory or formalism, as is demonstrated in the case of the history of the Higgs boson in the previous chapter.

Finally, to reiterate an earlier point, the three epochs of the novel are allegorical to the three variants of the double-slit experiment discussed in the previous chapter to show the gradation of increasing indeterminism, even as I am not suggesting that the fictional modeling of the double-slit experiments bear any resemblance to the novel. Rather, I am suggesting that they share conceptual filiations due to their individual relationship to quantum mechanics. In light of that affiliation, the novel highlights how all that underlies the determinate form of the media technology we consume in our mundane macro universe is the outcome of ontologies stripped of their subjectivities, and whose subjectivities have no direct bearing on our everyday. However, in the alternate world of Egan that conforms by different scientific laws, it is no longer possible to ignore the causal effects of such subjectivities. In chapters to come, one will see a two-prong realization of speculative experiment through the experimental search for the Higgs boson, and the narratives of two science fictional short stories and a novella.

*Scientific, objective truth is exclusively a matter of establishing what the world, the physical as well as the spiritual world, is in fact. But can the world, and human existence in it, truthfully have a meaning if the sciences recognize as true only what is objectively established in this fashion, and if history has nothing more to teach us than that all the shapes of the spiritual world, all the conditions of life, ideals, norms upon which man relies, form and dissolve themselves like fleeting waves, that it always was and ever will be so, that again and again reason must turn into nonsense, and well-being into misery?* (**Edmund Husserl in *The Crisis of European Sciences***)

# 5. Observation or What You See is Not What You Always Get: the Impossibility of Observing the Unknown

This chapter continues the conversation begun in chapter three that considers the critical historiography of the Standard Model Higgs boson through the perspective of speculative theory. As the story of the Higgs boson is the center of tension between new physics and the confirmed Standard Model, the choice of going either way becomes a choice over how a measurement is analyzed and interpreted; speculative theory provides the theoretical platform for that contestation to take place, as well as anticipation of crises that would then be experimentally resolved.

When measurement is enacted in a physical world, there has to be a decision made on what counts as an observable, and whether it is necessary to demarcate between an observable and an unobservable, or merely stick with what the available predictive models are able to reveal. Therefore, the degree of definability of the states being measured, or considered logically plausible, in a theory that could be compared to other overlapping or adjacent theories, is a pre-requisite to whether observational terms are considered as a direct correlation between a mathematical proposition and physical phenomena, or as inhabiting a separate world that brackets out theoretical terms as distinct from an observation-centered language. The reflexive connection between measurement and observation is where we begin considering the role of speculative experiment in consolidating or constructing new theories. Before that, let us consider a few quick examples that could demonstrate the role of observation during and after measurement.

One of the most obvious examples in particle physics is the observational term for denoting symmetry breaking in relation to the Higgs boson, and its points of parallel and convergence with the theory of the superconductor. This is because particle physics and condensed matter share a commonality: the transmission of energy between jostling (and oscillating) pairs of electrons (Cooper pairs) and phonons (considered a set of 'gapless' or dispersing modes that is a result of short-range interactions between bosons oscillating at a particular momentum), the violation of electromagnetic laws (through gauge symmetry breaking), and the Meisner effect (the expelling of the external magnetic field of the superconductor material once the temperature drops below a fixed critical level) that is similar to the gauge field obtaining a mass after the symmetry breaking of the gauge theory for electromagnetism. Another intersecting point between particle physics and superconductor physics is the 'energy gap' model - the difference between the normal energy and superconducting/excited energy mode, which Whitehead quaintly referred to as *quantum jumps* in *Science and the Modern World* - for describing energy at ground state in relation to other energetic states - so as to account for electromagnetic properties and demarcate the interactions of electrons and phonons within the confines of a system that operates by the rules of Fermi-Dirac statistics.[1]

Superconductor theory is useful for demonstrating quantum mechanical measurements because of the simplicity of its structure, and the simulation of supposed 'massiveness' of the photons to reproduce the effects found in the aftermath of the aforementioned symmetry breaking. The discovery of this dilemma anticipates a parallel problem in the relationship between the Higgs boson and the Goldstone theorem, as was discussed in chapter three. However, the seeming paradox becomes less paradoxical once we appreciate that there is a hidden symmetry that provides a logical description to such disjunctive behaviors in the particles.[2] What is observable in the measurement process is initially dependent on the configuration of physical parameters, including the latter's removal and addition from the boundaries delineated by the measurement.

---

[1] See Bardeen, J. Cooper, L. N., and Schrieffer, J. R. "Microscopic Theory of Superconductivity," *Physical Review* 106.162 (1957). 162-4.

[2] For an accessible account, see chapter eleven, "Nobel Dreams" in Sean M Caroll's *Particle at the End of the Universe*.

As the example above demonstrates, regardless of whether we relegate theoretical language to a more materially concrete syntax or a transitively predicated observation-centered language that maps experiential instances to physical behavior (the example above represents both instances), experimental epistemology is about making empirical associations regardless of the theoretical density of the language. Therefore, an observation language should include a capacity for computation while ensuring consistency in the language's descriptive capacity.

Therefore, I share with Mary Hesse the idea, as articulated in her essay "Is There an Independent Observation Language," that an observation language is not immutable regardless of the physical laws, and that a classification system is only useful if it can be modified to provide functionally useful descriptions of objects/subjects as new discoveries are made, with emerging knowledge of hitherto unknown properties causing one to question what should change and what could be retained. Hesse also brings up the problem of an inter-subjective language of observation, and whether such an observation language can be properly independent of theory-ladenness, an approach Einstein had apparently tried to move out of through an 'operational' approach not biased by "theoretical implications."[3] However, to suggest that one can retreat to an operationalist mode that is devoid of theory-ladenness, by excluding all references to any paradigms that might suggest the presence of a theory, is a naïve anti-realist position because it assumes lack of presuppositions governing the articulation of any scientific argument through a purely objective and 'blank slate' approach. Such an argument is unconvincing and improbable given that one could not design an experiment without assuming parameters that have built-in theories.

---

[3] See Hesse, Mary. "Is there an Independent Observation Language" in *The Nature and Function of Scientific Theories*. Pittsburgh, PA: University of Pittsburgh Press, 1970.

Even as the potential predicted by theory would drive early proposals for experiments to test the theory, experiments also construct their own theories, theories that were the aftermath of quantifiable measurements and observations direct or indirect. One might argue that there is an observation language independent of an all-encompassing Theory, but it is misleading to think that its own meta-narratives would have no theoretical consequences on the act of observation. In fact, the theoretical consequences of the meta-narratives of the theories could be analyzed through the historical genealogy of the formative influences, and considerations, behind the theory. Therefore, the elucidation of theory is full of contradictions and subjectivities at the level of formation when viewed only from a physical direction due to logical ruptures marked by the non-observability of certain processes that require a more metaphysical approach by way of further explanation of the seeming illogic. Also, the illogic is also the result of an emphasis on a particular mode of thinking over that of others, and it is that thinking that drives how certain theories are articulated, and how the experiments are then designed in relation to, and also in contradistinction, to the predictive theories.

According to Willard Van Orman Quine, attributing probabilistic certainty to events is an act of speculation anchored not on what is known but on what is believed. There might be an appearance of surface-level contradictions that can be finessed once the underlying ontology of the contradictions are uncovered, and therefore, resolved. But as Quine's web of belief does not allow for a consideration of contradictions, it is therefore insufficiently capable of dealing with a complex web of paradoxes within the Standard Model that are the outcome of our incomplete comprehension of the Model's ontology and lack of prescriptive capability. While Quine is right in expressing his concerns with the measurement of subjectivity, he does not leave any room for the possibility of adductive thinking that transcends the usual deductive and inductive modes of thought.

As he puts it, the observation language (and sentences) of science, as conditioned by strict evidential and positivistic correlation, could only point to what is externally obvious and verifiable by another witness but leaves no room for internalities. But what he is right about is that the element that makes science both hard yet possible is "…because it must build a coherent system from the diverse evidence gleaned and reported by people of different times, places, cultures, and interests…"(Ullian & Quine 29).

In the next section, I will discuss the searches that led to the final about confirmation of the Higgs boson by discussing the construct of the detectors involved, the design and triggers of the experiment, and the outcome of the searches that provide a composite profile of the Higgs boson. In the process, I intend to demonstrate how the discovery of the Higgs boson is an act of speculative experiment because of what it can and cannot tell us about new physics, and also because the analytic framework of its discovery is based on the current best understanding of quantum field theories that could still be further revised even if they continue to be able to make Standard Model predictions. I will then end by a brief foray into what this means for candidate theories, and their potential proofs, beyond the Standard Model.

## *5.1 The Fiction of the Higgs boson: Dimensions of Observability*

The Large Hadron Collider (LHC) consists of an agglomeration of different experiments occupying the different sectors of its mammoth complex. The four main experiments are ATLAS (A Toroidal LHC Apparatus), CMS (Compact Muon Solenoid), ALICE (A Large Ion Collider Experiment), and LHCb (Large Hadron Collider – beauty/bottom). Each of these experiments were constructed with goals, constraints, and methods of measurement of particle properties that will manifest a sort of physical materiality that contribute to the construction of the Standard Model, and possibly beyond. In this chapter, I will be focusing specifically on the ATLAS and CMS high-energy experiments because of how both experiments converge in their examination of different entities produced from the same high-energy LHC collisions. The experiments involve searching for the Higgs boson, whose mass the Standard Model does not directly predict. Indirect estimations are however obtained through the measurements of other particles, most notably the masses of the top quark and W boson with contributions from the Higgs boson.

Each experiment provides a multi-faceted approach to solving theoretical puzzles such as parity violation (an inversion process that violates the spatial reflection symmetry in weak interaction, a phenomenon not observed in electromagnetic and strong interactions), spontaneous symmetry breaking that is necessary for the Higgs mechanism, and a variety of foundational questions in relativistic quantum field theories. After all, one of the major goals of the LHC experiments is to explore vector boson interactions while figuring out the place of the scalar Higgs boson in the process of trying to achieve unification among all the available forces, finessing the relationship between matter and anti-matter, and understanding how different kinds of measurements can be enacted at the intersection of quantum interpretation and the actual experiments. As a general-purpose detector, the LHC tries to measure all of the particles of the Standard Model as best as it can. Whatever form the physics beyond the Standard Model takes, it would ultimately leave signatures from its decay corresponding to the known properties of the bosons, quarks, gluons, charged leptons, and neutrinos of the Standard Model.

The signals, signs, and codes are the fragments of clues that contribute to constructing a picture of the universe's ontology. The LHC, by way of colliding beams of protons, provides the starting point for enabling the different experiments located at varying points around the LHC ring to record all the possible outcomes from the energy produced during each collision. The inscriptive process 're-writes' the theories so that they might, in the end, converge or diverge from predictions. For a more thorough understanding of the role of instruments in performing measurements and observations, one must be cognizant of the kind of technologies built into each segment of the experimental apparatus to ensure accurate calibration and fine-tuning of the apparatus prior to the launching of the experiment.

For the purpose of this chapter, I will be discussing the detectors from a general perspective rather than speak to specific differences between the ATLAS and CMS detectors except where necessary, given that much of what they do is similar. At the same time, the description of the morphology of the detectors is also critical to considering how the different specialized sub-sections are important to the different speculative choices of theory that will then transfer into the experiments, as well as experiment's own limited attempts at working through the different what-ifs of available data for analyzing traces of the Higgs boson's presence.

The CMS and ATLAS experiments have similar yet different tracking systems: CMS uses a pixelated strip detector made of silicon whereas ATLAS uses a hybrid tracking system that consists of the pixelated silicon strip detector and a transition radiation tracker. Their respective assembly follows a strict linear order of customized instruments for performing specific functions given that a chain of evidence must be maintained for ensuring the credibility of the obtained data. Each of these specialty instruments will target specific properties located within the physics events passing through.

While the subsections of the detector served very specific purposes and do not deviate from doing what they have been programmed to do, what is interesting would be the choice of design and materials (also determined by the best that are available), which enable these preprogrammed actions to occur; actions that are the outcome of a history of theoretical and experimental choices, and the theory-ladenness of the hypothesis that inform the experiments, therefore producing, the transfer of speculative inquiry from theory to experiment. In other words, the sub-detectors are the sites that re-enact the physics choices that interweave calculated acts of speculation with increasingly confirmed (or disconfirmed) selections. All of these would be better illustrated as I explain the different tasks performed by the sub-detectors.

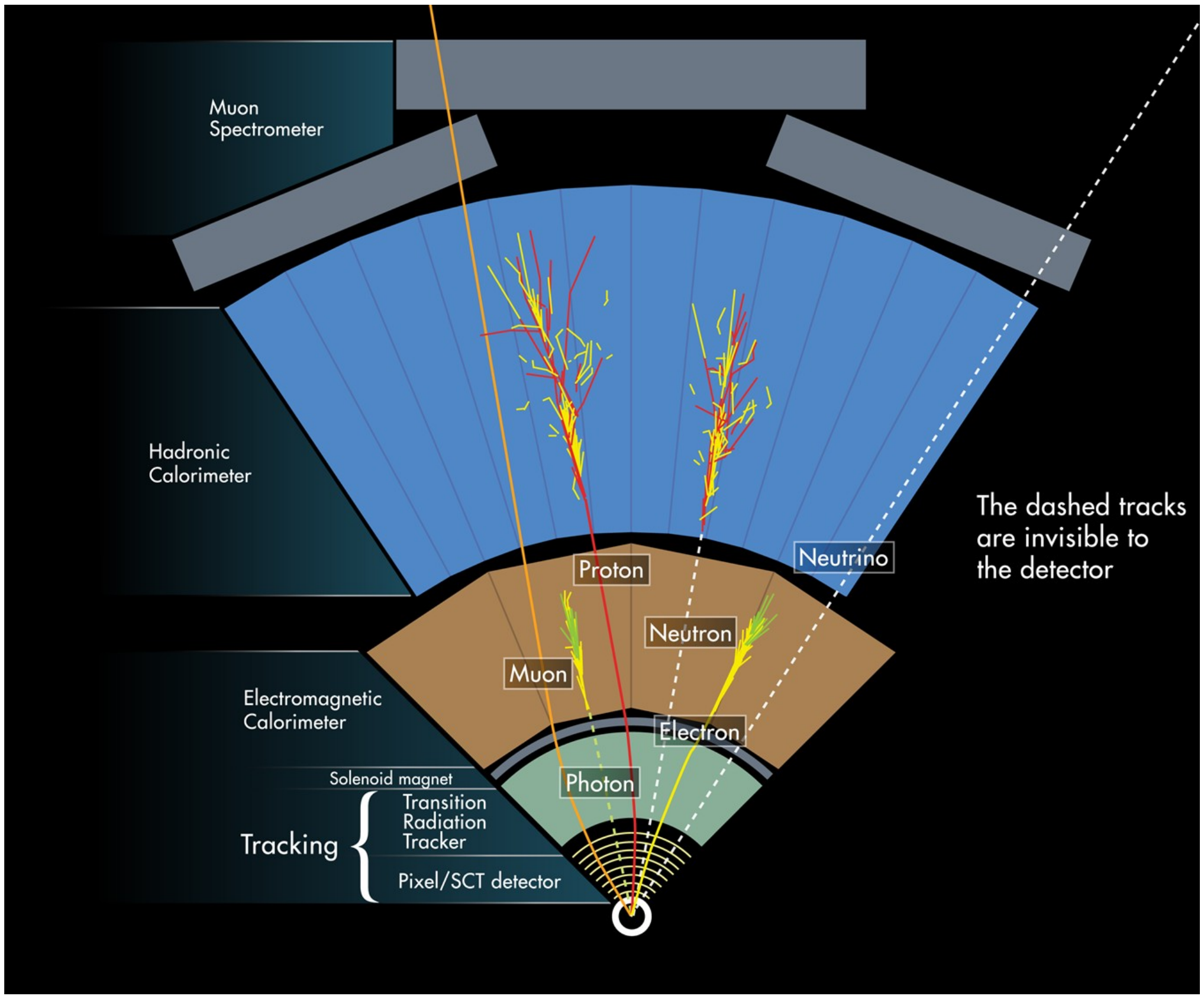

**Figure 1: A computer-generated image representing a cross-sectional slice of the ATLAS detector and how some of the typical particles measured interact with the various detector components. The colliding proton beams travel perpendicular to this view in the beam pipe as shown at the bottom of the image. The detector regions are cylindrically symmetric about the beam pipe. ATLAS Experiment © 2014 CERN**

For as the figure above illustrates, at the sub-detector levels, we have the:

1) momentum measurement of charged particles from reconstruction of the particle tracks through the inner tracker of the detector;

2) measurement of the angle of the particle produced at the point of collision in relation to the accelerating beam, also from the track reconstructed in the inner tracker;

3) electromagnetic calorimeter for electron and photon identification, and hadronic calorimeter for jets (experimental signatures of the quarks) and energy measurement;

4) muon identification through measurement of muon tracks in specialized detectors placed outside the calorimeter;

5) sufficient background noise rejection to provide a highly efficient trigger for high transverse-momentum objects.

The experimenters have also to bear in mind the tolerance level for radiation damage in the detectors. With the LHC experiments requiring fast, radiation-hard, and finely-segmented detectors, the readout speed of the detectors have to be such that they can resolve the time between two successive radio frequency (RF) bunches of colliding protons.

Additionally, one has to account for energy that might be 'missing' from observation, and this is where one has to be attentive to the kind of material structure of the detectors and compute the limit of their capacity for containing and transmitting the information on all the energy produced in the course of the interactions. Some particles, such as the 'neutral' neutrinos, do not interact with any detector components so as to leave the detector unobserved and to the 'dissipation' of energy that leads to the 'missing energy.' It is the calculations of the missing energy that provides the indirect detection of such neutrinos.

The calorimeter area of the detector is central for measuring the energy of electrons, photons, and hadrons encountered that was produced from proton-proton collisions. Photons and electrons are primarily detected in the electromagnetic calorimeter. A muon spectrometer surrounding the calorimeter consists of toroidal magnets needed for providing a large magnetic field, and precision tracking chambers. Moreover, the inner tracker is located closest to the photon beamline, consisting mostly of silicon pixel and strip detectors for high precision tracking of particles. The silicon detector subsystem also measures the secondary decay vertices of the heavy quarks (such as the beauty/bottom and charm quarks). This is particularly important for increased sensitivity to the Higgs boson given that one of the main states of its decay would be the beauty/bottom quark.

The scale of the transverse momentum (momentum-component in the plane perpendicular to the beam direction) of the particles resulting from the collisions that could be captured by the detectors is important for determining the accuracy of the measurement and the type of observational interpretation that could emerge. The accumulated raw data then goes through a process of reconstruction after the events are selected, to filter out what the physicists consider as uninteresting (though this is another point of subjective evaluation that is dependent on what the target focus happens to be at the moment) by tracking particular signatures of interest.

What goes on in the process of filtering and selection when it is done in the absence of a theory that can prescribe why one would select particular events for analyses over others; what criteria goes behind the decision on deciding which events will be able to contribute to greater knowledge of the Higgs boson? While certain historic high-energy physics experiments have to contend with discordant results, the search for new physics have so far been contending with marked absences; absences made all the more pronounced by the lack of further revelation from the already trillions data bytes that had been collected during the harvesting of informational signals at the cross-sections of particle interactions at the detector. Real-world data is complicated by the need for piecing together information garnered from millions of signals over multiple detectors even within the span of a single experimental run. Hence, the challenge created is unprecedented in big science and is the reason why statistical simulations and measurements of distribution are so important.

The design for the trigger and data acquisition from LHC experiments comes out of the manner in which selection is made so that rare physics signals can be extracted from overwhelming background processes such as is the common case for high-energy hadron interactions. For instance, there is ten orders of magnitude difference between the number of proton-proton collisions and number of Higgs boson produced. Hence, increasing the sensitivity of selectivity involves the addition of rejection power for 'irrelevant' material. However, the requirement for rejection power depends on which channel the selection is likely to take place, even if what is required of a hadronic interaction may not be necessary to other interaction channels. Therefore, the interaction rate that is being tracked has to commensurate with the ability of the technology to acquire and store the data being produced.

The process of data acquisition that features the configuration, control, and monitoring of data-taking operations require the sort of systemic coherence for high-level computational processing and switched networks, especially necessary for the global grid collaborations that are the nodes of labor processing through the data. Then, there is the problem of scalability in terms of the necessary computing power needed for dealing with high-level triggers. Once the data is accumulated, they have to be moved into more permanent storage before they are backed up.

The triggers are important to speculative experiment because of the former's responsibility in acquiring 'good' physics events for analysis, and the need for a minimization of dead time (latency in data recording) and accumulation of artifacts. A complex algorithm is included in the decision process and is, in itself, part of the determining factor for characterizing the best statistical method to be used on any sets of data. To illustrate the complex process of data acquisition informed by the triggering process, I lay out the stages involved in ATLAS data acquisition design, as described by L. Mapelli and G. Mornacchi in "The Why and How of the ATLAS Data Acquisition System" (398-99) in *At the Leading Edge: the ATLAS and CMS Experiments*, together with my own analysis of each stage of the design:

1) Factorization into main functions and partitioning of one system into multiple systems to ensure limited functionality and optimal operability. The latter is particularly important in the event-building stage for which batch physics could then be produced for analysis. Partitioning allows a detector to multitask so that a readout of the triggers could be performed simultaneously with data-taking, sort of like having a partitioned hard-drive where you can maintain multiple operating systems or operationally-directed processing in the background, then select the one you need to foreground when desired.

2) The minimization of the data movement to reduce the complexity and cost of the system. This means a sequential processing of events accepted at the first level and then the selection of readout based on the second level filtering of the events. Now this second part is important, since teams of individuals are involved in making that decision based on a compromise of what should then be kept. This means that the best argument for the case wins even if the weaker case might actually be the one that contains the hidden potential for unexpected discovery.

3) A uniform approach to data readout as a way of minimizing transfer loss. The uniformity is helped by the maximum adoption of hardware and software that maximally exploits advances in communication buses and links, networking and processing units, operating systems, programming languages, databases, and inter-processing communication packages.

4) During the staging period of the experiments (which is the pre-experimental stage), not only the physics but also the politics of funding and safety checks are involved in determining what would be the acceptable direction for data-collection, as well as the parameters to be set, such as the type of luminosities emitted (quantities measuring the ability of an accelerator to produce the required amount of interactions), and the final outcome of what could be detected through the data collected.

To make the data taking more efficient, tagging is used as a method of information classification so that event selection can become more efficient once one knows what to look for. This is after the data reconstruction and the analysis of the information have already taken place.

As Allan Franklin argues in *Experiment, Right or Wrong*, one of the most important parts of experiment has to do with data selectivity, and nobody who works with big data science would dispute that. While he uses the example of the analysis of the K meson (a particle composed of two quarks) experiment as an example of such selectivity, this parallels the process that goes on in selecting events that held the most potential for the discovery for the Higgs. Moreover, the choice that emphasizes the observation of one phenomenon can mask effects that could point to a phenomena operating at a scale that has been masked. Further, the story of the search for the Higgs did not begin at the LHC, but had been ongoing long before the LHC began data collection, so the decision for proceeding at the LHC rides on the earlier decisions made at the other accelerators and detectors, such as at LEP (a predecessor of the LHC) and the Tevatron (at Fermilab even if its high energy searches have since been discontinued). Without going into numerical details, the previous searches were later combined with the new searches at the LHC involving Higgs decay channels to vector bosons, diphotons, tau leptons, and hadronic jets (such as from beauty/bottom quarks).

A crucial feature of speculative practice in experiment, even if one might not consider it as that, is the incorporation of blind analyses into the kinematic region of the Standard Model Higgs boson to minimize the bias that may come about when correlating prediction to observation. While blind analysis is never entirely blind or objective given the aforementioned problem of theory-ladenness, it still points to a deliberate decision for imposing a particular conception of measurement in relation to the observables, and therefore, an epistemic choice for approaching ontology. A guideline for such an analysis was published in 2000 for the BABAR experiment

(that is of no relation to the Higgs experiments). To cite the guidelines (by way of Franklin):

> …There are a number of ways in which the Experimenter's Bias can infect a measurement which can be eliminated with blind analysis. First, the point at which the decision is made to stop working and present one's result can be influenced by the value of the result itself, and how it compares to prior results or predictions. In a blind analysis the decision to stop and publish is made based on external checks, and not on the numerical value of the result. After all there is no information about the correctness of a measurement in the numerical value obtained; a blind analysis enforces this separation. Second, choices about the data to include, or the cuts to use, can be subtly biased, if the effect these choices have on the result is known. [As we have seen, the bias is not always subtle.] Often changes in an analysis, which change the data set, can affect the value of a result on a statistically reasonable way. A blind analysis ensures that such choices affecting the data sample do not bias the result. Third, the values and types of cuts to use can be biased by knowledge of the effect of these cuts on particular events in the data. In particular, for rare decay searches or measurements involving small samples a blind analysis removes the possibility that cuts are chosen to include or exclude particular events in the data. In this case a blind analysis ensures a statistically meaningful result. (Selectivity and Discord 134)

However well intentioned the blind analysis method is for minimizing bias, the fact that the experiments are designed to produce a range of expected effects based on prior assumptions already 'biases' expectations to particular assumptions of ontological structures. In the final section, I will discuss the case of the supposed non-empirical materiality of the predicted supersymmetry and how its existence hinges upon a particular analytic approach to the Standard Model that is underdetermined. But before that, I would like to discuss briefly, the two important papers published by ATLAS and CMS in the aftermath of the announcement of the Higgs boson discovery on July 2012. These papers reveal the difficulties and uncertainties in putting together a deterministic picture of the Higgs boson in relation to all the other known elementary particles.

In the August 2012 article by the ATLAS collaboration containing an extended analysis of a post-discovery announcement, the four-lepton decay channel involving a Higgs boson decaying to two Z bosons seems the most promising due to good sensitivity over a wide mass range and the excellent momentum resolution of the final state particles (electron and muons from the subsequent Z boson decays) within the ATLAS detector (the CMS group also agrees on this point). The article indicates that a less direct decay to leptons, by way of bottom and charm quarks, electrons from photons conversion, and hadronic jets misidentified as electrons, are important backgrounds to the Higgs boson signal. Although only tiny excesses are noticeable, the Higgs boson to ZZ channel is relatively clean compared to its background processes and conforms to the Standard Model theoretical prediction for the Higgs boson within current experimental uncertainties. On the other hand, the Higgs boson to diphoton channel is problematic because of the greater background compared to the Higgs signal. In addition, there is also incomplete knowledge of some of the materials that were being measured by the calorimeter (6-7). Even so, the Higgs mass in this channel can be very well reconstructed and it has a good fit to the prediction of the Standard Model.

In the Higgs to WW channel, candidates for this channel's searches are pre-selected by requiring two opposite-charge leptons of different flavors, with the events classified into two exclusive leptonic channels depending on the flavor of the leading lepton. Electronic candidates are selected through a combination of tracking and calorimetric information, their criteria optimized for background noise rejection. Acceptable muonic candidates have to match the tracks at the muon spectrometer and inner detector. Sensitivity to the Standard Model Higgs boson is produced through the application of further selection criteria depending on the jet multiplicity, with the data divided into three different categories of jet search channels in accordance to the number of jets in the final state. The number of jets is vital for increasing sensitivity so as to be able to recognize the Higgs boson that is sometimes submerged in the background. I could go on with more examples of search channels, but the few primary examples I have highlighted, coupled with the more general but relevant critical descriptions I have provided above, serve well to foreground almost conclusively, that material evidences converge to produce a composite statistical profile compiled from different channels so as to be able to get at a statistically rigorous confirmation that an entity resembling the Higgs boson exists in nature.

There are a variety of systematic uncertainties in all analysis of the data because of the various (in)efficiencies, detector responses, luminosities, background sources, as well as data transfer factors from the Monte Carlo, for extrapolation into the signal region (more of the last point will be discussed in chapter seven). Statistical methods are used to interpret the significance for the Standard Model Higgs boson signal in the data. A likelihood test is done between the hypothesized values for the Higgs boson and a statistical profile that has been compiled based on likelihood ratios. In fact, as of today, ongoing analyses are working through some of the systematic uncertainties that abound in order to be able to consider details that might have been missed in the early stages of analyses while also incorporating latterly derived knowledge.

A comparable paper on the topic was published by CMS on August 2012 (in the same journal, Physics Letters B), although the paper spends more time on the history and future expectations of new physics than the more business-like approach of the ATLAS paper. However, the CMS paper provides a good point of comparison, given how one is dealing with the discovery from the vantage point of probabilities and uncertainties, even as the paper excludes what ATLAS includes in terms of the search within an estimated mass range, and vice versa. What the paper highlights is that the analyses published were re-optimized based on an afterthought, caused by new raw data reconstructions and savvier event selections.[4]

---

[4] The decision involved in when to publish, especially the strategy for representing results in an intensely high-pressure world of big science when the results are announced, is another fascinating subject for more in-depth study, and not merely because of the purported tendency of dressing up results or changing interpretation to suit epistemic, social, or economic expectations.

The analysis produced by CMS, using the blind analysis method for confirmation of experiment to theory, demonstrates how the data collected is correlated more closely to prediction than statistically expected because of an original experimental hypothesis that happens to be a statistical construct. The CMS collaboration was able to differentiate the spin of the Higgs from that of the other available particles, which was further finessed in the analyses they performed on the data in 2013. However, I will not be elaborating on the newer analyses since that would not add more to the arguments I am already making.

I need to re-emphasize that the non-prediction of the Higgs mass by theory has to do with how the theory of the Higgs mechanism has been derived and also that experiments are only able to access a mere fraction of all that is out there, with the potential of current interpretation being wrong even if one does not deny the existence of a material entity resembling the Higgs. As the case of supersymmetry reveals, its strong candidacy as the theory that is needed to extend the Standard Model, when coupled to the still not yet worked out reason of the how and why, merely illustrates the unresolved ontological problem of the Standard Model. Will the Standard Model become the be-all of our universe's narrative, or have our biases and localization within the site of observability barred us from ever having access to the ontology underlying the Model?

## 5.2 The Missing Supersymmetry (SUSY) and Going Beyond the Standard Model (BSM)

In a 2011 (revised from a 2010 talk) unpublished paper, J.D. Lykken from Fermilab considers the imprecise use of the phrase 'beyond the Standard Model' (BSM) by physicists. The phrase can be used to refer to experimentally verified physical phenomena that have not yet been accommodated by the Standard Model (such as Dark Matter, gravity, and neutrino oscillations) all the way to a deeper phenomenon that is accommodated by the Standard Model, but only under especially framed circumstances, including any relevant extensions to the Standard Model (1). It also appears that much of the theorizations are stubbornly embedded in our dimension of space-time and mathematically sensible degrees of freedom governed by the epistemic logic of relativistic quantum mechanics.

However, it is important to establish, foremost, that supersymmetry (SUSY) is not a BSM model, but rather a BSM framework that holds a multitude of models together. SUSY has been the favored theory among the other more exotic candidates[5] because of its simple premise: a single additional and seemingly natural symmetry and the range of problems it solves. However, until we can understand better the role SUSY plays in relation to the Standard Model beyond the computational finesse the former provides, it remains uncertain as to whether SUSY can provide a better ontological approach to understanding the mechanism of the Higgs boson.

---

[5] Technicolor variants, theories predicting axions, large extra dimensions, hidden valleys, little Higgs theories, and others, none of which I will explicate upon here beyond stating that they form various model-hypotheses for getting at the problem of unification, scales, and higher-order/multiplicity of dimensions not yet addressed.

Further, SUSY appears to have greater prescriptive power compared to the Standard Model, though the former still fails to solve the ontological conundrum surrounding the physicality of the Standard Model beyond providing more beautiful and elegant computational finesse. This is because SUSY and the Standard Model share the same quantum field theoretical delimitations, and therefore, their predictive capabilities are similarly constrained.

While there have been a number of literatures on what SUSY can achieve when searches are made for predicting a wide range of SUSY regimes, none of them address the heuristics of SUSY in any way that is external to the regime of discovery. The gauge theories that provide mathematical support to symmetry relations within electrodynamical, weak, and strong interactions, have not been physically explicated, yet the former theories provide the structural basis for framing the theories of supersymmetry. In addition, the transformation between the mathematical phases (such as from non-relativistic to relativistic mode) does not bring about any significant physical change, since it mostly revolves around the re-organization of positions without changing the frame of reference, and therefore, the theory-laden assumptions embedded in the frame.[6]

The non-existence of SUSY, however, does point to some gritty issues, particularly issues of verisimilitude. On the one hand, it points to SUSY as a non-counterfactual of the Standard Model since there is no other indication that the Standard Model is not right. At the same time, the existence of the Standard Model necessitates the existence of SUSY, or something comparable, since the deficit within the former requires an object that can at least fulfill the function that SUSY does in order for the Standard Model to exist. But to take a logico-phenomenological perspective, the problems of the Standard Model are not sufficient pre-conditions for confirming SUSY despite their sharing a similar mathematical ontology, and could therefore form a more complete theory together. Despite the many perceived justifications for at least a minimal supersymmetric extension to the Standard Model (MSSM), the lack of a strong experimental confirmation beyond the formalistic rationale of this theory points to the inherent difficulties of attaining a 'perfect fit' between the possibility of theory with the expectations of experimental justification and verification so as to be able to approximate the ontological foundation of the Standard Model more definitely.

[6] See for example Lyre, Holger. "Does the Higgs Mechanism Exist?" *International Studies in the Philosophy of Science* 22.2 (2008): 119–133.

At the experimental level, the most straightforward incarnation of SUSY requires supersymmetric particles to exist at around the 1 TeV mass range; any higher and the theory starts to run aground, particularly in the fine-tuning of the Higgs mass. In addition, the current lack of evidence of any supersymmetric particles at the LHC means placing limits on the former's mass when approaching the problematic level of 1 TeV. It is possible to consider SUSY as mainly a phenomenological tool, or what Dudley Shapere refers to as unobservable background information. The issue of whether SUSY can be observed brings to bear the question of whether its actual discovery has a role in the Higgs's measurement. Shapere also argues that the idea of what is unobservable (or an idea without observable consequences) can become accepted if it is mathematically or logically implied by another observable subject, and needed for maintaining consistency regardless of whether the unobservable is implied by the observable portions of the theory (Shapere 159).

While the LHC upgrades are now getting ready (the LHC will re-start operations in April 2015) to run at higher collision energy and luminosity, thus increasing the sensitivity to the existence of supersymmetric particles, the next stage would involve consideration of where data cuts can be made in order to single out any particular physics events that might increase sensitivity towards the manifestation of as yet unidentified particles that could be the best lead to locating SUSY. However, we also have to grapple with what it means to be making the right data cut in hope of landing on the right cluster, pending an empirical evidence of SUSY. Hence, what we end up with is a chicken and egg problem.

We know that there will always be problems with experiments because instrumental technology could not always catch up with theory for various physical reasons, and experiments are also dependent on the developments of other areas of applied material sciences. With the margins of uncertainty being wider than the preferred effects, attempts at excavating the physical footprints of SUSY within that range of uncertainty can be both informative and problematic. However, one would still have to search for these predicted elements in the region of 'uncertainty' that are parameterized by selection criteria determined by predictions and experimental triggers, which includes digging deeper into particles that are still not quite well understood by merely adjusting from existing methods. The questions that one should continue to ask are: how much of the uncertainty is due to measurement limitations and how much is constituted by an inability to deal with discordant results, issues of non-local observations, or speculative turns necessary for mediating between empirical evidence and theory? Speculating on possibilities that are ambiguous despite their basis on facts pushes us into considering the meaning of uncertainty, the duck and rabbit thinking that goes into the decision-making processes, and the discordant truth-claims that can bias the measurements.

This means a reconsideration of the cognitive process underlying the construction of models, proofs, and formalism such as that detailed in a thorough paper by Arianna Borelli "The Case of Composite Higgs: The Model as a 'Rosetta' Stone in Contemporary High-energy physics" when she discusses the problematic epistemologies underlying model formation such as is the practice in high-energy physics leading to the production of ad-hoc and expendable models that do not consider how these models can expand into more developed and rigorous theories. She also discusses the use of a hybrid mix of rigorous formalism and non-formalistic approaches utilized for the formulation of predictive models that could lead to either spectacular or disastrous results.

But more interestingly, she explores how physicists use models as an apparatus for constructing a fiction or toy model for heuristical explorations. She contrasts the non-rigorous model-building enterprise of particle physicists with the more deliberate act of theory-building that takes place through a rigorous invocation of mathematics that string-theorists engage in to get to a final theory, even if there are no available experiments for bringing those theories to life. She rightly points to particle physicists being more interested in the physical validity of the mathematics used for producing practically useful models. Moreover, she foregrounds the importance of combining a strong theoretical core with desired empirical features as a way of repositioning the role of the observational language within theory, theory that is constituted of a hybrid of narrative and empirical reference (or justification).

As the fictional and actual experiments discussed above have served to demonstrate, speculative experiment involves taking together all the supposed points of indeterminism and current theories that are ontologically problematic to reflect on the subjectivity of the experimental approaches while similarly considering emergent practices that could break the mold of continuity and contiguity for attaining new physics. One could also take the speculative practices in experiment as material for modeling more concrete approaches that foreground the ambivalent relationship between epistemology and ontology.

SUSY is the model of thought experiment and the unobservable that is ideal for demonstrating how theoretical speculation could broaden the concept of speculation for the exploration of the non-observable in relation to the observable. To re-iterate Friedrich Schelling loosely, an experiment is a self-construction of phenomena that can never transcend the forces of nature, so we can be assured that once we are on the right track, however uncertain our knowledge, the product will manifest itself, just perhaps in a way different than expected.

## 5.3 Conclusion

Does speculative experiment, in conjunction with BSM, represent a new crisis in the science of particle physics? After all, previous crises have led to unexpected discoveries that revolutionized particle physics in the 1970s, and more recently, the discovery of the Higgs boson that had once been on the receiving end of criticism in the same manner that SUSY now faces. Hence, it appears that the possible crisis, if reconstituted within the paradigm of potential revolution, is what is needed to encourage the reconsideration of the available and whatever else there is to be done still. One might even argue that experimental crises are crucial to speculative experiments because the crises expose the problem that can arise when the theories driving the experiments do not bring about a predicted result, or worse, lead to experimental 'crashes.'

In chapter seven, we will be revisiting some of the same issues that have been raised in this chapter with regard to how analysis is performed of the Higgs search data through Monte Carlo simulation used to produce test models for determining a distribution of probabilistic outcomes of unknown entities based on variable parameters. Taking a leaf out of the theoretical and historical issues raised by Borelli on attitudes to theory and model in high-energy physics, chapter seven will dig deeper into what is epistemically contingent and necessary about the simulation/statistical model and its implications for both speculative theory and experiment, as was discussed thus far. But before that, in the following chapter, I will venture into a discussion of how science fiction performs a speculative-experimental turn to demonstrate the other components of experimental practices that will complement the issues raised in this chapter.

*Yet, "the only reality capable" reflects an ambition to provide atoms with a mode of existence that, following Bruno Latour, I have associated with "experimental factishes"—beings that we are compelled to describe as having been constructed by us and simultaneously endowed with an autonomous existence…In other words, the "experimental factish" constructed by detection answers the questions addressed to it, but its responses, although veridical in the sense that they resist the accusation of being mere artifacts, are still not "truthful" if by this we mean the avowal a being can be led to make concerning its own truth.* (**Isabelle Stengers in *Cosmopolitics, Volume II***).

## 6. Fact from Fiction: the Impossible Affair of Observing the New through an Unknown

One of the more unique features, or considerations, of science fiction, beyond its ideological embodiment as scientific cognition of a secular age, is its expression of ontological curiosity in perforating the boundaries of the present and known to reach into the future and the unknown. Sometimes, the genre also extends into a past that has never been, or a past with a twist to its timeline. Science fiction is more than a creative mode for theorizing utopia and dystopia through the framework, or lens, of techno-gadgetry, socio-technological forecasting, and socio-political interfaces that could either be a part of a futuristic extension, another dimension of time, or an 'alien' culture (in a pulp fictional and metonymic sense). But, given that science fiction has its earliest incarnations in fictions, preceding the Enlightenment project, such as Johannes Kepler's *Sommium Sive Astronomia Lunaris* and Margaret Cavendish's *The Blazing World*, before being succeeded by post-Enlightenment (and therefore, a Romantic) critique of the hubris that came out of over-reliance on science and reason as demonstrated through Mary Shelly's *Frankenstein*, the genre is shaped by a categorical approach to epistemology that is Eurocentric in its teleology.

This is not to say that there are no earlier works that can be considered under this category, or that works from non-European cultures might not be considered, but rather that the development of certain 'canons' of the genre informed by particular epistemic sensibilities are what shaped our cognate of science fiction.[1] But most importantly, the genealogy of these works provides one of the earliest indications of fiction as philosophical and cultural inquiries into science by way of a systematic probing of a post-Copernican scientific culture. Observations and the interpretive acts following the observations were then catalogued and described, with fiction opening up imaginaries into the not yet observable for speculation.

According to Eric S. Rabkin in the May 2004 issue of the *PMLA*, the cultural system of science fiction "coordinates politics, technology, and more enduring symbolic concerns about gender (ants are female, nests are evil, but men have flamethrowers) and poetic space (aboveground is good; belowground is bad)" (463).

---

[1] Darko Suvim and Adams Robert have much to say about the genealogies of science fiction in their aforementioned respective books discussed in chapter four.

However, I am interested in thinking about science fiction beyond the usual critique of cultural valuations, aesthetics, and moral truth claims, while not neglecting the latter categories. I am also invested in thinking about science fiction as science studies (and social studies of science), such as is demonstrated in Sheryl Vint's "Archaeologies of the 'Amodern': Science and Society in *Galileo's Dream*" in the Winter-Spring 2012 volume of *Configurations*. However, we differ in our theoretical (and philosophical) approaches as informed by the subject of our study. Moreover, I concentrate specifically on hard science fiction that are focused on modern physics: science fiction is a platform for navigating some of the more subjective and speculative aspects of scientific epistemology, especially elements of theoretical sciences that are mostly speculative yet ground-breaking if experimentally confirmed.

Hence, I attempt to examine hard science fiction as a speculative account in science studies that problematizes non-quantifiable and ethically ambiguous ideals in the constellation of scientific knowledge. I argue that science fiction is a suitable platform for dealing with morality/ethical questions that span the internal and external values of science because of the genre's ability to model friction and problematic encounters between science and society, and even between the sciences. In addition, science fiction is useful for the production of thought experiments that amplify and defamiliarize ideologically and socio-politically invasive issues while crossing into the territory of 'extra-normal' science; the latter of which is hyperspeculative because of lack of direct connection to any established scientific theories. Extra-normal science can be deployed for asking hard philosophical questions about the metaphysics of reality as well as for complicating the usual run of physical uncertainties by compounding the perplexing problems involved in distinguishing the pseudo from the real. In this manner, science fiction is the platform for expanding metaphysical inquiries that critically assess institutional and epistemological demarcations of pseudoscience by employing the Schelling-ian argument of the ideal being explained through the real.

For Jameson, scientific and philosophical speculations are "ideological constructs designed to ground a particular political system in biological nature" (168). This connection from the ideological (quantum measurements) to the biological is illustrated in the dystopian subjectivity of Egan's *Quarantine*, and would become equally problematic in the works of fiction to be examined here. But the dystopia is also a utopia that, as Jameson argues

> …brings us to what is perhaps the fundamental Utopian dispute about subjectivity, namely whether the Utopia in question proposes the kind of radical transformation of subjectivity presupposed by most revolution, a mutation in human nature and the emergence of whole new beings; or whether the impulse to Utopia is not already grounded in human nature; its persistence readily explained by deeper needs and desires which the present has merely repressed and distorted…If absolute difference is achieved, in other words, we find ourselves in a science-fictional world…in which human beings can scarcely even recognize themselves any longer. On the other hand, if Utopia is drawn too close to current everyday realities, and its

> subject begins too closely to approximate our neighbors and our politically misguided fellow citizens…(168)

That struggle between differentiation and mimesis in the mundane permeates Benford's Gregory Benford's *Matter's End* and *Mozart on Morphine* at a much lower register than in the husband-and-wife (Richard and Nancy Carrigan) co-authored *Minotaur in a Mushroom Maze* (to be known as *MMM*), mainly because Benford's flight into the fantastical is carefully engineered to produce an imagined outcome unconstrained by physical realities (although not necessarily physical laws since the laws will exist regardless of our perception of them).

In the three fictional works to be discussed, I hope to make evident that the nomological fable penetrating the perforated lines of science fiction and ideological utopian fantasies are as much colored by Western centrism in relation to the 'other,' as they are by a desire for material agency. Even if one finds parallels between the physics that are the focus of the stories and the more concrete socio-political actions of the human characters, one would not be able supervene, or superimpose, one over the other, because, to put it in Nancy Cartwright's words,

> First, a concept that is abstract relative to another more concrete set of descriptions never applies unless one of the more concrete descriptions also applies…Second, satisfying the associated concrete description that applies on a particular occasion is what satisfying the abstract description consists in on that occasion (39).

While the abstract categories are pervasive and culminate into finales without arriving at an ultimate resolution as the concrete dissolves into the abstract, the incompleteness of what is observable and the difficulty in attributing direct cause-to-effect are reminders of all that is beyond plain sight: the missing petrons in *MMM*, the unattainability of higher dimensions of knowledge in *Mozart on Morphine* beyond the imagination, and an apocalyptic explosiveness in *Matter's End* that exhibits the impossibility of disentangling the epistemological from the ontological after the elusive temporal instability of the proton is pinned down (even if only fictionally).

Given that the entanglement of the epistemological with the ontological is also an entanglement between the physical and metaphysical, my support of Darko Suvin's argument concerning how the character of fiction in science fiction is an embodiment of estrangement does not extend to his attribution of non-relationality between science fiction and metaphysics. In the sense I am applying here, metaphysics is seen as an entailment, or counterfactual, of the physical (despite metaphysics's rather contentious history within particular strands of philosophy of science), while for Suvin, metaphysics is akin to myths, folklore, and the paranormal.

Moreover, science fictional works that take on the theme of the physical sciences intersect constantly with metaphysics: metaphysics often articulate theories of causation, or problems with causation; physical objects are utilized as examples while physics applies metaphysical conceptualization to articulate the ontological problems found in its interactions. Further, the model-theory relationship today has evolved tremendously because of the increasing sophistication in computational techniques compared to what was available for early twentieth-century predictions in quantum theory and cosmology. Available computational techniques enable the modeling of both discrete and continuous elements simultaneously, and are thereby able to foreground areas altogether deterministic and stochastic for big-picture analysis. This specifically means that there are increasing opportunities for creating computational thought experiments of metaphysically directed questions on physics.

Contra Suvin's argument that, "…To expect from SF more than a stimulus for independent thinking, more than a system of stylized narrative devices understandable only in their mutual relationships within a fictional whole and not as isolated realities, leads insensibly to the demand for scientific accuracy in the extrapolated *realia*" (28), chapter four wants to demonstrate that science fiction can be more than a stimulus for independent thinking or stylized narrative device, and the demand for scientific accuracy is both necessary and not insensible, but should be conceived as much more than what is defined by an empirical, and material evidence-based, mode of production. Science fiction, especially of the hard or mundane variety, has the potential for being a generative platform that encourages the exploratory potential of the novum for experimenting with the what-ifs of scientific speculation within, or beyond, the constraints of known physical and natural laws. Hence, to expect more from science fiction is not just to expect scientific accuracy, but also to expect a commitment in dealing critically with science performing ideology (or vice versa) and science's relation to difficult ethical questions. Science fiction simultaneously idealizes and reveals the power structures of science.

The works of fiction considered here contain literal and less obvious acts of scientific observations; some of these acts are transitively enacted with a theoretical goal in mind while others are involuntarily performed because of the manifestation of certain physical phenomenon. Nevertheless, observations are the outcome of measurements; if the act of measurement is more amplified in *Quarantine*, the after-effects of the performance of measurement, also indirectly alluded to here, will be further magnified by the relationships between the characters.

Whatever the outcome, it is obvious that measurement and observation are inextricably entangled. I begin by considering and re-theorizing the conceptual loci of hard science fiction in relation to the selected works before considering how science fiction can be a productive site for envisioning the scientific method and futures of science, especially physics. At the same time, I will be examining how the authors use scientific observation in their imagining of fictive discoveries in terms of the latter's implications on fictive universes paralleling ours.

## 6.1 Hard Science Fiction: the Ethics of the Speculative

When the potential for an unexpected discovery detonates a series of events that could change the course of our relationship to the world we inhabit (and not necessarily in a way that is comfortable or humanly safe), the ethics of dealing with that discovery is complicated by the differences between an abstract absolute truth and expedient truth: the former truth is governed by the desire of ontological access to the foundations of nature while the latter requires one to consider the consequences of, and primary motive for, unleashing that knowledge into the world. The scientist has to decide on whether the act of enacting an observation may require the amplification of particular physical properties of the entities that do not occur naturally in a form suitable for examination (especially in the sense of being easily observable), thereby bringing about catastrophic side-effects in the same manner that a mirror version of a drug chemical could produce a fatal outcome when ingested.

*MMM*, published in three installments over the summer of 1976 (May, June, and July) by hard science serial *Analog Science Fiction and Fact*,[2] brackets the discourse of the observable versus the non-observable as part of its plot device even if the act of observation is not always about advancing the cause of the physics. The novella parallels Benford's *Matter's End* in its deployment of the exotic features of particle physics. In the case of *MMM*, the search and discovery was for the petrons, which, according to the Carrigans[3] in their note preceding the story, were molded in the image of the magnetic monopole (a hypothesis that rode high on the waves of speculation at the time the novella was written)[4] and a very particular kind of elementary interaction known as muon catalysis.[5]

---

[2] Richard Carrigan, one of the co-authors of the short story, had authored an article in the Feb 1976 of *Analog* about the discovery of the meson through the Standard Positron-Electron Asymmetric Ring at SLAC.

On the other hand, *Matter's End* is centered on the discovery of the decay rate of a proton, a highly stable particle whose ability to decay into a neutral pion and positron is as hypothetical as the existence of a magnetic monopole. A real-life reconstitution of the fictional petrons and the actual discovery of a proton rate of decay will certainly break new ground for the Standard Model in a way the discovery of the Higgs boson has not, given the former discovery's potential for changing our understanding of matter completely.

Therefore, the unimagined effects of such discoveries that imply the instability of matter constituting our physical being is the reason behind the similar psychedelic (or nightmarish, some would say) finale of *Quarantine* (since unconstrained 'smearing' also denotes matter instability) and *Matter's End* as the authors tried to imagine an end of world sequence. The difficulties of imagining the not-yet-actualized is an important problem in speculative experiment, and this problem serves to explain why many of the predictions of the unknown are anchored on the known and how the dependence on epistemology derived from the observable delimits the range of experimental expectations.

---

[3] While Richard Carrigan used to be a physicist at Fermilab, his collaborator and wife, Nancy Jean Carrigan, is an artist and poet who had found much inspiration from her indirect association with the scientific world. See http://home.fnal.gov/~carrigan/Nancy_carrigan_cosmology.htm.

[4] There was an article about the magnetic monopole by Margaret Silbar in the November 1976 issue of Analog.

[5] Incidentally, Richard Carrigan had also written an article ("The Discovery of the Gypsy", February 1976, *Analog*) on the discovery of the J/psi elementary particle that heralded the November 1974 revolution birthing 'new physics.' As one would see in Appendix A, the muon catalytic process was part of the chain of events that led to the eventual discovery of the J/psi.

However, one might wonder if the almost similar endings of the two-abovementioned works are merely a confirmation of a niche practice in the genre as stated by David Brin in "Running out of Speculative Niches"

> It is betraying no fraternity secret to say that to some degree hard SF writers write for each other. That is, in addition to wanting their works to be good stories for the sake of the broad audience (and even critics), these writers also tend to be aware of the other hard SF authors as they work out the details of their plots and universes (9).

In the same anthology where Brin's essay appears, Gregory Benford, in "Is There a Technological Fix for the Human Condition," states that fidelity to the physical universe is demanded of hard science fiction because of its capacity for detailed prediction as well as falsification by experiment, and is thus a more reliable indicator of future possibilities.

According to Benford, even fantastical tales that fall under the rubric of hard science fiction abide by the logic of scientific verisimilitude in the same vein that the future technologies envisioned would appear 'magical' to the world prior to actualization. However, he argues for a science that is deeper than humanity's concerns, "…remorselessly deterministic, uncaring of our personal preoccupations, and yet capable of revealing wondrous perspectives. It can either encase us in the indifference of the universe, or liberate us" (85), Benford uses as an example of such a case, the story "The Cold Equation" by Tom Godwin, the one that John Huntington, in "Hard Core Science Fiction" that also appears in the same anthology, insists as hovering between "two incongruent ideas of SF: one promises liberation by means of an ingenious solution to what seems an irresolvable problem. The other promises a rigorous holding to the rules of plausibility (51)." Huntington goes on to suggest that the linear narration of the story that leads to an inevitable end riffs more with the perception of what the hard sciences do than what they are about, given the limitations of our ontological access to the science and the insistence on the idea of 'scientific rationality' in tales of hard science fiction.

The arguments Benford and Huntington have advanced with regard to the present science as being unhuman and unconnected to ethical complications differ from the narratives of the three works explored here, even when the characters voiced their trust in the science to the exclusion of other social considerations. The reasons are two-fold: 1) barring natural catastrophes unrelated to human activities, certain artificial conditions have to be created for the manifestation of a physical phenomenon in a manner that is observable and analyzable by the scientists 2) the scientists, as are other humans, are embedded within the consequences unleashed by any physical phenomena that have directly impacted their world, regardless of whether the phenomena are induced or natural. Even if the scientists who set the course for a catastrophic outcome might not be directly impacted by the consequences, such as Benford's evocation of India's biotechnological environmental disaster in *Matter's End* purportedly caused by first world scientists, the scientists still face the consequences through other means, such as the wrath of the public and the lost of general good will and trust. Despite setting such an event in a distant land, the urgency of Benford's not-so-subtle political commentary is not lost on his American readers.

Philosophers and sociologists such as Allan Franklin (who is also a physicist), Ian Hacking, Andrew Pickering, and Harry Collins, of course, have discussed external but directly impacting issues in relation to how one constructs the material position of experiments and the outcomes projected (including the lack of resolution), even if they do not always agree on how best to define the truth-claims of experiments in relation to the realities produced.[6] However, as Knorr-Cetina has noted, and whose argument I am extending, epistemology constructed from experiment is marked by the level of access that one has, and the access is not purely from the relationship between the physicist and the instrument, but also the privilege that allows more access to certain groups of physicists than of others, such as those from particular research institutes or universities. In fact, the setting of the discovery of the proton decay in India, with its 'primitive' technology and apparent distant location from the main artery of 'cutting-edge' research in the US, foregrounds, rather sharply, the privilege of access within the communities of physics. Benford's Indian scientists are at pains to gain recognition from the 'control-center' of science, as when US protagonist, Robert Clay (the centrality of his character is made manifest by the fact that his inner life is the only one the reader has any insight into, while his Indian colleagues remain caricature-like), was flown in and brought over to their humble facility to validate their findings. At the same time, one gets a sense of resistance in conforming to the epistemic expectations of their western colleagues. Prestige and recognition that comes from changing the course in the world of science were never far from the minds of all the physicists involved, whatever their nationality, as were also rivalry and communitarianism (both of which were noted by Traweek and Knor-Cetina)[7].

---

[6] See *Selectivity and Discord* by Allan Franklin, *The Mangle of Practice* by Andrew Pickering, and *The Golem: What Everyone Should Know about Science* by Harry Collins and Trevor Pinch. One can also see Franklin's summary and response to the arguments advanced in the introduction of his aforementioned title.

---

[7] More historical studies about doing science, and national to international collaborations in particle physics, can be found in *Fermilab: Physics, the Frontier, and Megascience* by Hoddeson et al. and the 3-volume *History of CERN* by Hermann et al. Also the first five chapters in Galison and Hevly's *Big Science* documents the history of particle physics collaborations of mostly US big national laboratories, with the exception of CERN, from the 1920s to the 1970s.

Speculating about the potential for grand discovery means that we must reconsider our relationship to the ontological foundation grounding the laws of physics governing our mundane world; speculating through thought-experiments-made-real-by-way-of-a-fictional-universe is the leitmotif of all the works of fiction discussed here, selected especially for demonstrating the balance between factual science and the imagined extensions of scientific possibilities through fiction, even if not all the works take equal advantage of the affordances of science fiction to push harder at the epistemic boundaries of the science even if the science is a primary object of interest to the plot. Further, the representations of scientific productions (including observations) that one finds narrated in the stories, follow from a tacit assumption of scientific expertise that the reader is expected to accept almost at face value. However, authors and readers meet in the difficulties they face in imagining the unknown, therefore providing this chapter with an ontological cliffhanger that segues well into the penultimate chapter's consideration of the important role of generative simulations for piecing together 'big' data, of many orders of magnitude, to work out the what-could-be instead of merely the what-ifs, of the unknown.

## 6.2 Configuring Scientific Observation in Hard Science Fiction

*MMM* is set in the aftermath of the 'November revolution' of 1974, a revolution touted by physicists as changing the face of particle physics through unexpected discoveries. Richard Carrigan uses his own experience in the field to conjure a quite realistic plot of the scientific industry: from the job market right to the actual experimental processes while not hesitating from injecting bits of the fantastical. Compared to the more surrealistic tone in Benford's short stories, *MMM* is decidedly an Indiana Jones-like adventure tale. While lacking the same level of literary finesse and philosophical reflections found in the other two short stories, *MMM* can be read as an embodiment of a pragmatic (male) experimentalist's gaze of the world he inhabits. That said, all of the chief protagonists of the stories discussed here are male, with the co-starring female characters holding mainly peripheral roles, with substantial amount of character development for the latter taking place outside the immediate purview of the reader. The smartest woman in the story is portrayed as a villain, even if a complex and unreadable one, who upheld dangerous beliefs incomprehensible to white bread Americans.

The discovery of the 'petron' particle is read analogously to the then feverish excitement in the world of high energy particle physics of the 1970s: the confirmation of the existence of quarks and the discovery of stable exotic particles able to produce enormous amounts of energy through fusion-like recombination of specific properties that govern matter and anti-matter. There was no reference to any particles of Higgs-like quality although the existence of the boson had been predicted more than a decade earlier when the story was published. But then, the importance of the Higgs boson, beyond that of the theoretical spectrum, was not quite realized.[8]

At an age when big science was, and still is, integral to physics experiments, there were echoes of resistance and conspiracy through the art of espionage, and a call to return to an earlier pre-big science era that is not necessarily a return to the pre-technological. Big science also represents the coming together of technological triumphalism with the ontological status of scientific realism; the former cannot lead to a resolution of the latter. This is particularly the case since observation is instrumentally mediated, and therefore, the noetic referent (intellectually abstract signifier) of what is observed depends on the conceptual judgment that is performed. But then, what is experimentally achieved could not, usually, correlate with theoretical predictions directly, if because, as philosopher Evandro Agazzi puts it, the

> physical reality we propose are not visual (or the like), but conceptual, and here the autonomy of the noetic world may pose problems. Even with visual representations – we have seen – we are free of combining them in a fictional picture that represents only a possible world of referents with no concrete counterpart. With the conceptual representations we have an even greater possibility of constructing 'theoretical models' whose degree of abstractness has only certain weak limitations of mathematical and logical character. ("Observability and Referentiality" 56)[9]

---

[8] See Ellis et. al, "A Historical Profile of the Higgs Boson."

[9] An essay in *The Reality of the Unobservable*.

Therefore, what this chapter intends to demonstrate, through the cognitive apparatus provided by science fiction, is that the conceptual need not be limited to the mathematical. One might want to consider what other areas of epistemology that cannot be mathematically reducible to unequivocal formal logic, and which might be better demonstrated through subjective, yet politically loaded, interactions. This is particularly true in dealing with how one tackles the consequence of making a choice when operating through the scientific method.

Further, the Carrigans juxtapose the tale of the Cretan Minotaur (the figure of the mythical bull takes on multiple figurations here), the labyrinth (that points to a physical underground structure and convoluted experimental paths that led to the petrons), industrial intrigue (financial gain to be had from control over cutting-edge renewable energy resource), and the more ancient (or timeless) form of deism. The story begins at a bar by a wharf not far from the real-life Brookhaven Laboratory is located; the wharf and the bar are where the reader is introduced to some of the characters who catalyze the events of the novella: Bull Tauroman, a man of supposedly inscrutable pedigree but of an ambiguous background, is the central antagonistic force; the physicists Mario Petronelli, who claimed to have predicted the discovery of the petrons, and D.A. Silverman, who was instrumental in bringing the main protagonist, John Leigh, into the story.

The real action began when some petrons were found stolen from Petronelli's laboratory. Leigh was requested to leave his 'cushy' job at a laboratory in Illinois (and comfortable non-relationship with a flight attendant who happened to be his neighbor) to re-enter the 'job market' at the American Physical Society so as to track down who might be holding on to the stolen petrons and would therefore be in need of an accelerator physicist to build the necessary generator for producing more of the said particles. It appears, in the novella, that Leigh holds a double professional identity as a physicist and a 'secret agent' of sorts. In the process, he met Daydala Pandarou, an alluring yet mysterious physicist with hidden ambitions claiming to be of Cretan descent but whose clan had been residing in Morocco for generations. She had hired Leigh, on behalf of Tauroman, under the pretext of building a coil needed for sterilizing compost used in underground mushroom growing. She was a brilliant theoretician and phenomenologist (due to her ability in straddling the practices of experiment as acutely as theory) who claimed priority in the prediction of the petrons (that she referred to as the dayons), and also a worshipper of the Earth Goddess, together with her band of male Moroccan downlinks. Further in the novella, it would soon become evident that the dayons/petrons served a larger purpose for her than just science.

Leigh was initially enamored of her, as was evident in their near-intimate night of dinner and clubbing at a nightspot, but their relationship cooled immensely after he moved into the mansion, the Hall, housing Tauroman and his factotums that also included Daydala's cousin, Alexi Pandarou, another enigmatic figure and Tauroman's secret paramour. Leigh became increasingly suspicious of Daydala and her intentions independent of whatever Tauroman was planning. Further, it was during his stay at the Hall that he met and formed an alliance with the stepdaughter of Tauroman, a skilled sculptor alternately referred to as Ariadne and Abbie, with whom Leigh shared more cultural resonances.

At the same time, the novella contains a subplot involving a securities agent Nathan Hunter (the only minority character who is not 'othered'), who had been tasked with investigating a possible money laundering activity involving terrorists even as he was personally bent on investigating a potential financial scam going on in his old hometown of West Virginia. The reader would soon find that the separate subplot is connected to the mushroom growing and stolen petrons, with the main link being Daydala herself, especially after Hunter was unceremoniously murdered by Tauroman's kennel man, Hole. Thereafter, much of the novella is preoccupied with Leigh trying to recover the stolen petrons from Daydala and her men while outwitting Tauroman. Even as he was doing so, he tried to 'protect' Ariadne (despite her being a very capable and strong-willed woman) while figuring out the ultimate plan that Daydala had for both the petrons and Tauroman's downfall.

The authors represent the fantastical features in the novella as constituting indeterminate features of the world that juxtaposes the 'exotic' qualities of quantum theory against the greater determinism of classical physics. Although the science that forms the core of the adventure is rigorous, the authors did not seem to do more than rehash the dominant narratives about discovery and experiment in physics. At the same time, there were also some issues with gendered stereotyping and Hollywood-derived masochism. The possibility of modeling the outcome of the discovery of the petrons through a macro-micro interaction within a macro-universe has not been taken advantage of.

Nevertheless, there are fascinating references to the analog and digital technologies that were available at the time, such as a tape recorder that can encrypt messages, and an early version of a search engine with analog features reminiscent of ARPANET. Since the story was written prior to revolution in personal computing and desktop supercomputers, the reader is offered a glimpse into the writing and debugging of codes that required hand-coding onto punched cards before being processed through the computer; computer time was costly since it had to be shared among multiple users, one user at a time. But as the next chapter would demonstrate, the employment of computational simulation to aid in experimental heuristics had grown in parallel with the development of big science.

Within this story, which one will also witness in *Matter's End*, is an underlying gesture to phenomenology (in the form of the experiential and the ontic), where the experimental and the theoretical come together to produce an explosive outcome through the deployment of the rationally inexplicable (though it could be a mere illusion hiding the science behind), such as how Daydala and her men were able to vanish quickly without a trace despite being holed up in a cavernous mine. As we will see later, both stories can be read as parables of ontological pursuits of dimensions not unequivocally constrained by direct physics-to-mathematics correlations, and therefore, need only adhere to the constraints imposed by the aesthetical needs of the narrative.

At the same time, both stories draw on narratives of big laboratory sciences that two of the three writers were acquainted with in their capacities as physicists. When *MMM* was written, the social studies of laboratory sciences were only just beginning, and most discussions of the latest sciences in this region are more likely to appear in hard science fiction magazines such as *Analog*. Though only gestured to in the stories discussed here and not directly explored in the chapter, the organization of big science had implications on the decisions and choices pertaining to experimental directions and design. Nevertheless, it is not only the politics that fund and grow the science that are at play, but also market forces, as the technologies produced are seen to provide potential profitable avenues, as was highlighted in *MMM*.

*Matter's End* is the more straightforward of the two Benford short stories discussed here but the thinking behind the sociology of physics, and of scientific knowledge, is much more sophisticated than that found in *MMM*, possibly the desired effect of learning from science fiction history, given that this story was written almost twenty-years later. Benford is interested in providing the necessary scientific edification to his readers so that they could appreciate the intellectual flavor of the story: the implausibility of the decay of a proton due to its inherent stability and because it is the lightest and most fundamental of the baryonic (three-quark) particles.

Moreover, the decay of the proton would change how the universe is viewed while also enabling a major stride in envisioning a symmetrical feature beyond the current Standard Model, and therefore, a new model that is either derived out of Grand Unified Theories or something completely unprecedented. There are intricate maneuverings involving chirality (such as whether the field occupied by the particle has left or right-handed spatial directionality), and also the imbalance between matter and anti-matter caused by the explicit breaking of the baryonic number symmetry, therefore ending in a non-zero sum total. The story challenges the assumption of logical materiality with regard to the unexpected and speculative in a high-energy experiment, as was presaged by his mystical-minded Indian physicists with their rudimentary instruments but unbending will.

Probably, as a way of upping the notch on political intensity that could culminate in the fulmination against scientists in the finale, Benford includes a background of third-world disenchantment, and hostility, towards first world science, that is equivalent to the ecological narrative of the dumping of products of questionable provenance onto impoverished and less stringently regulated low-income countries. Interestingly, Benford also illustrates how interpretations derived from semi-mystical philosophies can have actual physical ramifications, a ploy possibly used to illustrate an aesthetical counterpoint to a materialistic form of scientism.

Here, the laws of nature became the point of contestation when what was impossible becomes actualized. Whatever the physical constraints, creative interpretations would be able to extend beyond the physically observable in the same manner that quantum theoretical interpretations could work around the emergent qualities of matter that appeared paradoxical. There was even a reference to the "implicate order" first advanced by Bohm to reference the deep-seatedness of the unknowable.

Benford's decision to set the fiction of the discovery of the proton's lifetime in India, a country with a complex history in physical and mathematical sciences that preceded modern science, allowed him to engage in some metaphysical speculation in relation to the physical. Metaphysical speculation does not entail changing the ontology of physics, but rather, changing the perception of its laws, particularly when an unexpected discovery is made regardless of the degree of expectation. As one of the Indian physicists tried to point out to Clay, nature responds in accordance to our choice of theoretical and experimental approach. However, nature's response is as limited as the approaches used, therefore representing science's imperfect access to the ontological, for we have no omniscient relationship to the deep-time of history.

> "You are an experimentalist, Dr. Clay, and thus – if you will forgive my putting it so – addicted to cutting the salamander." Patil made a steeple of his fingers, sending spindly shadows rippling across his face. "The world we study is conditioned by our perception of it. The implied order if partially from our own design."
> "Sure, quantum measurement, uncertainty principle, all that," Clay had sat through all the usual lectures about this stuff but didn't feel like doing so again. Not in a dusty shed with his stomach growling from hunger. He sipped at his cup of weak Darjeeling and yawned.
> "Difficulties of measurement reflect underlying problems," Patil said. "Even the Westerner Plato saw that we perceive only imperfect modes of the true, deeper world."
> "What deeper world?" Clay sighed despite himself.
> "We do not know. We *cannot* know (271)."

What is not revealed, at any point in the story, was how speculation is deployed experimentally, whether by the following of a hunch or hypothesis that constituted the fictional discovery of proton decay. Benford did not bother to go into the details of what steps the Indian physicists took to have envisioned the breakthrough, because the very fictionality of the event made the steps opaque guesswork to him; even if he were able to imagine them, not all of his readers would care enough to know. Nevertheless, if Benford had wanted, it would have been possible for him to draw on the store of countless other experiments of a similar nature that had been performed, by working backwards from the point of discovery. However, it would be hard to justify the literary value of such extended descriptions of the experimental method unless the narrative can be sublimated. All that we have, in the story's finale, is the perceived effects of the breakthrough by reconfiguring his characters into a fugue-like state. This is understandable, for it is difficult for a scientist to go back to his/her science and imagine something that has not yet happened in actuality because that act of imagining would reveal a bias that is a result of his/her web of belief, such as I have discussed in chapter five.

If a collaboration, and tension, between theorists and experimentalists, as well as their divergent attitudes, were demonstrated in *Matter's End*, *Mozart on Morphine* is a semi-autobiographical but fictional account of a mathematical physicist that alternates between works of intellectual creation and the 'annoyances' and obligations brought about by the quotidian, including life-altering real-world events such as an illness, that no amount of rigorous probing at ontology could change. But the story's core thesis is that, however much major life events might shift the narrator's attitudes and priorities work-wise, the objective realities of the physics are immanent and will continue to be regardless of how he perceived or interpreted the scientific episteme.

In *Wholeness and the Implicate Order*, written in the 1970s, Bohm argues that the prevailing trend of modern physics is against thinking in terms of a divided wholeness, where any aspects of relativity and quantum theory that suggest otherwise are de-emphasized and shoved into the dense lines of mathematical calculus. He suggests that the trend is towards a "traditional atomistic notion that the universe is constituted of elementary particles which are 'basic building blocks' out of which everything else is made" (14-5). This is not an inaccurate account of the tension between a particle and field-centric discourses of microphysical events, as I have pointed out in chapters two and three, as well as Appendix A. The 1970s constitute the height of particle-centrism because of the kind of discoveries emerging from particle physics, many of which were highly unexpected and have changed our fundamental relationship to our physical universe. Even the construction of Higgs profile mediates between field-like and particle-like dispositions.

India, epitomized by the depiction of spiritual scientists and smoldering representations of the exotic unknowability of pungent smells, organized chaos, and intoxicating rhythms, is put into contestation with the clean and definable logic of the Western physicists; India is supposed to sit at the binary opposite of logical positivism and exaggerated demonstration of theoretical underdetermination. With that in mind, Whitehead's concept of process as a way of synthesizing the not always harmonious narratives of matter across the supposedly unconnected physical properties can be repatriated for understanding how process operates in speculative experiment. Moreover, his philosophy of organism presents the organism as a community of "actual things" inseparable from process that is repeated from phase to phase but accumulates information with each new phase so that a semblance of epistemological completeness, and therefore ontological attainment, becomes apprehensible. Probably, fiction as process philosophy can be the bridge to the seemingly irreducibility of epistemological dualism that attempts to reconcile the theoretical unknown with an actual world.

In addition, Benford's representation of the Indian physicists' view of the universe in *Matter's End* bears a resemblance to Bohm's discussion in the latter's aforementioned book. Bohm argues that in the East, such as that represented in Benford's rendition of India, exists a view that perceives nature as representative of wholeness and denies fragmentation and division. Bohm further adds that the measurable is less fundamentally important than the immeasurable, the latter of which cannot be named or understood through conventional reality. What that is measurable is false and deceitful, and Benford evokes this very end in the story when he conjures a semi-apocalyptic outcome of the acts of measurement, for as Bohm states "the entire structure and order of forms, proportions, and 'ratios' that present themselves to ordinary perception and reason are regarded as a sort of veil, covering the true reality, which cannot be perceived by the senses and of which nothing can be said or thought" (23). In other words, one can observe more directly the effects of a physical action but not so much the causality of that action, which would also allow one to capture the ontological quality of a physical event; what could be measured is the effect of the cause. Therefore, one is dependent on an interpretive theory as a surrogate to observation of the cause.

The fictive what-if representation of the proton decay in *Matter's End* parallels the historical narrative in the scientific tale of the muon and beta decay in the weak interaction of the Standard Model, even if the reader is not told the details and has to take the author's word that a discovery has definitely been made. While the experiments are solid and the data indisputable, the calibrations and selections that led to the collection of raw data, the reconstruction of the data, the statistical analysis, and the final interpretation, all operate through approximations that allow a measure of constrained speculation to seep in. However, as speculation is wedded into decision-making paths, they are not consciously viewed as separate from the acts of hypothesis-forming or theory construction. Yet, it is clear that many of the metaphysical postulations that Benford engages in over the course of the story are a result of his indirect deployment of speculation to discuss hidden causalities.

The question of whether there is uniqueness to the scientific method in relation to scientific observation comes up in *Mozart on Morphine*, a meta-narrative that deconstructs the cognitive processes of a single physicist even as *Matter's End* puts an individual physicist in relation to other physicists of very different cultural backgrounds despite their sharing of an unassailable faith in the common vocabulary of the scientific method: measurement, observation, and interpretation. But in both stories, Benford is bent on showing, with as much authenticity as he could muster, how the scientists (who are the physicists in this case), would behave when left to deal with the professional world they inhabit, something which Benford claims that certain science fiction novels featuring scientists as protagonists fail to do, such as Le Guin's *The Dispossessed* (*Is there a Technological Fix* 89). But there is no fixed natural law governing how scientists behave other than by force of habit, epistemic socialization, and their relationship to the environment they navigate, as my second chapter has indicated.

*Mozart on Morphine* highlights a multi-layered act of measurement: the measured dosage of morphine delivered to the narrator in his sick bed, the measured rhythms of the classical symphonies he listened to, and the precise rendering of mathematics for explicating the problem of quantum theories extending beyond the micro world to one that transcends the unrelenting routine of our macro-worlds. For the narrator, there are depths of chaos and irreducibility in that grey area, not yet totally embraced by either classical or quantum physics, so that the point of irreducibility can mean a difference between corporeal life and death. That point could be none other than an act of everyday violence that is nevertheless too difficult to be mathematized

> I saw the teenagers scattering and the scrawny man in his twenties poking the small silvery gun at them, yelling something I couldn't quite make out. I assumed as an automatic axiom that the gun was loaded with blanks; certainly it wasn't very loud...The man started swearing at a kid near me, who was moving to my right. I was still doggedly running so when the second shot came I was just behind the kid and the found went tssiiip! by my head. (10)

The narrator was not shot but the rush from the event activated his neuronal circuits and unblocked the pathway to inspiration for untangling the tangled equations he had been working through, for as he continued

> I had been pursuing a model for the universe which did not begin with any assumption about its dimensionality. We are used to our cozy three spatial directions plus ever-flowing time – four dimensions in all. When God made everything, was this choice forced? Could the deep laws governing matter work well in, say, six dimensions? Twenty-six? (11)

The sense conveyed is that the depths of our unknowability is so great that other forms of ignorance that are merely expedient for maintaining social relations are too petty in the grander scheme of things, yet so absolutely necessary for social cohesion. At the same time, what Whitehead argues as elements that can extend beyond the body, in this case the body of the narrator, are the straight lines and planes, which, when read metaphorically, represent the physical scientist's conditioned response to the world. According to Whitehead, there is a problem in identifying the distinction between a 'presented locus' that is a geometrical relation of the body that reacts to certain presentational immediacy relating to antecedent occurrences external to the body that are then interpreted for relevance to the main body; and the 'union of becoming' that is dependent on actual entities and their qualities, locus, and duration in relation to each other. These are what the narrator strained against in his projection of the implicit order between the events of his trip to Alabama to visit his folks, the car accident, his hospitalization because of an appendicitis, and his calculations that hinted on a theoretical merging between the as yet disjunctive world of gravity and the other forces of nature, in preparation for the emergence of quantum gravity.

Benford's deliberate use of stream of consciousness, which grips at each temporal frame as they pass without imposing a rigid categorization of each moment, showcases the tension between the striving for an elegant prescription of the physical world and the inevitable chaos that reigns in the everyday of a physicist's life. Then, there are the direct or indirect consequences on the narrator's attitude and sense-perception regarding the events of his life in relation to the calculations he was trying to solve, and how those events had, or not, shaped how he thought about his science. The reader is only belatedly informed that the story he/she is reading is a speech at a Nobel Prize award ceremony. The underlying significance of the narrator's explication is revealed in the finale, as he speculates on the connection between mind and matter, as well as that of quantum choices, none of which are connections that can be broached with certainty. But the source of his obsession that is also a topic receiving a revival of interest among metaphysicians, philosophers of physics, and theoretical physicists currently, is temporality and the structures of time, for this is where philosophical and theoretical physical speculations could range most freely beyond the limits of experimentally imposed constraints

> One of the deeper implications is that there is another kind of time. In that system, our truncated space-time forms a surface in the more general, superstring space-time. To that world, everything we perceive would seem like the surface of a soap bubble, wobbling in air. The bubble has no edge, no boundary – and so we will never, in that higher coordinate system, plunge through an abrupt juncture. This implies that time, in the larger sense, is never-ending. Of course it is not our time, but rather, the duration defined in higher spaces. The existence of this generalized time is perhaps the most startling deduction of mathematics. But what does this mean? We search for a completely unified theory, curling the fragmented forces of our hobbled universe into the Ur-Force. Still, even that is just a set of rules and equations. (19)

The quotation above relates to a long-standing interest in the connection between string theory and particle physics, with gravity and the dimensions of time as mediators. With the confirmation of the Higgs boson, interest in revisiting the potential of string theory as a logical connection to supersymmetry, which is considered as the most promising among the exotic candidates for extending beyond the Standard Model, has been renewed as described in the penultimate section of chapter 5 (in 5.2). Despite its fictional quality, one could read this story as scientific auto-ethnography.

Nevertheless, as the narrator (and Benford) both admit, the story does not fall into the usual category of an oral history produced by a scientist in response to questions about his/her scientific achievements and serendipitous acts that culminated in the breakthrough in his/her work. However, the story demonstrates how remarkably difficult it is to define matter as it performs in accordance to, and beyond, the multitudes of physical laws that are meant to constrain the actions of matter, but which, in actuality, fail to control for non-causal metaphysical events.

Further, the mind-matter interaction in the quantum space demarcated by *Quarantine* that relates molecular neuroscience with acts of collapsing quantum states is also taken up by *Mozart on Morphin*e when the latter addresses the connection of fundamental essences to humanity, not unlike Lightman's *Einstein's Dreams* discussed in the second chapter. That essence cannot be readily explicated; all attempts at describing the essences tend to be impoverished and require substitutions by other substantial signifiers. The aforementioned problem parallels the difficulties of envisioning new physics materially when one is uncertain as to whether the current analyses are unequivocally correct, or if a major revision of a paradigm that reconsiders cases of discoveries going back more than five decades is required in order for a new, yet unforeseen, model, to be reconstructed.

### 6.3 The Morality of the Scientific Method: Fictive Realities

There are three common scientific features, or values, embodied in the three stories. The first has to do with metaphysical morality, or the abstract conditions where ethical questions arise: do we try to save the world, as we know it, or allow a domino chain of events that can lead to an unpredictable outcome just for the purpose of advancing science. In *MMM*, Leigh had to choose between making sure that the mine where the scientific operations and mushroom farming took place would not explode and kill everyone in it; or recovering all of the petrons, including the ones that Daydala had filched from the leaking vat, because of the potential for greater catastrophe should the petrons be inappropriately let loose. This was the same ethical question that confronted the narrator in *Mozart on Morphine*, though at a more conceptual than corporeal level (despite his having been very ill), on the question of human experience in relation to the science and the effects of amplifying certain physical choices or experiences.

In *Matter's End*, one is given a stark and literal representation of quantum observation that riffs off the Schrödinger cat and wave-particle duality paradox[10] of probabilities and knowability. By announcing the confirmation of the rate of proton decay, Clay and the team in Bombay made the entire world complicit in their own unraveling, now that the latter were made conscious of the instability of matter.

[10] For the uninitiated, the Schrödinger cat story was used as an example by Schrödinger in one of his papers, *Die gegenwärtige Situation in der Quantenmechanik* (The Present Situation of Quantum Mechanics), where he fretted over the ontological incompleteness of the quantum mechanics in its ability to describe but not to predict. In this original thought experiment, which is about the superposition of states (the state of the cat being alive or the cat being dead), a cat is secured in a sealed steel chamber (supposedly with a device that would still bring air to the cat), together with a Geiger counter that contains bits of radioactive substance. Should the atom decay, the cat would then be dead after about an hour. However, there is no way in which the observer would know unless he/she is able

> Mrs Buli stretched lazily, as though relaxing into the clasp of the moist night.
> "So we have proven the passing nature of matter. What fresh forces does that bring into play?"
> "Huh!" Clay spat back angrily, "Look here, we just sent word out, reported the result. How –"
> "So that by now millions, perhaps billions of people know that the very stones that support them must pass."
> "So what? Just some theoretical point about subnuclear physics, how's that going to – "
> Who is to say? What avatar? The point is that we were believed. Certain knowledge, universally correlated, surely has some impact –" (*Matter's End* 286)

The concern for the safety of the world and the danger that scientific experimentations could unleash had been part of the public rhetoric in the days leading right to the re-starting-up of the Large Hadron Collider (LHC) in 2010 after

---

to unseal the chamber and look inside. Until then, the smeared condition of the cat is that it is either dead or alive (though not both), 'smeared' out in equal possibility. The smeared, as translated in the paper, is "…part of the inner law of the concept that it should change in a given manner, that is, if left to itself in a given initial state, that it should continuously run through a given sequence of states, each one of which it reaches at a fully determined time. That is its nature, that is the hypothesis, which, as I said above, one builds on a foundation of intuitive imagination (152)." One can find the translation of this paper in part 1.11 of *Quantum Theory and Measurement* edited by John Archibald Wheeler and Wojciech Hubert Zurek. Princeton, NJ: Princeton University Press, 1983. The wave-particle duality concept also draws on the same concept as that found in the Schrödinger cat paradox, except that the indeterminism concerns how the performance of observation at a quantum or classical level could lead one to observe matter as behaving as particles or waves. There are both dominant and alternative quantum theoretical interpretations of this problem that I will not get into here.

its unceremonious shutting down in 2008 due to a problem with some of its giant magnets. Prominent biochemist Otto Rössler continues to insist on the danger that the LHC experiment poses to the rest of the world.[11]

Then, there is the second feature connected to the first, in that the scientific observations through the performance of experimentation and calculations could lead to the gradual revealing, in tandem with the unraveling, of the properties of entities being analyzed. In *MMM*, scientific rivalry reveals the problems of precedence and ownership, and the different point of views and knowledge that rival groups can have of the same entity, as laid out here with pun intended.

> "Petrons!" she exclaimed. "I presume you mean the charge. If they must have a nickname, I prefer to use dayons. I predicted them years ago. Petronelli only found them."
> Leigh almost felt like laughing. Even at this charged moment the injured vanity of a scientist was fighting for recognition.
> "I really don't care what you call them – dayons, petrons, they are still stolen property. I'm afraid you have no right to them." Still holding the gun on Daydala, he moved to the petron vat and with one hand began to uncouple it from the machine.
> …
> "Nonsense, Daydala. I know you can't blow the mine up. Petronelli says that the boiler might make a minor explosion but it would take a while to develop."
> Daydala laughed scornfully "Petronelli has not the knowledge I have. He doesn't know the parameters. Put your gun down or I shall throw the switch." (*Minotaur in a Mushroom Maze*: *Conclusion* 91-92)

[11] See an example of one of Rössler's statements on the danger posed by the LHC at the Lifeboat Foundation website, https://lifeboat.com/blog/2012/03/cern-cannot-continue-the-lhc-experiment.

In *Matter's End*, the usual work of turning scientific observation into a reportable event with solid consistency and logic

> Consideration of each candidate event, his checks and counterchecks, vertex corrections, digital array flaws, mean free paths, ionization rates, the artful programming that deflected the myriad possible sources of error. He could feel tension rising in the room, as he cast the events on the inch-thick wall screen, calling them forth from the files in his cubes. Some he threw into 3-D, to show the full path through the cage of iron that had captured the death rattle of infinity…And at the end, all cases reviewed, he said quietly, "You have found it. The proton lifetime is very nearly $10^{34}$ years." (274)

does not necessarily work with the geometrical properties we are used to working, though it could lead us to question our 'normal' perception of geometry as solid and non-transient.

> Smooth glistening forms began to emerge from the rough, coarse earth. Above the riotous, heaving land the moon was now a brassy cube. Across its face played enormous black cracks like mad lightning…Quietly the land began to rain upward. Globs dripped toward the pewter, filmy continent swarming freshly above. Eons measured out the evaporation of ancient sluggish seas. (288)

The description above is reminiscent of the amalgamation between consciousness, the mind, and quantum physics that we have witnessed in chapter three in Egan's *Quarantine*. While *Mozart on Morphine* takes a more sober approach, the description by the narrator of his relationship to the world that is shaped by his dependence on what the mathematical equations tell him merely made him acutely aware of the depth of his ignorance

> Still, there emerges now evidence of mental processes at work on many level of physical reality. We may be part of some larger act. For example, perhaps we contribute remotely to the universe's thinking about itself… We probably cannot know this with anything approaching scientific certainty – ever. The

> recent work of myself and others suggests, though, that higher entities affect our times in distant but profound fashion…The equations can only hint, imply, describe. They cannot tell us why. (22)

This brings us to the third scientific feature, which is the oriental Other as antipodal to western scientism in its emphasis on material rationality. In *MMM*, a conspiracy subplot involving a pantheist/deistic religion and goddess worship of the oriental other is constructed to heighten the tension of the unknowable, but Benford's fiction discussed here also deploys the trope of the exotic oriental to amplify the scientific speculation into exotic forms of matter that is core to the narratives while engaging with the politics of scientific knowledge production within the context of the Third World. The binary representation of scientific knowledge in Western and Eastern mode of thought within these stories are marked by an American exceptionalism of the 1970s and mid-1990s that turned the promise of international scientific collaboration into an international race for scientific ascendancy due to the Cold War, and the unending rhetoric concerning the threat of the non-allied Other.

At the same time, there was a technological optimism, in the sense of the enormous scientific complexes from space to nuclear programs, lasting until the early 1990s that marked the official end of the Cold War (though Jameson, whose remarks I am appropriating here, has a different take on technological optimism from myself. He considers technological development primarily from the context of industrialism) that was "brutally effaced by the neo-conservatism revolution and its accompanying effects – the debt, population explosion, the failure of modernization – in the Third and later in the Second Worlds" (Jameson 153). The rise of reactionary neo-conservatism that has been displaced from the US onto the foreign Other is explicitly detailed in *Matter's End*. In *Matter's End*, the Indians wreaked havoc by breaching an (un)secured scientific facility (that was hosting a US citizen) and planned an insurrection against what they perceived as a desecration of sacred beliefs.

This is where Whiteheadian process philosophy becomes useful in helping us understand the modes of scientific thoughts underlying the narrative process, even if I am modifying his original reading slightly to suit the current object of examination. According to Whitehead, there are two kinds of processes: macroscopic and microscopic. The macrosopic process represents a transition from actualization to the process of actualizing ("actuality in attainment") while the microscopic process involves the conversion of conditions postulated as real into determinate actuality (214). In the two sentences, Whitehead reconstitutes the meaning of the real in the same way that the real of the universe is transformed (though not definitely actualized); the theoretical methodology resonates with the sequences of events culminating into acts of discovery and validation in *Mozart on Morphine* and *Matter's End*. In the former short story, the formal actualization is kept subdued while the narrator gives free reign to speculating about the 'other' hidden potential that his theory might not have predicted because his theoretical model could only extend from the observable, even as he realized that most of the processes of nature are non-local to our dimension, and therefore, hidden.

The three stories converge onto a theme of discovery and a feeling for discovery, where the subject-superject[12] underpins the foundation of a responsibility that one feels in one's relationship to knowledge, and the knowledge process originates a set of feelings that strives for the ultimate act of creation that is always negotiating between the actual and the ideal. With the discords involved in the narrative of discovery in particle physics, epistemic unity is broken once a seemingly marginal problem, or a 'hidden variable', comes out of the woodwork and messes up an otherwise 'neat' interpretation of an observation.

The concrescence, which is the process of prehension for achieving ontological cohesion, attempts at integrating between multitudinous entities that might have values in conflict with one another in order to achieve momentary ontological unity. Concrescence governs the final pronouncement of a discovery that contests a linear methodological approach: one that excludes the potential for speculation. In this case, knowledge could serve a definite goal of resolving particular issues, however indisputable and continuous the chain of custody of scientific evidence, if only because disruptions that do not have to answer for macro-state consequences can be remaindered out.

[12] A superject represents Whitehead's attempt at countering the limitations of a subject that stops short of the creative.

In the case of *Mozart on Morphine*, Benford tries to show that there is a macro-level consequence if a particular theory has been proven true, and that consequence stems from the change in one's primary feeling, or physical feeling, after realizing that the familiar universe one occupies is about to change. As Whitehead puts it,

> A simple physical feeling is an act of causation. The actual entity, which is the initial datum is the 'cause,' the simple feeling is the 'effect,' and the subject entertaining the simple physical feeling is the actual entity 'conditioned' by the effect. This 'conditioned' actual entity will also be called the 'effect.' All complex causal action can be reduced to a complex of such primary components. Therefore simple physical feelings will also be called 'causal' feelings. (326)

Even if the physicality of that feeling is exaggerated in the story, it enables the presentation of an important point made in *Process and Reality*, which is his attempt at foregrounding the conceptual development of the superject and concrescence, whereby the presence of an entity, or the employment of a particular action or sets of feelings, can stimulate the generation of more entities (that could be either actualized or merely idealized).

The extension, or projection, into another feeling or set of interactions is what experimentation entails. In other words, the interpretation of the observables (or absences/non-observables) in the experiment constitutes following a direction because a specific interpretation of a high-energy event focuses the observer/experimenter's attention inward on a theory with sufficient point of preoccupation to the exclusion of other consideration. But, should such acts of concrescence become discrete and almost-independent propositions by disinterested parties involved in the act of crosschecking, with each party independently forming their own hypotheses of the said proposition. The nexus of these acts would be cumulative for each individual concrescence to converge into a point of heightened predictive capability.

In all the three stories, acts of observation and interpretation of the measured effects raise abstract but ethical questions that are implicitly connected to our macro-world but are still speculative, with too many moving-parts in the potential resolution for an ultimate conclusion. Moreover, the very abstract quality of the ethical questions lead to responses that are insufficiently explicit for producing any definite resolution as that would require one to have more access to the ontology that sets off the questions than one already has. At the same time, through the marshalling of the fictional form, the authors are able to write a poetic history of science that allegorizes the problem of enacting intellectual histories when much of what is understood are extrapolated from facts constituted of the best logical options. In other words, one does not always know which layer of the multifaceted socio-political dimensions can provide the most direct influence on the knowledge produced.

## 6.4 Conclusion

The physics portrayed in each of the story is a demonstration of a crisis in particle physics that had been ongoing even to this day, with no final word in the horizon. Hard science fiction is suited for such explorations of crises and for throwing together all the vexing paradoxes to see whether one might create the ultimate 'grand' paradox that unveils the rupture in that fabric of cosmological ontology, or if the paradoxes will smooth out the way wrinkles on fabric do.

At the end of the day, science fiction is located within a 'classical' dimension of actions, replete with its own language codes, but well poised for elucidating types of knowledge located in dimensions alternate to our physical ontology. Therefore, knowledge aimed at exploring a dimension beyond the scope of our linguistic capability would require the localization of that knowledge within a comprehensible rubric. In the process, areas of knowledge that cannot be as easily localized takes on a speculative flavor.

*Information Science must refer to the actual domain of code, even if the theory could generate completely different models (and should). And despite the will and belief of the code's developers, decodings are just as possible as they are rare.* (**Friedrich A. Kittler in *Protected Mode***)

## 7. Simulating Nature: Monte Carlo as Speculative Methodology of Approximating the Ontology of Particle Physics

Had you been a fan of the *CSI* franchise, or watched a number of episodes from this, or similar crime-busting TV shows, you soon find that in many cases, we start with an unknown subject (the unsub) who had perpetrated the crime. In some cases, the identity of the victim might not even be known. You will then have teams of forensic specialists, from crime-scene investigators to a forensic pathologist, combing through physical evidence and the body, to piece together the cause of death, to locate the primary crime scene, and uncover the motive for the murder (often, the lack of a motive can just be as informative). Without any proper lead, or idea if the evidence is intact, what one gets are almost random pieces of a puzzle that are part of a crime or crime scene.

The detectives, at their end, would be trying to obtain as much paper trail as possible, on both the victim and unsub. In some of the more exciting episodes, one could have so little evidence to piece together, including the occasional lack of an actual body (even if the identity of a possible victim, and perpetrator, might be known), or the body was so badly decomposed, that the forensics team and detective(s) had to rely on computer simulation to piece together the last few moments, as well as identity, of the victim. The acts of evidence gathering, simulation (whether through manual crime scene reconstructions or computer generation), and organization of collected evidence to create a composite profile, converge into a heuristic method of testing, analysis, and interpretation.

The scenario depicted above that is most analogous to the Monte Carlo simulation in high-energy physics are the acts of simulation, specifically, the simulations that involved the crunching or processing of data to predict the likelihood of the how and why of an event occurring (which would be the crimes in the aforementioned scenario). Likewise, in high-energy particle physics, important to any analysis of data is an understanding of the different physics processes in relation to their detectors, such as at the Large Hadron Collider (LHC), which could only provide partial information. This is where the Monte Carlo simulations come in to help fill in the blanks. However, the Monte Carlo as depicted in high-energy physics is a strange beast, for in that field, the implementation of the Monte Carlo is more an accident of history than an intentional progression. It is also what makes its operation a little different from what might have been understood in terms of the Monte Carlo's development within war operations research and nuclear physics.

In this chapter, I will discuss how the Monte Carlo acts as a phenomenological bridge in the divide between theory and experiment in particle physics, as illustrated by macro level measurements of entities that go by quantum rules. Monte Carlo mediates the more speculative aspects of the new propositions in theory with the ultimate act of confirmation through experiments that had also been originally speculative. The development of the philosophy of Monte Carlo computation has as much clarifying effect on the scientific method of the physical sciences as it would have, surprisingly, in other theoretical developments in humanistic studies, such as in media theory, critical code theory, and the technology of film. Further, I will demonstrate Monte Carlo's synthetic position within physics and computing by locating it within cybernetic theory; it so happens that one of the pioneers of the Monte Carlo method, John von Neumann, was also a pioneer of cybernetics. That said, I attempt to go beyond Galison's positioning of the Monte Carlo method as a philosophy of pseudo-random generation, which though an apt description of the method, is a too limited reading as it does not sufficiently account for information creation, recirculation, and epistemic discords that can arise when speculating from the what-might-be within statistical data analyses.[1]

The Monte Carlo method evaluates what would have been rather complex multiple integration practices in calculus by reducing (or 'normalizing') the computation to probability density distributions. In other words, the Monte Carlo method "provides a method of simulating experiments and creating models of experimental data. With a Monte Carlo calculation, we can test the statistical significance of data with relatively simple calculations, which require neither a deep theoretical understanding of statistical analysis nor sophisticated programming techniques" (Bevington and Robinson 76).

[1] See Peter Galison's "Random Philosophy" in *The Reality of the Unobservable.*

There is a finiteness of range whence the random samples are selected, and each calculation starts from the simplest expansion series of a distribution function that are rendered only as complex as the computation process needed to obtain the required level of resolution and detail. Then, in the process of 'random' selections, parameters are set and values that fall outside the boundaries of the parameters are rejected. However, as will be explained later, having pre-determined parameters for initiating cuts on data of presupposed interest is problematic. For as Bevington and Robinson note, rejected events "do not improve the statistical accuracy and every effort should be made to reduce the time spent on calculations that lead to 'misses' rather than 'hits'" (92).

However, unlike the forensic scientists who get a semblance of corporeal access at the end of the day, the particle physicists would never achieve direct contact with the particles they study. In high-energy particle physics, the most direct contact the physicists could ever have with their object of study is statistically and digitally mediated, through the deployment of simplifications and approximations. Imagine trying to discover one's family history and only being able to do so 'virtually,' where even contact with living family members must be mediated. Further, the digital affordances that the Monte Carlo simulations deploy further augment the problem of undecideability because every legible fact and established theories are inbuilt with varying degrees of indeterminism.

When Karin Knorr-Cetina wrote *Epistemic Cultures: How the Sciences Make Knowledge*, the sensitivity of the detectors was not what they are today; the detectors' abilities to scan background events for locating faint signals of unknown particles were limited. Hence, the reference that Knorr-Cetina makes to the term 'smearing' is in relation to the distortion of physical distributions because of the detector's inability to resolve, or give distinguishable responses, to pairs of particles interacting too closely together. This kind of smearing is also a form of data 'smearing' (rendered imprecise) that comes from the addition of random variables to sets of computations to simulate what comes out of a finite measurement of errors (Bevington and Robinson 93). Smearing works at a much higher macro level than the micro-level smearing depicted in quantum level operations as represented by mixed quantum-classical states and the Heisenberg Uncertainty Principle's depiction of kinematical events at a relativistic energy level.

Further, there is another type of distortion, such as in the case of misidentifying one category of particles for that of another, which would result in more misinterpretations down the line, even when these misidentified particles are interacting with other properly identified particles. Such potential for misrepresentations stemming from misidentifications are also referred to as 'fake events' that are usually background to the signal events of interest. However, the still fresh-in-the-memory disaster, between September 2011 and March 2012, involving the OPERA detector in France that boasted of having detected a superluminal neutrino only to find that the supposed discovery was caused by flaws with the experimental equipment, kept the physics communities cautious about making stringent checks on every result before they are willing to concede that something new has actually been discovered. Even the detection of the Higgs boson was couched in the careful language of statistical probability, so that every other explanation unrelated to the discovery can be considered.

Nevertheless, one might want to consider the possibility that the so-called 'fake events' might indicate the limits of certainties in our searches, therefore highlighting the need to re-strategize how analyses and interpretations are performed; in other words, what techniques of instrumental, or even analytical, calibrations of experimental instruments must be mustered to anticipate the unknown within the finitude of the statistically known? While the Higgs boson was an expected discovery (in the sense that its discovery has been statistically calculated within viable standard deviations above the material background of other known elementary particles), a number of the discoveries that had preceded the discovery of the Higgs boson *had been unexpected*, and the physicists had, at one time or another, thought that the discoveries were artifacts of experiments that merely needed to be finessed, such as was the case with the discovery of the J/psi and charmed mesons (that are made up of two quarks).[2]

Even if the arguments proposed by Knorr-Cetina on the epistemic cultures of the high-energy physics still hold true, particularly the technical-epistemic core informing the narrative of discovery, analysis, and interpretation in standard quantum physics, we have to be mindful that certain assumptions, based on the limits of current knowledge, would have to be revised as new facts emerge, with the possibility of a scientific revolution à la Kuhn as and when more revelations flood in. Further, instead of arguing that a measurement is 'meaningless,' in the tongue-in-cheek manner of Knorr-Cetina; because of lack of direct correlations between the object being examined, how the measurement is performed, and the location of the measuring apparatus; one might want to consider measurement as not about approximating precision, or the most accurate depiction, of a material phenomenon.

---

[2] See Goldhaber, Gerson "From the Psi to Charmed Mesons: Three Years with the SLAC-LBL Detector at SPEAR" in *The Rise of the Standard Model*.

Instead, one might want to consider measurement as constituting the reality where certain theoretical interpretations can work in relation to that phenomenon, and then be prepared to capture the outcome of the interactions between the phenomenon and the theories involved. Such a process of measurement has already been at work in the description of quantum-like features, but requires a shift of the macro-observer's perception, and attitude, of its relationship to its physical world.

In fact, the development of simulation generators,[3] and the Monte Carlo method for testing pseudo-data generated from reconstructed experimental data, is what we refer to as modeling-as-measurement, whereby the data is used to build a composite representation of an entity, of groups of entities, without insisting that everything must converge into a sharp profile to produce representations of an unknown entity. At the same time, data that are considered ill-fitted are rejected as long as the recurrences of rejection do not become too frequent, to the point of having to reconsider one's entire epistemic framework. It is through the act of constructing models that the points of speculation within theory and experiment come together to demonstrate the problem of quantifying uncertainties.

There are two parts to the Monte Carlo simulation of which we have to be aware. First, there is the Monte Carlo dataset obtained through an event generator that provides the parameters for the momentum and trajectory of millions of individual particles, created from high-energy collisions that produce a specific process of interest such as the proton-proton, or the top-top (quark) events. The Monte Carlo dataset is used to determine what the process being generated would look like in the detector so that the physicist knows how to measure, or search for, event choices of interest. The simulation includes the modeling of the process itself, and how the particles from the simulated process subsequently interact with the detector. The raw data is then reconstructed and re-inserted into the Monte Carlo process for a second round of analysis and generation to produce another set of Monte Carlo data for comparison to experiment as part of the global-fitting process. An important aspect of the process elucidated here is how one might enact the process of data cuts out of the millions of events available, and to use the selected events to construct a physical (or microphysical) picture that has not been predicted prior to the process of construction.

---

[3] I will sometimes refer to simulators as generators and vice-versa as they are the same thing. But for the most part, I will use 'generator' as a more specific referent and 'simulator' to denote a more general one.

Knorr-Cetina highlights an important role that the Monte Carlo performs for excavating signals with similar signatures from background noise. However, she did not adequately address the assumption of errors, specifically errors that either came of accidental inclusion of artifacts, the background interference of faint signals, or the misidentification of certain unknown particles because of ignorance of a potential interaction. Further, the consistent appearance of errors may point to the incompleteness of a theory or model at hand, and potential misunderstanding of what a phenomenon entails.

Monte-Carlo as a scientific tool blurs the distinction between experiment and theory. As Peter Galison notes, "…the Monte Carlo appeared experimental; by eschewing the bench, it appeared theoretical. The categories themselves had begun to slip, in what must have appeared provocatively oxymoronic" (*Image and Object* 701). Galison and Knorr-Cetina agree to Monte Carlo's capability for expanding beyond the reach of the analytic numerical method, but that capability hinges on computing power rather than manual prowess. In recounting the history of Monte Carlo's development for detecting muons (from neutrinos) and neutron scattering, as well as in the calculations that led to the development of the A-bomb, Galison recounts the historical twists and turns that come from attempting to reconcile experiment to theory, especially when the theory does not provide sufficient descriptive guidance that could indicate conclusiveness.

One example was the neutral-current experiment; neutral-current is an oxymoronic term because it implies a current that is neutral (rather than charged, which is usually the case), and the very characteristic of its neutralness complicates the detection process. The neutral current is produced through the weak interactions mediated by neutral Z bosons; the search for the neutral current parallels that of the Higgs boson except that neutrons and neutrinos were involved and the experiment was conducted at a much smaller scale.[4] The phenomenological characteristic of the Monte Carlo is affirmed by its position at the intersection of theory and experiment. As Galison asks, "How should we class the type of data-to-model (inversion)? As experimental theory? Theoretical experiment? Is it a case of induction from data? Deduction from theory?" (747).

For the mathematicians, the Monte Carlo can be a measure of the changes within the state space and stochastic processes that are due to temporal variations as well as probabilistic representations. For the statistician, (s)he obtains the sampling methods that correlate to physical processes. For the particle physicist seeking to provide controlled reconstruction of a certain phenomena, both these different features converging at the Monte Carlo are important. As we will see in greater detail in the following section, Monte Carlo provides the background calculation that can be calibrated for greater sensitivity to new physics, thereby providing the model by which any background excesses can be calculated for new physics detection. Hence, Monte Carlo mediates between old and new physics by setting the condition for the latter to emerge, potentially, from the former.

---

[4] See *How Experiments End* chapter four. The earliest discussion of the neutral-current experiment can be found in *Reviews of Modern Physics*. 55.2 (April 1983): 477-511 as "How the first neutral-current experiments end." In *The Disunity of Science*, Galison discusses the historical role of Monte Carlo in early developments of nuclear thermodynamics and particle physics in "Computer Simulations and the Trading Zone," which I will not get into here. Suffice to say that the development of the Monte Carlo technique coincides with the growth of big science, where the combination and transmutation of multiple, and simultaneous, causal events produce details beyond one's expectation.

However, Galison brings up an important point regarding how Monte Carlo changes the way physicists deal with theoretical uncertainty and predictive precision. There are definite aspirations towards tracking errors to attain 'correct' expectation values and to decrease the inexactitude of measurement during searches. Furthermore, statistical investigations of the empirical data require the development of computationally efficient algorithms to generate pseudo-random numbers to fit the distribution of the generated data with the experimentally produced data. Monte Carlo constitutes an alternate reality, what Galison refers to as the tertium quid intersecting theory to experiment, that I ascribe to a phenomenological constitution of the epistemology of particle physics. The Feynman diagram provides the blueprint whence specific algorithms are developed for processing selected interactions, especially in calculating crucial corrections to the quantum field theory of the hybrid electroweak force.

Even if the Monte Carlo simulation process is embraced as unequivocal to refining experimental calibrations, the uncertainty embedded within its predictions, through the generation of pseudo-random numbers, had created uneasiness among physicists used to an analytic approach with well-defined resolutions. For the physicists, it means ceding the tight reign of control they have over the experiment and having to contend with not knowing where and when an uncertain quantity might appear in their data dump. Dealing with heuristically-obtained uncertainty also requires a cognitive shift in how one understands the acquisition of scientific facts, the construction of algorithmic and experimental modularity, and the epistemic identity within the nebulous intersection of experiment and theory.

Additionally, the data generated through the Monte Carlo is also vulnerable to what Knorr-Cetina refers to as "negative knowledge," knowledge at the limits of knowing, of the mistakes made in the process of knowing, and also the existence of a barrier (expected or unexpected) that interferes with our process of knowing. As methods are developed to decrease, and suppress, errors caused by random number generation, Monte Carlo shows up the delimitations one encounters when trying to develop algorithms out of theory whose ontological robustness is questionable, such as quantum chromodynamics (QCD), which is a short-range strong interaction that is intense yet highly unstable above a certain threshold and therefore very complicated to simulate. But it allows physicists to skirt the irresolvable questions of the theoretical simulation of a jet, an experimental signature of a quark in the LHC detector production that is important for simulating the energy range that the Higgs boson is expected to appear in.

In the following section, I discuss how the stochastic and probabilistic quality of the Monte Carlo techniques can be read in conjunction with cybernetics, and what this would mean in terms of code, algorithm, and digital epistemology. However, I will not be entering into the specific formalisms that shape the development of the Monte Carlo, as they do not contribute to the core arguments of this chapter.

## *7.1 The Cybernetic World of the Monte Carlo in High Stakes Physics*

The stochastic process that dominates the Monte Carlo routines might eschew temporal histories epitomized by the Markov chain (a stochastic computational method) but the Monte Carlo has a Husserlian eidetic quality because its mathematical narratives and statistical excesses enable one to approximate ontic unity, the very embodiment of 'quantum memory,' which has a ring of conceptual incredulousness given that the quantum states, in their pure conditions, do not maintain information of past events after the events had taken place. The idea of quantum memory, as was proposed by cybernetics pioneer Heinz von Foerster at the Sixth Conference of Cybernetics in 1949, simulates both the transference of remembrance from one 'cell' to another, through the process of 'impregnation,' but does not destroy the information stored in the original cell. The quantum state-like explication of neural networks is reminiscent of the concepts Egan had invoked in *Quarantine* with regard to the energetic quality of information transmission.

In fact, von Foerster was trying to draw parallels between what was understood about the behavior of the energy (and particles) at the micro-quantum universe with what was known about neurons and molecules in biochemistry and neuroscience at that time, to produce his idea of a "biophysical backbone" that draws heavily on the philosophical version of phenomenology and early simulation processes afforded by developments in nuclear science.[5] The samplings and data counts conducted through the Monte Carlo converge into unity; unity has to be determined purely through the essences (that are both the data and effects extrapolated from the data) provided by the experiences. All the actions embodied within the Monte Carlo contribute to a physical knowledge that will then feed back into the redefining of the experiential known as particle physics phenomenology.

However, this means that Monte Carlo becomes unitarily irreducible, phenomenologically in the philosophical sense, because of a need to consider secondary and tertiary interactions that might not be reducible. In other words, particles are not only produced through the interactions of a single particle pair of partons (that are 'point-like'), but also through additional partonic (hard) interactions and softer interactions from beam remnants, all important for fitting the experiment with the Monte Carlo generated statistical models. Therefore, the models are representative of the

> …empirical consciousness of a self-same thing that looks 'all around' its objects, and in doing so, continually confirms the unity of its own nature, and the essence and necessarily possessing a manifold system of continuous patterns of appearances and perspective variations, in and through which all

---

[5] See Foerster, Heinz von. "Quantum Mechanical Theory of Memory." *Cybernetics: Circular Causal, and Feedback Mechanisms in Biological and Social Systems. Transactions of the Sixth Conference March 24-25, 1949, New York, N.Y.* Ed. Heinz von Foerster. New York: Josiah Macy Jr, Foundation, 1950. 112-145. Print.

> objective phases of the bodily self-given which appear in perception manifest themselves perspectively in definite continua. (Husserl 131)

Husserl appears to have anticipated the same reasoning that enters into the formulation of Monte Carlo algorithms, with its own fundamental perspective based on the bounds of its sampling quality and measure (including the margins) of statistical error. The Monte Carlo generative practices of simulation and reconstruction represent that manifold system of "continuous patterns of appearances and perspective variations" (131).

Even as the Monte Carlo generates events that focus on slices of the moment, without the baggage of past recollections, the fine-tuning of the algorithm is based on systemic eidos, and therefore, of a recollection that has already been synthesized (and processed) rather than merely accumulated. The datum of Monte-Carlo reality, where the reality is a more sublime reconstitution of a physical feeling for nature, can actualize metaphysical subjectivity while still attending to the demands of physical reality. Hence, the eidos of the physical is re-articulated by Whitehead when he states that "simple physical feelings embody the reproductive character of nature, and also the objective immortality of the past. In virtue of these feelings time is the conformation of the immediate present to the past" (*Process and Reality* 238). Even if not intentional, the preceding quotation can be read against the affective measure within quantum mechanics where time does not play a central role even in time-dependent equations, and which has a tendency to metamorphosize into a spatial quality due to the geometrical equivalence factor. Therefore, quantum mechanics can only predict the moment when the calculation is performed, with no innate sense of historicity or futurity.

However, the quantum mechanical memory of Foerster works on preserving memory against the coefficient of forgetting by utilizing the classical properties of the physical world, with the Monte Carlo presenting a form of eidos (the memory system) that compiles and profiles the quantum physical reality of particle-events by conserving the memory of what it was to prepare for the prediction of what-might-be. Galison has referred to Monte Carlo as an inquiry into natural philosophy because of the latter's ability to offer insight into a reality previously missing from a constricted view afforded by analyticity. In other words, the structure of the Monte Carlo is able to break away from a default assumption of null hypothesis to build relationships between seemingly unconnected and differentially manifested physical phenomena, through extrapolations between comparative datasets. Thereafter, new analyticity can emerge from the consolidation of these new connections.

As previously argued, Monte Carlo can be considered as a cybernetic construction of computational media (of the Whiteheadian concrescence) whence the flows of discreteness/digitality and continuity/analog become the discursive measure of the content within the bounds of the simulation and generation practices. The flows are sensitive to potential mis-measurement, and to the encoding of potential biases onto the observation process because of the observer's 'special' relation to empirical and theoretical expectations. A Monte Carlo model is largely probabilistic and stochastic, and derived of theoretical models that are mostly indeterminate, mainly descriptive, and insufficiently prescriptive. Whitehead's genus of conditions for novel consequences of an enduring/eternal object are aimed at ensuring the preservation of the original identity of a physical subject, while providing sufficient contrast between the subject at the "ground" and a stochastically conceived model. Moreover, Whitehead's theory of extension, where he posits that

> …decided conditions are never such as to banish freedom. They only qualify it. There is always a contingency left open for immediate decision. This consideration is exemplified by an indetermination respecting 'the actual world' which is to decide the conditions for an immediately novel concrescence…some actual entities may be either in the settled past, or in the contemporary nexus, or even left to the undecided future, according to immediate decision. (285)

The quotation above anticipates the same epistemic attitude as the extension theories of Standard Model, where even the physical revelation of the Higgs does not denote any epistemic certainty with regard to experimental choices. After all, the creative emergence that the Monte Carlo method supports can be imported into the "physical feelings" of actual "pseudo-determinants" that transmute into a singular nexus from an array of intensities, valuations, and eliminations; the physical feelings arise from mediated rather than directly physical connections to the various strains of nature. Monte Carlo actualizes physical representation to be as congruent as possible to the microphysical computational processes being modeled, a process Whitehead refers to as the conversion of conditions from the merely real (which represents the potential for physical realization) into determinate actuality (even if the microphysical events can only produce an approximate outcome).

As someone who has pioneered foundational mathematical thinking and cybernetics, John Von Neumann, through his *The Computer and the Brain*, provides insight into the human-computer interface that shapes the development of the Monte Carlo by considering the process of memory (the Husserlian eidos) as digital and stochastic. According to Von Neumann, access to memory is dependent on the general state of the machine at the point of access, and not all points can be accessible at the same time, as provisions are needed for prioritizing when each access is allowed. He also warns of how errors may get into the system through the process of computational iterations that lead to divergences brought on by the amplification of errors.

By demonstrating the structural parallels between mathematics, computation, and what was understood of the neuronal sciences in the 1950s, von Neumann provides a purview into a total expression of problems and intentionality within each sequence of the control points in relation to their connections to several organs (organs of the body, or, in the case of the LHC, the various sub-detectors) for the simulation of more than one operation. The Monte Carlo simulators available today can be read analogously to the historic differential analyzers and integrators of mid-twentieth century physical computers considered as more economical and efficient than a purely arithmetic processer; the latter requires non-trivial but rather arduous steps when attempted electromechanically until the same steps could be sublimated in digital-electronic circuitry. That is why there are multiple modular codes available for breaking down event processing, whereby the codes ride on the idea that specialization will improve and increase the rate of output.

Moreover, as the previous chapter indicated, the search for a singular manifestation involves the differentiation of leptonic from hadronic decays, and the constitution of these interactions within electroweak and strong fields. Then, we have to attend to the transformation between the initial, intermediate, and final states of the particle processes. Given that the interactions simulated represent strong interaction particles as point-like, there have to be considerations of how the final statistical reconstruction can be interpreted in relation to the search, such as the Higgs boson in this case.

While early variations of the Monte Carlo methodology are much more physical due to the use of analog computing, their gradual transformation to the purely virtual can be read against Whitehead's theory of feelings, such as the process of transmutation that attempts to find unity in the disparate actualities of feeling (from the primitive physical to the noematic) and the oscillating movements that transmit between different categories of feelings (physical and conceptual, objective and subjective). The Monte Carlo process, despite its location within the virtual, is primarily a physical causality that has undergone transmutation at multiple points for the production of mathematical approximations that von Neumann refers to as vehicles in the transmission of information. Moreover,

> …a system of logical instructions that an automaton can carry out and which causes the automaton to perform some organized task is called *code*…If the machine is to solve a specific problem by calculation, it will have to be controlled by a complete code…Such systems of instructions which make one machine *imitate* the behavior of another are known as *short codes.* (*The Computer and the Brain* 70-1)

The algorithm controlling the Monte Carlo forces the computer to simulate, as authentically as possible, the process of data production at the detectors, before the data (with point-like and field-like interactive representations) are reconstructed and analyzed, to facilitate the process of passing judgment and locating of congruence between experiment and theory. This is to construct a procedural approach used for dealing with problems relating to independent trials and types of replications that could be either positive, negative, or producing of neither truth-claims nor final verifications. Since the Feynman diagram structures perturbative (approximative computational modeling) of quantum field theories and contains the rules of operation for describing the branching acts of fermionic and bosonic decay, the former becomes the ontology for the algorithm that underlies the short code for the computer to imitate the real-time behavior of the microphysical components of nature.

When thinking about the role of code in prediction, determination, and stochastic processing, let us reconsider the cross-sectional interactions that are the focus of the acts of simulation, and the Standard Model particulars that are of interest within the simulations, now that we have arrived at the cross-road in the aftermath of the unveiling of the Higgs boson, with the upgrade plans of the LHC to higher energy levels underway before the second phase of experiments begins. Additionally, the stochastic process of the Monte Carlo is able to trace and reconstruct the moment-by-moment effects of the decay represented within the different scales of the Standard Model gauge (force) interactions, with the weak interaction particularly important for describing the aforementioned Higgs mechanism's role in mediating the Higgs boson in relation to the other known elementary particles.

Furthermore, when particle collisions take place, they produce angular rotational momenta needed to calculate the handedness of the coupling processes effecting sensitivized polarizations that can then be computed. During the process of computation, loop corrections (the aforementioned quantum field corrections) are included because of their numerical relevance to Monte Carlo's statistical processing. Primary-stage corrections at the vertex (the primary site where particle interactions at the initial point of collisions are calculated) contain couplings that are flavor dependent (the quantum number properties of isospin and strangeness) for the bottom/beauty quark mass but with corrections added from the top quark channel. The corrections expand the top quark sector, thereby increasing the probability of obtaining information about the Higgs boson from background signals.

An important aspect of the aforementioned particle physics search involves the detection of very faint effects of new physics in the background of all new physics searches that make the calculation of the Standard Model contribution challenging. In the search for new physics, Monte Carlo provides directionality and legitimization to the process through its ability to engage in precision measurements required for constructing the parameters of the Standard Model: the pre-computed electromagnetic fine structure constant (characterizing the strength of the electromagnetic interaction) at zero momentum transfer, a Fermi constant that is extracted from muon lifetime decay, and the mass of the Z boson. All of these features are important for the evaluation of the different aspects of the Standard Model weak interaction needed in the decipherment of the Higgs boson. The effects are measured from the low and high end of the energy, whereby exotic new physics with slightly different structures can produce particles with lower energies than that born out of typical processes.

Therefore, a check for global consistencies of data obtained of relevant particles requires a fit from data to model by accounting for all correlations while leaving an unknown parameter ‘free’ to float about the fit; this parameter represents that degree of uncertainty in data to model fit. To give due diligence to experimental uncertainties, measurements of basic characteristics of the particles, such as the mass of the Z boson and top quark, are simultaneously included in the dataset and left unbound in the simulated fit-to-data. The precision data modifies the data either at the level of Born approximation, a mathematical physics method used for perturbative calculation in weak scattering, or loop corrections in electroweak and strong interactions. Moreover, a targeted interference with the Standard Model can increase sensitivity while deviation from the asymmetry measure of the Standard Model could signify non-Standard Model new physics.

As previously mentioned, the hadronic partons behaving as point-like objects in the Standard Model, can be addressed by scaling, a process that ensures the scattering cross-section does not depend on kinematic or trajectory values. This is particularly useful for describing the hard scattering process (that happens for strong particle interactions); in fact, the gluon was discovered through the measurement of the jet rates (produced through the fragmentation of quarks and gluons via collisions at a point of convergence of that interaction, also known as vertex) during the interaction. However, the perturbative theory does not work for obtaining precise calculations in all hard-scattering cases. Instead, a partonic distribution function is employed to compute fractions of energy (of the strong interaction quarks and gluons) too complex for perturbative calculations, with the algorithmic aid of leading orders (and higher) variable calculus. Of course, asymptotic freedom is given, and always assumed, for any strong interactions.[6]

The parton energy distribution in the hadrons is distinguishable by the flavor of the quantum numbers. The parameterization of the parton distribution from global fit (the general big-picture fit) accounts for the effects of experimental measurement limitations, including uncertainties within the parameters of the fit. Moreover, the large-distance effects of the strong interactions are decoupled from the original hard reactions and do not distort the measurable picture too much, therefore predicting the presence of initial partons as collimated streams of hadrons. Early measurement of the event shapes allows the differences in the modeling of multijet production to be studied, which provide valuable input to the Monte Carlo generator tuning.[7]

The next section will engage in a close reading of the Monte Carlo contribution to the search for the Higgs boson, including an analytical consideration of the generators important to simulating the required events for Higgs boson analyses. We will also consider how the quantum computer, in its ability to simulate and predict datasets that are significantly more complex and multivariable in their dimensionality, can impact the future capability of the present-time Monte Carlo method. The quantum computer, and the promise of new knowledge it embodies as an outcome of simulation practices, is a superjective (more than a pure subject because of its latent promise of unimagined possibilities) potential occupying simultaneously, physical/material and conceptual/intellectual states of becoming so as to produce novelty and new scientific idealisms.

---

[6] In the case of asymptotic freedom, it helps us understand how energy suppression happens for quarks in terms of the quark and gluon coupling strength that can be calculated. The next to leading order, and beyond, help fine-tunes the calculation for collinear and low-energy (infrared-safe) event shapes that are the result of electron and positron collisions; all of which are then combined with other hard scattering calculations to improve the sensitivity of the latter. These are processes that are built into the Monte Carlo algorithms for dealing with the hard scattering processes involving strong interactions.

[7] On this site is a gif file of a graph illustrating the datasets for signals and background in the Z bosons decay channel for the Higgs. The datasets are data obtained through the two-fold process in experiment and the Monte Carlo https://twiki.cern.ch/twiki/pub/AtlasPublic/HiggsPublicResults/4l-FixedScale-NoMuProf2.gif

It is also crucial to bear in mind the role of prediction within the phenomenological and experiential that enables the point of convergence between theory and experiment. We can begin to understand the role of Monte Carlo as a speculative superject because the latter can extend beyond the observed subject and add a creative dimension to data's emergence in and out of a virtually constructed world. In understanding Monte Carlo as a speculative superject, we are more able to comprehend its order within the epoch of speculative spatio-temporality. This is important if we consider all the independent and dependent objects located within the global fit of the data generated. The mechanism of feedback and circulation between determinate and indeterminate points in the stochastic processes, derived of objects in the global fit, requires thinking about knowledge production and transmission as distributed and circulatory, with points of recollection and feedback as found in cybernetics.

## 7.2 Monte Carlo as Speculative Heuristics of the Higgs Boson

I want to state explicitly, that the Monte Carlo simulation of experimental data is separate from the algorithms (and software) used for event reconstruction (and particle identification) of experimental raw data even if there are points of intersections; without the event reconstructions, the Monte Carlo simulators would not be able to access a coherent and analyzable version of the datasets. Regardless of which sub-process a particular Monte Carlo generator focuses on, the subprocesses are about simulating the final states of a high-energy (or low-energy) collision, right down to the properties of individual stable particles and their individual momentum.

However, what is interesting about the simulations is the basis of their statistical analytics: reconstructed data. In the process of reconstruction, the collected raw data are digitally sifted and selected so that our most immediate connection to the data is always mediated. The actual materials for reconstruction are from the raw data that have been especially triggered to produce events of particular interest. Then, reconstruction is done with the aid of algorithms that provide probabilities for the detection of the particles produced during collision, and after decay. At the end of the process, the generators will then evaluate as to whether the reconstructed data can produce sufficient information for commensuration with initial theoretical predictions.

There are multitudes of generators, which are basically code modules that produce new transformations with the intention of attaining novel techniques for the analyses of the Higgs boson to improve prediction capacity. The generators of the Monte Carlo are not similar to the generators of the microphysical features one might find in quantum field theories, but can still extend the capabilities of the latter by providing new factorizations and normalizations for easier calculations within a statistical framework. The relationship between the different generators lead to the coding of the algorithms used to further substantiate or disprove theoretical conjectures.

There are a number of similarities in the Monte Carlo generators employed by both the ATLAS and CMS collaborations in discerning the probability for observing the Higgs boson. In fact, the statistical methodology for processing the data leading to the prediction of the Higgs boson was developed by both collaborations through the collaborations' involvement with the LHC Higgs Combination Group as part of the requirement for simultaneous analysis of data, from separate search channels, to control against all statistical and systemic uncertainties. The methodology has been developed as a result of 40 sub-channels, each of which could, in combination, contain a total number of distributed events that converge onto a mass for the Higgs boson within the range of 110 to 600 GeV, quite close to its latest energy-mass of 126 GeV.

Most of the uncertainties come from our knowledge of how detectors react to particles that transverse the former: jet energy, for instance, contains the largest uncertainty. Among the uncertainties that arise, confirming my earlier arguments, are "theoretical uncertainties on the expected cross sections and acceptances for signal and background processes, experimental uncertainties arising from the modeling of the detector response (event reconstruction and selection efficiencies; energy scale and resolution), and statistical uncertainties associated with either ancillary measurements of backgrounds in control regions or selection efficiencies obtained using simulated events" (Chatrchyan et al 28). The combination methodology involves the introduction of a signal modifier for increasing the signal strength of the expected Standard Model Higgs boson, a method for evaluating the Higgs boson's background yields, and a likelihood estimation for evaluating the probabilities of selected events through the statistical frequentist method of counting.

For testing the hypothesis on the production of the Higgs boson, a test statistic (such as the chi square) is constructed out of information encompassing observed data, expected signal, expected background, and all uncertainties associated with these expectations. Therefore, one is able to rank all experimental observations based on whether they are more consistent with background only, or background + signal, hypotheses. Finally, to infer the non-presence, or seeming absence, of the Higgs boson signal in the data, the observed value of the test statistic is compared to the distribution values in background only, or background + signal, hypotheses.

The pseudo-datasets generated by the Monte Carlo are taken from the vast library event samples already available for evaluating the likelihood of the background only or background + signal hypotheses (29). The methodology comes with techniques for quantifying absences and excesses in the signals obtained but I will not enter into the technicality here; suffice to state for now that they will be important in the generation of pseudo-datasets and for dealing with the expectations of observation when the time comes for reconstructing evidence for the Higgs boson.

What the above explanation illustrates is the importance of signal data as part of the search data, which is then compared to the simulated events sharing common parameters for producing enriched individual background contributions, so as to develop techniques for excavating the signal created from background processes that share a similar signature with the detector. As was discussed in chapter five, the search for the Higgs boson signal means performing a maximum likelihood fit to one selected final state of Higgs mass across different channels with multiple events constitutive of the state. To do so, one enlists the computational capability of Monte

Carlo generators with multipurpose and specialized functions.

PYTHIA is one of the major general-purpose event generators used by both ATLAS and the CMS. The generator contains the Lund string fragmentation model (a model that propagates individual quarks produced from the collision that leads to the formation of the aforementioned 'jets') used by other programs that link to PYTHIA for the hadronization phase of the simulation. The 'string-like' effect at the hadron is basically caused by the long-distance self-interactions of gluons ('glue-like' particles) that make field lines attractive to each other, with the strong interaction field lines compressed into string-like tubes. PYTHIA, which is linked to other sublevel generators, is important to the modeling of final states at hadron colliders, such as the parton shower emissions and jets produced out of multiple partonic interactions. Originally programmed in Fortran, PYTHIA's later versions had been ported over into C++ to take advantage of the object-oriented modularity, particularly for the convenience of being able to call subfunction libraries internal and external to the code.

However, the problem with PYTHIA engaging with too many sub-processes is the lack of concentrated resources for dealing with problems found in higher order precision. Moreover, because of other complicated dynamics involved with color correlations (a case particular to strong interactions because of the 'flavors' of the gluons), the implementation of PYTHIA is more complicated and lugubrious than that of the HERWIG generator, another popular general-purpose event generator, because the latter does not have to deal with complicated and non-trivial color correlations. Nevertheless, both require other more specialist generators, such as TAUOLO and PHOTOS, for describing tau lepton decays and photon propagators respectively, as well as other specialist generators that focus on parton distribution function calculations.[8]

To demonstrate how the different specialized generators fit with the general-purpose generators in deploying different precision measurement methods for the simultaneous analyses of signals and backgrounds across different channels, I consider examples of sample events and simulators used in the CMS and ATLAS's searches for the Higgs boson. For both the collaborations, once events have been generated with either PYTHIA or HERWIG, the particles produced are then propagated through a detector simulator based on the Monte Carlo program GEANT4 that is used by all the detectors at the LHC. The simulation includes a realistic modeling of the (beam) pile-up condition observable from collected data. Corrections obtained from measurements in data are applied to account for the small differences between actual data and simulated ones. For example, large samples of W, Z, and J/Ψ (J-Psi) particle decays are used for deriving factors for lepton reconstruction probabilities and enabling greater efficiencies in particle identification. The description of the Higgs boson signal is obtained through Monte Carlo simulation such as special matrix element generator like POWHEG (taking advantage of the method involving the leading orders) that is interfaced to PYTHIA. One of the more important work of code that is built into the generators and data-fitting processes is the capacity for making material cuts that would have implications on what data get processed through the call functions of the generator, thereby allowing certain conditionals to be fulfilled and measurements to be made to meet the expectations of the experiment. I take as an example given in figure 1 below, a snippet of code written in C++ that is focused on a particular final state containing two leptons (dileptons). While a large segment of the 'introductory' section of code involves functional, variable, and library declarations coming from non-publicly accessible databases and other sub-processing modules of the code, there are also modules within the code that specialize in making 'cuts' and producing selections from the datasets of the samplings.

---

[8] For examples of the sets used for the parton distribution functions, see page 2 of the ATLAS Collaboration article "Observation of a new particle in the search for the Standard Model Higgs boson with ATLAS detector at the LHC." http://dx.doi.org/10.1016/j.physletb.2012.08.020.

The cuts are made for excavating the background and signal rejections, determining mass delimitations of the selected particles, producing a match between the target particles, adjustments and allocations (through weighting) of data samples, obtaining of acceptance levels for particular events, tagging of jet events (for quicker retrieval in the future), and making determinations for particle trajectory.

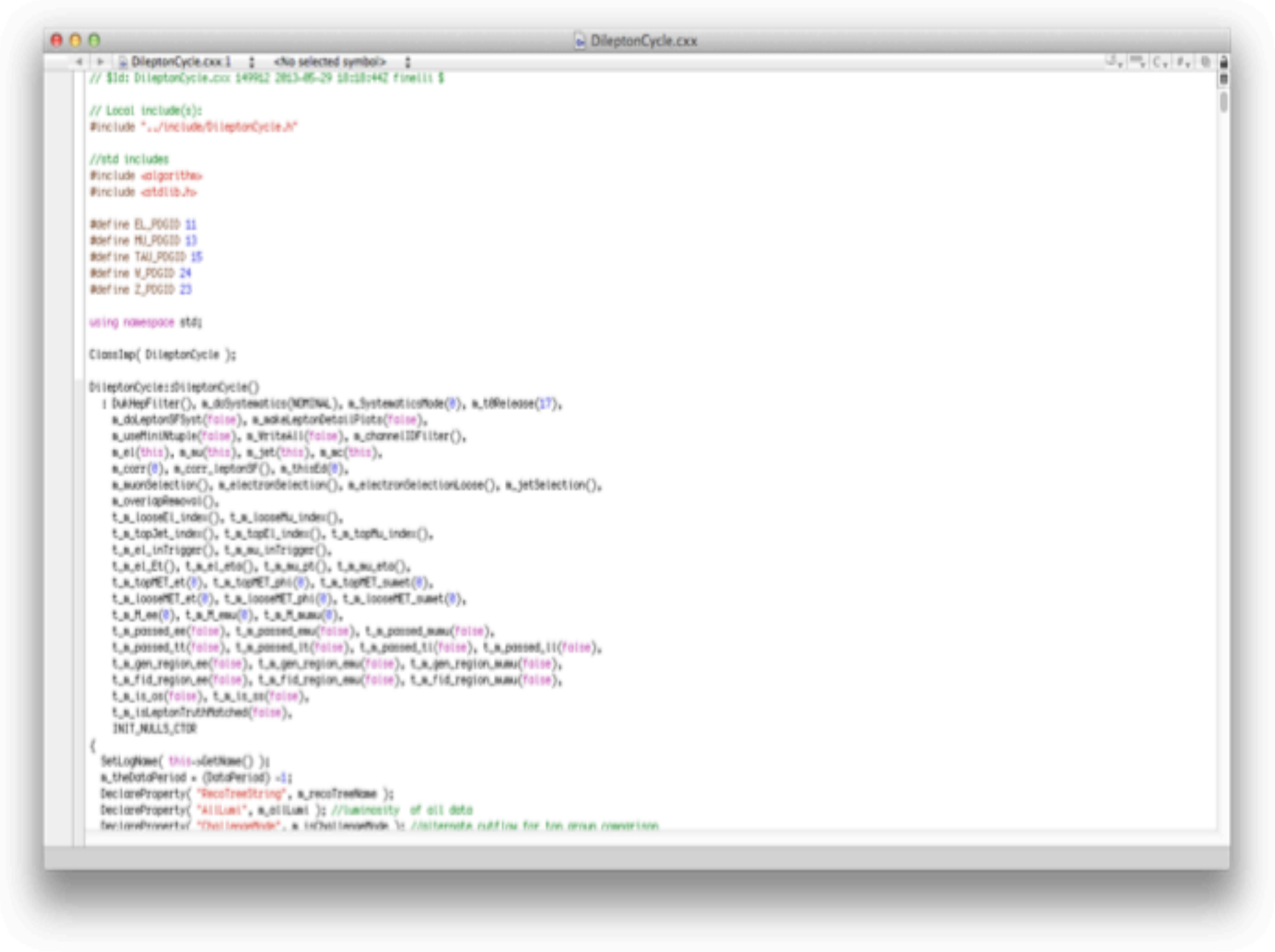

**Figure 2: A snippet of a specialized code for analysis of Monte Carlo events in the dilepton final state by recently graduated Duke PhD in particle physics (2013), Kevin Finelli.**

Furthermore, there are multiple sub-functions involved with how the cuts should be projected; each represents a problem of decision-making in that choice between flexible exploration and probabilistically saturated selection. They could be constituted as Whitehead's "Category of Objective Diversity" that asserts the essential self-identity of an individuated entity (225). In this case, the entities are the multiplicity of cuts denominating the concrescence of each consequence that comes of making choices, and the superject of a satisfactory outcome is obtained through the inversion of that identity. When the entities that were theoretically conceived become actualized, one has to be wary of potential duplications in order to ensure that no double counts are made of the background material and foregrounded signals.

The data cuts, as enacted by the algorithmic parameters of the aforementioned code, is definitive yet speculative; definitive in the arbitrariness that is involved when Monte Carlo analysis suggests a 'cut' on the experimentally obtained data for further processing. At the same time, the experimental triggers are themselves another set of 'cuts' enacted when they are made with the aim of producing specific sets of interactions. The speculative, on the other hand, is the indeterminateness of what gets included, excluded, and occluded during the process of exchange and realization of any sets of knowledge (scientific in this case). While it is highly possible that the realization could merely reinforce the theory that shapes the stochastic process without rupturing the theory's sense of security, it is still possible for ambiguity to seep in as the cuts are made out of best available knowledge. Therefore, further testing and replication of the events can reveal a potential weak link.

At the end of the day, what is actualized becomes public, and the private language of the code known to the technically initiated can produce interpretations that express the experiential core of nature, even if that experience operates from different levels of existential onticity. The discursive conjugation between code and the popular image of the double-slit experiment (that has been discussed in the past few chapters) gives rise to a new form of computing that is the technological equivalence of quantum entanglement. In other words, the quantum idea of the double-slit experiment, by way of the Aspect experiment[9], becomes, at this point, embodied in the form of a technologically sophisticated quantum computer, or quantum supercomputer, that draws on the theories of quantum entanglement as its operational directive.

As we sieve through an exponential amount of data (a huge amount of data accumulated never got addressed due to labor shortage and limits in computing resources) and become intensely aware of making cuts that could preclude one from making a discovery of a lifetime, the affordances of a quantum supercomputer, once sufficiently developed, could change how physicists make decisions about experimental triggers (experiments that are conducted due to an earlier decision made) from currently narrow parameters. However, to optimize a quantum computer for future particle searches requires knowledge over how to manipulate

---

[9] This is the experiment that tries to prove Bell's Inequality Theorem and quantum entanglement. The experiment was conducted to resolve the conundrum the came out of the Einstein-Bohr debate, and the EPR paper of the 1930s. The experiment was conducted using two channel polarizers, with spinning photons or protons. For more details, see Aspect, Alain et al. "Experimental Realization of Einstein-Podolsky-Rosen-Bohm *Gedankexperiment*: a New Violation of Bell's Iequalities." *Physical Review Letters* 49.2. 12 July 1982: 91-95.

the computer's built-in strengths and weaknesses.

In *Quantum Computer Science: An Introduction*, N. David Mermin states that a quantum computer is one "whose operation exploits very special transformations of its internal state…The laws of quantum mechanics allow these peculiar transformations to take place under carefully controlled conditions" (1). He also emphasizes the greater sensitivity to potential disruption in operations of a quantum computer should there be any unexpected physical interaction that has not been pre-programmed into the system, resulting in a decoherence that fatally disrupts the transmission of informational bits. In quantum computing, the macrophysical scale becomes less relevant and a microphysical atomic scale system is decoupled from a macrophysical environment.

Just like the famous story of the Schrödinger cat involving the determination of quantum states, the Alice and Bob sender-and-receiver narrative about quantum entanglement and cryptography is a staple in the history surrounding the development of quantum computing.[10] However, I will not enter into a discussion of the role of cryptography, but would instead, concentrate on an aspect of quantum computing with greater relevance to the discussion of Monte Carlo: the measurement gates, which is a hardware representation of the Born rule (a law in quantum mechanics that links the squared magnitudes of amplitudes to hard values that one can read off during the process of measurement); quantum searches (also known as the Grover iteration method) that uses the principle of (unitary) state transformations at the quantum level to obtain an almost definite recovery of a targeted value; and quantum error corrections that can be done without knowing the cause of the corruption. Imagine being able to conduct multiple levels of searches without worrying about a buffer overload or storage capacity issues.

[10] The Alice and Bob story involves the sending of information encoded in qbits or quantum bits. Alice will randomly encode her message in a specific photon into either one of the polarized states available to her then transmit the string of qbit encoded messages to Bob. Bob will then decide whether he wants to, first, transform the message with the help of a special operator or just pass it directly through a measurement gate. Alice tells Bob over an unsecured channel the type of the qbits sent but not the state in which it is located while Bob tells her which of his measurements agree with Alice's choice of information type that she has sent over. The reason for this is to avoid interception by an eavesdropper, whereby the actions of the eavesdropper would then be detectable by Alice and Bob by the alterations made to the system. However, this is only possible because of the no-cloning theorem of quantum

Quantum computers have to deal with a measurement problematic that alters the state being measured due to unpredictable disruptions stemming from the process of interaction and entangling between states, as well as the non-discreteness of an error that can increase exponentially over time! Due to the stochastic quality of quantum computers, it shares parallel proneness to error as that of Monte Carlo data-fitting, given that much of the microphysical features of the events Monte Carlo work with are largely of a non-macrophysical, and therefore non-classical, nature.

---

computing that renders it different from classical computing, in that there is no memory retained of the selections or changes that have been made. See Mermin for more technical details.

Hence, what we learn from dealing with the state of coherence and decoherence in quantum computing can be applied to dealing with error corrections in the Monte Carlo. Another aspect to consider is the amount of data that can be simulated because of how quantum memories work (the no-cloning theorem where no memories are retained of the selections that have been made) and the transmission of information that does not require a classically induced communication channel that has to deal with the issue of pile-ups and energy inefficiencies.

The still unfinished narrative of what quantum computing can do to contribute to the simulation of multipath events is of interest here, as it speaks to the same problems faced by physicists who have to deal with way too much data so that too many cuts have to be made, sometimes to the detriment of getting sufficient data to produce truly global representations of nature. How would the developments of quantum computing be productive to changing the current practices in statistical modeling would also come from changes in how one develops the modeling techniques, as well as a reconceptualization of data, within more dynamic spatial-temporality in relation to existing categories.

## 7.3 Conclusion

Finally, the constitution of the Monte Carlo generators and simulation practices within the backdrop of media technicity requires a more thorough development of machinic phenomenology, one mediated and buffered by the detectors of the LHC that bear witness to microphysical events by tracking the signals during quantum level interactions with macro-detectors operating under the rule of classical physics. The presencing of media, as epitomized by the relationality between the detectors, data, and Monte Carlo simulators/generators, is one of absence and presence, of immediacy and delayed manifestations.

The LHC requires extra-sensory extension, but it is not quite certain that the Monte Carlo simulators could provide that extension unless the former can be reconstituted as a speculative machine that finds wriggle room beyond the formality of its responsibility to the Standard Model experiments. Therefore, the media question here is the question of the cybernetic, a molar organization whence we can locate the "power center" that Deleuze and Guattari refers to as "defined not by an absolute exercise of powers within its domain but by the relative adaptations and conversions it effects between the line and the flow." In fact, their respective concepts of the molar and the molecular most aptly define the relationship between the detector events and Monte Carlo events because it is the system of reference to the events that are important rather than size, scale and dimension (239). The molecular in this case is one that is expansive and flowing, not demarcated by rigid segmentations and lines (hence the Monte Carlo events) while the molar (the detector) is organized to resonate within the delimitations of a nature it can access.

This brings me back to the speculative writing performed through the Monte Carlo, and the modes for reading the events in terms of the analytic output and more expansive interpretations performed. Should we be successful in producing a speculative media space not only for constituting the epistemic bodies of scientific objects, but also for consideration of the visuals, sounds, and the tactile, what new aesthetical analytics can be invoked in this bridge between the media arts and sciences?

For as Joanna Drucker argues with regard to speculative computing from the intellectual tensions of digital humanities, where speculative computing is defined as pushing, "subjective and probabilistic concepts of knowledge as experience (partial, situated, and subjective) against objective and mechanistic claims for knowledge as information (total, managed, and externalized)"(5), the speculative computing that grows out of a Monte-Carlo and LHC detector hybrid creates a connection from the mundane delimitations represented by the interaction between the detectors and nature, into a virtual world that is artificially induced and contains a propensity for cinematic display. In addition, through speculative computing, the ontological can be better approached when one shapes a methodology around the unexpected instead of merely the fulfillment of existing criteria.

The connection between high-energy particle physics (including quantum theory) to the cinematic happens at multiple levels but I will only discuss the two most relevant to the chapter. At the first level is the rhetoric of reading the science of the genome (in terms of the cross-fertilization and hybridization of genes) against quantum entanglement (that is a part of development in quantum informatics), which are then used to produce a biophysical thinking about information transmission and embodiment, such as what the aforementioned essay by von Foerster attempts to do. Even if physicists are using physics theories to study biological subjects, such as in a recently published paper on the migration studies of population that applies the stochastic method of equilibrium to model the study in ways similar to the Monte Carlo statistical analyses[11], the actual merging of the quantum with the biological has not gone beyond the level of the fictive and the speculative. It is in imagining the infinite ways by which we can represent the techniques of biophysical hybridization, departing from the limits of science, that the combined techniques of the science fictional and cinematic storytelling become increasingly cogent. The end product could bring about the explicit actualization of the virtual-real relation even as critique is simultaneously performed. But the audio-visual-tactile representation of the end product remains to be seen.

At another level, there is also the actual filmic realization of the significance of the LHC, the activities at CERN, and their relationship to society and science. While there have been multitudes of animated documentaries produced by CERN (and related national laboratories) about their experimental set-ups and actual experimental processes as part of public outreach, the year 2014 sees the launch of LHC inspired-documentaries addressing larger issues of science and society that are not specifically science communication infomercials; more significantly, the two documentaries are released more than a year after the announcement of the Higgs boson even if the production of the documentaries coincided with the summer of 2012, which was when the announcement was made.

---

[11] See Lombardo, Pierangelo et al.

*Particle Fever* is specifically about the science of the particle physics experiments and the people involved in it. The exhilaration and euphoria of the scientific production (and the imagination that derives from speculating at the edge of the unknown) is foregrounded while the actual scientific grind is mostly submerged. The same goes for another documentary, *The Circle*, which is less about the science that comes out of the Large Hadron Collider, but about the reaction (and speculation) of the surrounding community in the south of France to what is going on at CERN, as well as how an artist appears to be trying to interpret what goes on there. However, both documentaries emphasize the different acts of speculation that are involved, using filmic techniques (and animation) to demonstrate the multi-layers of speculation at an epistemological and even ontologically existential level.

How do the affordances of digital media bring us a step closer to the ontological, to that point of ontic genesis, and towards the attainment of unitarity?

The idea of the simulacra, such as that embodied by the experimental work and simulation surrounding the search for the Higgs boson, appropriates Baudrillard's hyper-reality to probe at the presentation of ontology during the process of simulation; what paradigm of the observable is deployed in the epistemics of simulation practices that also informs the epistemic consumption (and thus reading) of the output of the practices? If there is one Baudrillardian claim of the simulacrum that epitomizes the spirit of this chapter, it is that closing of the gap between the imaginary and the real through the absorption of their distance between each other, so that there is room for a critical projection of the reconstitution of meaning behind an experiment, the experiential, and the virtually-real.

# 8. Epilog: Evaluating Speculative Physics

I started out by appropriating speculative physics as enunciated by Schelling, but through reading him not as being antithetical to empiricism in his metaphysics; rather, I read him as encouraging reflection into how one deals with the relationality between product and productivity, and the prehension (and extension) of nature, a philosophical belief also shared by Whitehead. What I had done in this dissertation is less about attempting to force Schelling's idealist philosophy onto any scientific philosophy, but to update the his philosophy for developing a more encompassing theoretical methodology for reading modern physics, and its philosophical conceptualization.

However, it is by taking Schelling's original conceptualization of speculative physics, and then rewriting it against new identities conceived about nature that one comes by the revelation of quantum laws, of the speculative philosophy of Whitehead that was born when the new fields of physics was coalescing, and as algebraic geometry underwent a transformation from the time of its inception in the eighteenth century. There is an idealist spirit shared by both Schelling and Whitehead with regard to nature and the mind in the seeking of the proto-causes of nature that had been much neglected in philosophies overly wedded to empirical concerns and logical truism. If there is a take home lesson from speculative physics and process philosophy, it is not to assume that one can neatly carve out a 'bare' physical organism ('bare' in the sense of the ontic) into purely quantifiable or discretely measurable packages. As the previous chapters had indicated, the quantifiable are to be managed next to the unquantifiable, to potentially positive outcome, and that the process points to behavior of physical elements that are not evident at the initial and endpoint of a physical process.

Therefore, we cannot assume that a seemingly precise scientific argument within a linearly logico-empirico format will progress onto an intuitive or certain end. Foremost, we cannot claim certainty over the epistemological framework through which logic operates and must be ready to accept that the calibration and adjustment done by the rules of that epistemology are always flawed. Moreover, given our problematic access to ontology, the supposed physical laws governing the structure and purpose could be the cause for destabilization when a paradox erupts.

As Ackermann states in *Data, Instruments, and Theory*, "…Logic cannot coerce directionality in which projection from a basis to anticipated future data should take place" (5). As he rightly argues, the underdetermined undecideability and constraints of logic mean that the empirical uncertainty in data may never be adequately resolved. Further, despite wariness in ascribing scientific practice as speculative, there is plenty of evidence to prove that acts of speculation exist at different points of the decision-making processes in science, as the chapters and also Appendix A have sought to demonstrate. Even with the most precise of measurement, the measurement can only be as accurate as the descriptive limits of its epistemic effectiveness, or the type of instrument deployed, as the relationship between the knower and knowledge is dependent on the apparatus used for delineating the properties of that relationship. The oft-cited example is the Schrödinger cat, whereby one assumes that the state of the measured is changed after interaction with the observer if only because it is difficult to disentangle the state of the observer from that of the observed due to their simultaneously, and commonly, 'smeared' conditions. But as Ackerman notes, and which I agree, it is possible that the object studied is unaffected by the observer/knower; the observer/knower could be the one to have changed because of exposure to transformative (game-changing) events that alters his/her/its perception, such as was illustrated in *Matter's End*.

Hence, to counter that lack of autonomy, which speculative physics gestures towards, one should view scientific praxis as a methodological emergence that can only be consolidated over time as data accumulates and as our understanding of what to do with the data improves. However, we must still be cognizant that it takes a long time to get sufficient data to demonstrate *ceteris paribus* truths. Even so, that is a hard accomplishment because of the multiple levels of considerations involved during decision-making in order to know which model to invest in (in the form of one's labor and funding) and the realization that any exclusion of data could result in the preclusion of a potential discovery. If the narrative of the Standard Model, especially the Higgs boson, is any indication, it is obvious that the theory of justification and discovery are speculative, even if differently so. In the former case, one has to obtain empirical justification on why a particular model is the best ontological explanation there is, with explanations defined as syntactical or logical relationships to putative theories (in Ackermann's words) of the observable behaviors, and properties of local symmetries that undercut the interactions of the Standard Model. For instance, the supersymmetry, or a version of it, has to be sufficiently justified so that there is an unequivocal determination of its role in balancing between the theoretical prediction and experimental manifestation of the Standard Model Higgs boson.

Nevertheless, as was stated, it is possible for justification to break down when it turns out that what is expected becomes unexpected, and something else that has no known version in a model is discovered in its stead. In other words, poles of singularities and ruptures can prevent one from following through with even the most elegant mathematical logic. This is where one gets into the mode of discovery, a mode even more speculative than the justification of theory, because there is no formula for determining how discoveries are made, and the parameters for that determination can be as broad or as narrow as the steps required to manifest the object of discovery. But it is in this potentiality of discovery that speculative physics is most interested in working out, because this is where new knowledge can happen.

The consideration of the scientific model as a fable or fiction is not novel, as other philosophers of science, such as Nancy Cartwright, Joseph Rouse, and even Hans Vaihinger (his notion of fictive judgments), had considered how fiction can be used for understanding epistemological and ontological accounts of physical theories, such as the relationship between abstract to concrete sets of descriptions, and for satisfying a concrete description by way of satisfying an abstract description (Cartwright 39). We consider first the differentiation of fictional construct at the level of a hypothesis and theory, before moving on to consider fiction and semi-fiction. Fiction, in the sense meant here, is the use of narrative to model particular groupings or clusters of ideas that could be scientific, or not, while semi-fiction is a model that enables expediency during inference making. For Vaihinger, there is an added valence of difference between semi-fiction and hypothesis, with the former limited to the realm of the metaphysical and the latter having a semblance of scientific value. A good example of a semi-fiction would be the *gedankenexperiment* used to provide interpretive heuristics to aspects of modern physics that could not be empirically demonstrated while a hypothesis would be an initial idea experimentally set up for testing and repetition, at multiple levels.[1]

From chapters two to seven, I have been building an ontology of fictions that unshackle the restrictions of all kinds of praxis-coupled fictions; fictions such as the laboratory fictions that fall within the rubric of how one can characterize particular distinctions of an object that falls within the embrace of a scientific domain, thought-experiments that serve to illuminate physical phenomena within familiar constraints while gesturing to the need for resolving paradoxes, and experimental systems that are conceptual arguments of pre-determinate judgments. Instead of following the common Hempelian model of articulating deductive inferences that draws on theor(ies) and the praxis of known facts, or even inductive logic, I have tried to show how all features of the theoretical, experimental, and fictive can be coherent yet not necessarily similarly interpretable, or even be mutually consistent at all times. In the sense that contradictions can happen in a system, the same contradictions are merely representations of observables interspersed, and intercalated, among hidden logics of the unobservable variables of the system. The appearance of illogic is because the observer only has a partial view of the ontology.

---

[1] Interested readers can check out Hans Vaihinger's *The Philosophy of 'As If': A System of the Theoretical, Practical and Religious Fictions of Mankind*, Trans. by C. K. Ogden. The first edition was published in England by Routledge and Kegan Paul, Ltd. in 1924 but there have been multiple editions made by other publishers since, the latest being in 2008.

The above connects to the seeming illogicality of juxtaposing precision measurement and determination with that of the speculative. As the Monte Carlo critical narrative informs us, tight control over inputs and the global scale of the data can be summarized into more precise outcomes. But this is riding on the assumption that the parameters are unproblematic; in reality, we are uncertain as to whether predictions always end in fulfillment. But, there is also another logic represented by the null hypothesis that argues for a non-correlation between the tight controls of input with the precision of output in the sense that one cannot control for determinate outcome from the point of input. A good example of such a case is the oft-discussed double-slit experiment and its variants. Nevertheless, this does not mean that precision could never be obtained; it just means that some adjustment would have to be done in order to get at the desired measurable. To use the double-slit experiment example again, this means adjusting the setting of the experiment so that we might be able to determine which slit the beam of particle will be entering into at a specific slice of time, or whether the information obtained from the which-path experiment can be erased after the measurement has been performed, so that initially 'suppressed' interference pattern can return (Barad 311-12).

The null hypothesis could also arise from the underdetermination between a measured effect and interpreted causality. In other words, in the case of causal effects involving experimental data, one cannot be unequivocally certain of direct cause-and-effect correlations between a particular set of data and a physical phenomenon, and therefore, the cause of that phenomenon. For instance, the inclusion of hitherto new parameters unrelated to the original data could make no difference to what is already observed; or, the new parameters could sufficiently complicate the original observation so that one can no longer differentiate the cause from the effect, or even determine the relation between cause and effect. One can approximate the truth but not draw direct conclusions from the former.

All the acts of theorization, experimentation, and analysis have to do with acts of reading, writing, and interpretation that parallel, but do not directly correlate to, measurement, observation, and modeling. In thinking about how readers interact with a variety of media, whether physicists reading of the values provided by the console screens connected to the detectors, reports, preprints, and published results; or a literary scholar reading through the core of poetic corpora, the process of reading (a cognitive inscription) can be constituted as surface level (epistemological) reading in relation to a deep-level reading that tries to get right to the ontological layer.[2]

Epistemological reading can then be constituted as a form of comparative reading, where one compares between the multitudes of epistemic content derived from a range of subfields of knowledge. Epistemological reading can also be used to compare between variable readings of a physically similar phenomenon. I have brought up such an instance in chapter three when I discussed the parallels between the superconductor and gauge field theoretical developments. Another example is the relationship between neural networks and quantum states demonstrated multiple in the chapters, beginning with Egan's *Quarantine*. Were we to read these knowledge fields as 'texts,' we can then apply Wolfgang Iser's phenomenological account of reading, "…one text is potentially capable of several realizations…no reading can ever exhaust the full potential…the text refers back directly to our own preconceptions…the potential text is infinitely richer than any of its individual realizations" (280). While there is a whole load of difference between a 'literary' text and raw data mined and processed through an array of scientific instruments, both are similar in that the degree of their potency, and potential, are the result of the mode and choice of interpretations performed in the readings of these disciplinarily inflected 'texts.' To sum it up, the connections formed between the subfields require a particular imaginative synthesis that converges metaphysical realization with the physically observable.

---

[2] I have a deeper discussion of this in a conference paper published in a Proceeding. See Lee, Clarissa Ai Ling Lee "Speculative Reading, Speculative Physics: the Ontology of the Large Hadron Collider" in EPJ Web of Conference, vol 28. <http://dx.doi.org/10.1051/epjconf/20122812033>

Ontological reading, on the other hand, reads into the lines of tacit knowledge to confront the subsumed ideologies embedded in the 'cell structure,' the code, of the foundational structure of knowledge. While epistemological reading means working with hyper-mediated or distantiated datum that screens us from the 'raw-ness' of an event, ontological reading requires dealing with unmediated affectivity and the subject as it comes into being. Not much of what goes on with regard to the events of the Standard Model has involved ontological reading because of the 'distance' between the scientists and the triggers.

However, the Monte Carlo code can narrow the gap between the ontology of the scientific object at hand and the physicists' epistemic relation to the object. Beyond quantum theory, epistemological and ontological readings could also be applied to the critical readings of other scientific, humanistic, and artistic subjects. Moreover, I have been suggesting, in all my chapters, that the performance of speculative physics has the purpose of bringing epistemically disparate subject-matter together for creating new and hybridized perspectives. Speculative physics enables intervention into supposedly epistemically inscrutable and politically unreachable subject matter.

In a sense, speculative physics updates the extensive reading on process, the abstract method, and concrescence promulgated by Whitehead in light of the new ontological problematics raised in quantum field theories and even potential new physics that could extend, and further reveal, the holes and self-contradictions in current theoretical-physical interpretations. Nevertheless, a more extensive development of this critical study, of which this dissertation is only a start, would have to come later. At the end of the day, it is my aspiration to attempt the development of a methodology that could be applied for working with linguistically and epistemically incommensurable disciplinary fields by drawing foremost on the highest order of their conceptual abstractions, which is the 'summit' that these disparate disciplines are most likely to meet. However, this 'summit' is largely unattainable by most interdisciplinary scholars who might only have a tenuous grasp of the disciplines external to their home disciplines. In light of that, greater collaborative conversations are in order for improving the development of interdisciplinary discourse that goes beyond gesturing from the sidelines.

## *8.1 Extending Speculative Physics Beyond Physics and Science Fiction*

As the concluding section of chapter four suggests, speculative physics could be an emergent approach for reconstituting the essentials and substances of speculativeness, of a concrescence where the relationship between the environment of the media and the content projected is more deeply embedded within the affective nexus of the organism, such as is the case with Nick Stavrianos in *Quarantine*, where all forms of sensory consumption is configured and determined by increasingly well-calibrated, and precise mods that are of the nth orders of cybernetic biological-to-digital-and-back controls. But is it merely a façade of control, or is there an emergent form of life that takes control of the mind's relationship to the world? Whose *eidos* is being constituted: the mods or the organic brain? But then again, one can also argue that the distinction between the intellectual and sensual is blurred, when the strategy for feeling appears to be regulated by the mind.

D.N. Rodowick, in the *Virtual Life of the Film*, has re-imagined the medium of new media technologies in juxtaposition to the old celluloid of the cinema to reconfigure the medium of film, by way of a contestation between the analogous and the digital medium. The Monte Carlo machinery has also, at one point, developed from an analog computer for simulating data collected for war strategies. The same statistical methods are then deployed, in the aftermath of World War Two, to the study of the big science of space and nuclear physics, but with much mediation and modification to suit the contingent requirements of the object of study. The problem with the analog computer is that its operative mode is only at the level of the macrophysical, something which Von Neumann also alludes to, less directly, in *The Brain and the Computer*. Therefore, many of the discrete and macroscopically paradoxical processes at the quantum level become too complicated to be computed onto a macrophysically oriented computer. However, this does not mean that the analog computer cannot be programmed to crunch data from quantum interactions, such as is illustrated in *The Minotaur in the Mushroom Maze* novella in chapter six. It is just that the outcomes are semi-classical approximations constrained by the mechanical limitations of the computer that means the exclusion of higher orders of complex calculations, and therefore, of more complex representations and inscriptions.

But how can one begin to read epistemic and cultural objects speculatively and politically; to understand that there can be abstract and concrete, as well as absolute and relativistic, entanglements that take place within the architecture of information mingled with the ideological, the factish, and the unanalyzable? The unanalyzable represents knowledge that could not fit any analytic logic, or contains too many higher order terms that require greater sensitivity of instrument (combined with the right algorithm) before access becomes feasible. The unanalyzable, as Babette E. Babich explains, is couched in the expectancy of clarity that is attained by definition or a fiat, aiming to dissolve the state of perplexity, but that could also reduce a problem not readily reducible without causing repercussions; such repercussions can include changing what might have been (33). In light of that, we might also consider the theory-ladenness of the gaze upon the object of analysis that is dependent upon the situatedness of the subject; the subject acculturates the interpretive process emerging from the gaze. With that, let us consider what is that extra which speculative physics brings to critical science studies.

## *8.2 Critical Science Studies and Speculative Physics*

Critical science studies have existed as multiple variants, especially in the form of the social and cultural studies of science. All of them are interested in the operation of politics in knowledge, including the politics that circumscribe the various stakeholders of knowledge. In *Science Studies: An Advanced Introduction*: David Hess explains that the term "critical" is as ambiguous as it is complex in the context of contemporary science and technology studies (STS).

> In the humanities the term 'critical theory' usually refers to a theory of literary or cultural criticism, in other words, a theory that helps guide the interpretation of texts. In the social sciences the same term often refers to the Western Marxist tradition associated with the Frankfurt school and post-Marxist researchers influenced by the school, such as Jürgen Habermas…In STS the term is sometimes used to describe the confluence of research traditions that includes feminist/antiracist studies, critics of the technological society, radical science researchers, and various other scholars who are concerned with issues of social justice and democracy. The category of critical STS therefore overlaps with but is not continuous with cultural studies of science and technology. (113)

However, what is of concern here is what critical science studies could do, in a radical sense, in dealing with the uncertainties and destabilized knowledge categories, including one's knowledge of the self and others, to produce an approach to knowledge that is empowering even to the minority stakeholders of the knowledge structure, albeit those who came late into the game.

Moreover, given speculative physics' interest in how one can fit the unobserved in relation to the observed, the former is interested in comparing between the different interpretations and propositions for synchronizing the determinate with the indeterminate. At the same time, speculative physics is aware of the complication involved in separating dominant from subordinated categories of knowledge, as well as in contending with the difficulties of self-reflexivity and epistemic doubt. Such is the case when the qualities of some micro-entities are not as well delineated, with uncertainty over how one might speak of the politics of the invisible. To give a better demonstration of my argument, I will draw on a favorite example of philosophers, taken from Nelson Goodman's concept of projectable predicates.

Imagine an object A that contains a property allowing it to emanate the color blue at time $t_1$. But, with additional data and interactions over time, or maybe, the unleashing of an unpredicted event, the object A now seems to exude a different hue, perhaps green, at $t_2$. The blue-green, or *grue*, of object A has not actually changed the characteristics of the object; on the surface, the characteristics merely look different to an external observer (remember the double-slit experiment). This is actually Goodman's deployment of lawlike versus non-lawlike generalizations (no underlying characteristics of object A has changed in going from blue to grue) that enable (or not) the making of forecasts and predictions, assuming that there is direct continuity from the past to the present, and into the future, and that we do not need different calibrations, methods, or instruments to trace that timeline of continuity.

Perhaps one does not assume that intellectual continuity is necessary to insert oneself convincingly as a subject that matters in making interventions to otherwise inscrutable topics; or the marginality of the observer position itself can be an advantage in injecting non-dominant epistemic contributions that could potentially break the cul-de-sac in current scientific ventures so that such contributions become invaluable.

I am reminded of Traweek's example of an international postdoctoral fellow who came from a less developed country and had to contend with the classist and social biases of his colleagues while working hard to prove his competence in a world that privileges the mastery of technological gadgetry. Couple that scenario to Galison's historical review, in *Image and Logic*, of the contribution of the lesser known Marietta Blau whose 'primitive' gadgetry had brought tremendous contributions to the developments of nuclear science (and is one of the earliest work on nuclear level interactions), one seriously ponders the role of a knowledge producer and actor located at the margins, or who were made invisible. What form of epistemic subversions could they potentially perform from the so-called periphery, to be able to impose, more solidly, their epistemic matter onto an intellectual tradition that wants them forgotten?

Nevertheless, an epistemic memory of an ontological potential is not predicated on the certainty of what went before. Rather, it is built of historical traces that continue to be carried by objects progressing through time. That is the essence of speculative physics.

# Appendix A: The Standard Model of Particle Physics: From Prehistory to History

Knowledge starts out as bare life with radical potentialities. As layers upon layers of the potentialities and information build up and gather, we break through the walls of established facts (or the factish, where agency and dependence on interpretations occur simultaneously) to obtain fresh insights. At the same time, there is also a perspective that one has to deal with, such as the possibility of a 'specious present' that Isabelle Stengers refers to in *Thinking with Whitehead*. She argues about "how much the notion of appearance is a powerful attractor" for inducing confrontation between two stereotypical categories that can lead to the choice of one possibility over that of another (60).

However, there is no such thing as an autonomous objective fact in scientific knowledge since objectivity is based on the premise that the self of the knowledge producer can be effaced during the narrative of knowledge construction, and that knowledge can attain absolute non-subjectivity by mechanical, or instrumental, means[1]. Since knowledge formation takes place at the intersection of subject-object relation, and is defined by its relationship to theoretical paradigms selected precisely for being able to produce the most convincing explanation/description (an act of subjective agency), factual objectivity can never be determinate. Moreover, pure objectivity does not account for the more speculative elements that are included into the building blocks of facts.

---

[1] See Daston and Galison's *Objectivity,* particularly its chapter one that introduces the history of mechanical objectivity in knowledge production through instrumentation and other mediating apparatuses to provide the beginnings to the aspirations of objectivity in science.

The mechanistic paradigm influencing the developments of theoretical and experimental physics in the areas of thermodynamics, optics, and electricity, between the $18^{th}$ and $19^{th}$ century, converge in classical electrodynamics. Classical electrodynamics is the study of electromagnetic field through ontologically determinate mathematical formalism set within a bounded area of the electrical and magnetic experiments performed by major scientific players such as Faraday, Oersted, Weber, Ampere, Biot, Savart, and Thomson, among others.

It was around this time that the trope of action-at-a-distance emerging since Newton and Leibniz took a more practical turn whereby non-touching bodies are able to interact in such a manner that would produce observable physical effects, whereby these effects may occur through scalar (non-directional magnitudes such as temperature or the potential between two charged electrodes) and vector fields (that have directional magnitude such as momentum and force). It is important to note that the paradox of non-local mechanistic interactions is not restricted to epistemic developments in physics, but has application to manner in which the *techne* of media is conceived, such as what one might conceive of the calculating machine of Babbage, the magic lantern, early forms of photography, and the automatons. In fact, many of the developments in optics and electrodynamics that fed back into the development of these objects are early precursors to digital media objects.

Through the developments of wave theory via geometrical optics as well as field theories through the aforementioned work in electrodynamics, nineteenth-century natural philosophers-turned-scientists were able to piece together a more complete picture of the science of energy via a more thorough understanding of the mechanisms of heat, power, and the transformation and conservation of energy, since all of these observed physical phenomena of electricity, magnetism and thermodynamics could now be seen as manifestations of multifaceted properties of energy at different spectrum. All were the result of extensive works by natural philosophers/scientists such as Rankine, Joule, Carnot, Clapeyron, and Kelvin. In fact, the work done on transformations between forms of energy, and their conservation, becomes more crucial once physics move from the realm of the macro to the micro-world of the subatomic particles, and particularly after quantum mechanics, with its challenge on absolutism, enters the story.

The development of scientific fields in the nineteenth and early part of the twentieth century were not only influenced by mechanistic and dynamical worldviews, but also by the direction in which specific theological beliefs and political philosophies such as Marxism, in active production and circulation at that time, were disseminated among, and absorbed, by the scientists. These ideologies, while not influencing directly how the scientists were doing their scientific work, can have a less indirect influence on certain choices that the scientist might make with regard to the sort of interpretation, or even mathematics, that they preferred.[2] Moreover, specific ideologies embraced by the philosopher-scientist enabled him (the dominant figure in most history of science narrative) to speculate on the best explanation for his theory composed of his physical worldview of nature, as well as the multiplicity of representations on realism, including a realism that may be veiled due to an inadequately prescribed path to causality. In physics, realism is the empirical real delimited by the ability for current scientific models and experimental actualities in explicating its observable and non-locally observable properties.

Some of these ideological speculations took on a more formalistic outlook and is advanced for the interpretation of physical phenomena. One such theory is the vortex theory of the atom, conceived as a solution to an age-old problem of the constituent of matter. The vortex theory was one of the earliest attempts to finesse the paradox of the split between the atomistic and the continuum worldview of matter, and hence, an early predecessor to the concept of the field that I have discussed in the previous section. The theory itself was advanced through the work of William Thomson, later to be known as Lord Kelvin, for whom the standard of absolute measurement of temperature, kelvin, was named.[3] The same group of people who were initially interested in the vortex theory of the atom were also the people involved in the development of thermodynamics where questions on freedom and constraints, relating to the question of positionality and movement through time, stand for the point of departure in thinking about the arrow of time, the directional flow of the energy, and the kinetic motions of the molecules.

---

[2] It was also around this period that one sees the rise of functional analytics for dealing with the question of conduction and transmission between different physical phenomena, as well as that of finite elements and statistical mechanics. The previously more geometrical dimensions of theoretical physics was also beginning to take on a more algebraic form with the development of fields and particles, bringing about the need to correlate and demonstrate the relationship of different indeterminately related properties and variables to each other, especially within different physical states. Marx himself was interested in the development of algebra in mathematics, and had written a paper on the expansion series and the Lagrangian (see "Marx's Mathematical Manuscripts 1881." Web. 10 Mar. 2013. < http://www.marxists.org/archive/marx/works/1881/mathematical-manuscripts/index.htm>.

While there is no evidence that the vortex theory is embraced widely, it was certainly a part of the materialist concept of science that would have some impact, at varying degrees, on the sciences they were aware of. However, it would appear that much of the interests at the time included how one may quantify phenomena such as heat, electricity, magnetism etc. It is the attempt to quantify that led to the first version of a unified field theory through the work of scientist-mathematician-philosopher, James Clerk Maxwell.

In his article "On Faraday's Lines of Force," Maxwell is interested in working out a more mathematically systematic way for speculating on the formation of the

---

[3] Kragh, H. "The Vortex Atom: A Victorian Theory of Everything." *Centaurus* 44.1-2 (2002): 34-113.

laws on electricity.[4] His earlier work contained a geometrical outlook that conformed to the more dominant mathematical discourse for privileging geometrical thinking over analytical subtleties.[5] His later work, influenced by the concept of ether and its formulation as vortices, was drawn, I suspect, from work on eddy currents, current measurements, magnetic fields/forces, and from Weber and Gauss's work on electrodynamics for measuring current flows of positively and negatively charged particle flows.[6] The latter work, built on a mechanical viewpoint driving the dynamical model that led to the formation of Maxwell's four famous equations of electromagnetism, are connected to changing electric forces and the displacement of electrical current observed in the work of Faraday, Ampere, as well as Biot and

---

[4] See "On Faraday's Line of Force" in Thomas K. Simpson's *Maxwell on the Electromagnetic Field*. New Brunswick: Rutgers University Press, 1997. 55-138.

[5] See Richards, Joan L. "The Geometrical Tradition: Mathematics, Space, and Reason in the Nineteenth Century" in volume five of *The Cambridge History of Science: the Modern Physical and Mathematical Sciences.*

[6] This has its quantum analogy in the form of Dirac's sea, which is Dirac's theory of the 'hole' that predicts the existence of holes or 'positrons' as a positively charged equivalence of the electron, within a sea of negatively charged particles. The positron was predicted before its eventual experimental confirmation in 1932. Nineteenth century scientists such as Weber were already speculating on the possibility of the electricity current as a double-stream of positive and negative charges, flowing in directions opposite to each other. Coulomb's law on electrostatic attraction and Ampere's law of current elements supported Weber's theory. See more in "Electrical Theory and Practice" by Bruce J. Hunt in the same volume of the *Cambridge History of Science*.

Savart. These equations were first articulated in his *Treatise on Electricity and Magnetism*[7].

Beyond the interacting force fields came parallel developments in electrons, protons, and neutrons taking place in Britain around the late 19th century, such as the famous 'plum pudding' model of the atomic structure of J.J. Thomson, who had discovered the electron before the nucleonic structure was known. Over in France, radioactive physics grew through the work of the Curies. By then, developments were made concurrently in algebraic geometry in Germany, and the Germans, through under-acknowledged women scientists such as Marietta Blau, extended work already underway that pushed for thinking about radioactive elements as particles, such as the then already discovered protons and neutrons, by developing heavy nuclear emulsions to detect and measure them. The nuclear emulsions were also used as photographic techniques to observe nuclear disintegration due to the cosmic rays.

The cosmic rays, as we will see in the next section, represents the turning point in physics whereby particles of properties not seen before are constituted through forms of experiments that can be considered as speculative mainly because the scientists constructed an experimental apparatus to investigate physical entities that they could only vaguely conjecture. The entities were detected because of unexplained anomalies that occurred with one of the other instruments that had been constructed for a different experiment. The properties of these unknown objects could not be comprehended due to an almost non-existent understanding of the connection between the particles inhabiting the microscopic world with physical phenomena manifested extra-terrestrially.

---

[7] However, Maxwell's work was less well understood during his lifetime, and it was only after his death that his work was furthered by a group of younger 'Maxwellians,' many of whom made important contributions to developments in field theory, especially how measurements (such as length and velocity), can be made in the theory of relativity developed by Einstein. H.A. Lorentz, for instance, had the hypothesis that matter contains enormous numbers of tiny charges able to move freely within conductors, but is bounded by material dielectric (Hunt 324).

## A.1 Cosmic Rays and the Particles of the Standard Model

Cosmic rays are high-energy charged particles that have their origins in extra-terrestrial and extra-galactic sources. However, upon entering the atmosphere of the earth, showers of particles are produced through the interaction with the atoms of the atmosphere, which are then captured by ground-based detectors. Nevertheless, there is still a layer of mystery about the rays that physicists are attempting to understand better, although their knowledge is continually improved through deep studies of the data obtained from research in high-energy astrophysics and particle physics when these data are re-contextualized against new information.

Neutrinos can come from cosmic sources such as the supernova of dying stars. Interest in neutrino physics came about as the result of physicists' attempts at resolving the enigma surrounding the conservation of momentum and energy, and neutron decay to protons and electrons. Physicists found that the laws of conservation of energy and momentum were apparently violated because of the possible existence of a third particle that carried away some of the energy. Enrico Fermi named this new particle, which exhibited zero mass and zero charge (a type of 'virtual' particle at the time) so as to counteract the 'shortfall' that would have resulted from the proton and electron not being consistently emitted as a 'neutrino.'

Such experiments involving the disintegration of neutrons are a result of beta-decay (which is the decay of any 'parent' particle to 'child' particles that can include an electron). Chadwick discovered that the electrons he was analyzing from neutron decay showed a continuous energy spectrum and that a 'neutrino' might be the possible cause for it. Neutrinos are produced as a result of the electroweak interaction, an interaction arising from the unification of electromagnetic and weak forces.

The work on neutrinos led to interest by physicists to contemplate, further, the possibility of colliding beams involving the 'anti-electrons' (positrons) and electrons, in the 1950s and 60s, despite the technical difficulties since the colliding beams would have to be injected from directions opposite to each other.[8] However, it was an important realization of Fermi's dream project, and of the Italians, via the Laboratori Nazionali di Frascati that is 25 kilometers from Rome. The experiment has to do with the exploration of collisions at the center-of-mass region of the collider and certain physical theories in circulation at that time such as the dispersion relation of particles stemming from their collision-based interaction.[9] More importantly, the development of colliding beams in the form of electron-positron was itself a form of speculative endeavor for there were many reservations concerning the plausibility of its design, especially since some changes in the design would be required to allow one beam to be coming from the opposite direction from the other beam while still able to track their collisions.[10]

---

[8] Another experiment that interested physicists but which they were skeptical of is connected to the positron, which is an outcome of work on the Dirac hole, symmetry, and the concept of the neutrino. Burton Richter of the Stanford Linear Accelerator Center articulated some of the problems involved in his piece "Colliding Beam Experiments" in the proceedings of the *International Conference on Theoretical Aspects of Very-High-Energy Phenomena, Geneva 1961*, pp. 62-3.

The development of the electron-positron annihilative interaction (seen as an interaction between matter and anti-matter) enabled the experimentalists to move from these lighter leptonic[11] collision experiments to experiments involving collisions

---

[9] See Bernardini, C. "AdA: The First Electron-positron Collider." *Physics in Perspective (PIP)* 6.2 (2004): 156–183. Accessed 1 Jan. 2013, for more details on the development of the earliest electron-positron collider.

[10] According to Burton Richter in the same report for the International Conferences on theoretical Aspects of Very High-Energy Phenomena in Geneva in 1961, the (electron-positron) beams must still be prevented from colliding head on for the same reasons as in the electron-electron scattering experiment. RF cavity and the interaction region must be shifted so that they are opposite to each other. A new steering and inflecting system to control the beam must be constructed, and a new counting system required. Positron-electron experiment was incompatible with the then available electron-electron experiment. More understanding about the storage of the beam was required.

[11] Leptons are subatomic particles forming part of the Standard Model, which are involved in weak interactions. While the electron is the most common and also representative of the first generation (or

between heavier hadrons, thus providing more energy during the collision.[12] Through searches involving heavy leptons, physicists were able to connect the differences in the length of the chain of decay of the leptons from that of the hadrons: there are not too many steps in the chain of decay between the very light electron and its leptonic counterpart, the very heavy muon. Nevertheless, the physicists may have to deal with the possibility of misidentification of the lepton in the process of working out the relationship between the electron and the muon.[13]

There are some important lessons to be learnt here about how speculation takes place in the lead-up to the discovery of the tau lepton, the heaviest leptonic particle, at the theoretical and experimental level. The discovery of the tau lepton completes the theoretical prediction made of the entities in this particle group and was accomplished due to advancement in the fine-tuning of the experimental apparatus. Initially, physicists had to make use of what was already known about

---

flavor) of the lepton, the leptons constitute three generations; electronic leptons, muonic leptons, and tauonic leptons.

[12] Hadrons are subatomic particles constituting the Standard Model that is made up of baryons (particles with three quarks) and mesons (particles with two quarks). There are important for strong interactions and for later developments in quantum chromodynamics involving gluons, which are used to explain that boundedness of quarks and the non-existence of a free quark.

[13] By working out the events produced in that annihilation between the positron and electron, physicists attempt proper identification by positing that the positron-electron interaction leads to both the production of the electron-muon pairing (which include the neutrino) and the muonic and hadronic pairing within the same energy spectra. This led to the production of the heavy lepton that was later known as the tau.

the lepton-lepton and lepton-hadronic interactions which also meant working with the possibility of misidentifying hadrons for leptons that would then had to be noted, just in case, as these are regular occurrences in the 'anomalous' hadronic interactions back when the existing infrastructure was not sufficiently sensitive in detecting the differences. It was only later, when a muon tower was added into the existing SLAC-LBL detector, that misidentification could be corrected and the crude detector was vastly improved.

By the end of 1978, it was confirmed that electromagnetic interaction produces tau lepton while the weak interaction is what produces the decay.[14] One of the interesting aspects of the experiments, and similar future experiments, is the need to produce theories that could possibly account for the excess of events not easily explicable within existing known interactions. Hence, as is often the case, speculation takes place by setting aside the 'excesses' of the known and then modeling them within the parameters of what is known, even if that meant locating their presence rather imprecisely. What matter is that care be taken when noting the problem of that 'excess' and their observable characteristics, particularly when the latter are located within impoverished explanatory frameworks.

---

[14] See Perl, Martin. "The Discovery of Tau Lepton" *The Rise of the Standard Model: Particle Physics in the 1960s and 1970s*. Eds. Lillian Hoddeson, Laurie Brown, Michael Riordan, and Max Dresden. Cambridge: Cambridge University Press, 1997. 79-100.

## *A.2 The Metaphysics of the Gauge Entities*

Now, we should delve into the theoretical structures undergirding the events just described. One of them is the development of quantum electrodynamics (QED), which represents the final vestiges of the 'old' physics theories: it is basically the quantum version of the previously discussed classical electrodynamics of Maxwell. QED is one of the primary and earliest constituents of the Standard Model. Schweber's *QED - And the Men Who Made It: Dyson, Feynman, Schwinger, and Tomonaga* provides a more thorough explication of QED's extensive epistemological and political history. For the purpose of this chapter, I would just state that QED is important for thinking about the interaction processes between photons, protons, electrons, and other leptons. It is also the theoretical-predictive framework for describing the movement of electromagnetic particle-waves from a four-dimensional perspective of field theory.

The double-slit experiment of subatomic particles provides a simple and elegant explication of the development of QED. However before going into that, I would like to discuss the photoelectric effect, which, due to its connection to the history of the electromagnetic field, is a good entrée into understanding the background of the 'paradox' of particle-wave duality demonstrated by the double-slit experiment. The photoelectric effect is an observable phenomena of photo (light) emission caused by light-waves pushing through the surface of a solid to free the electrons at the surface. This has been observed since 1887 by Heinrich Hertz, and was later further studied by J.J Thomson. While it is not hard to explain why the light of particular frequency that is shined onto the surface of a solid with loosely bounded electrons could lead to their being freed, what could not be easily explicable is the level of threshold frequencies required of the light-waves for the electrons to be tipped over and out of their bounded regions, and how the threshold is different for different sets of light waves even when the energy-frequency for all the light waves across the spectrum remains the same. In other words, in a curved graph, they all have the same slope, showing how the energy-frequency relation is constant, and this constant is the Planck constant.

The photoelectric theory, as explicated by Einstein in 1905, had its background in the development of thermodynamics (such as the blackbody radiation that posits that light can only exist as discrete bundles of energy) and the particulate model of the atoms, which posits that light contains particle-like properties in the form of the photons. This is because the energy of the ejected electrons is proportional to light frequency, even as the ejection energy is independent of the total energy of illumination.[15]

---

[15] More can be found at <http://physics.info/photoelectric/> and <http://hyperphysics.phy-astr.gsu.edu/%E2%80%8Chbase/mod1.html#c3>.

In the double-slit experiment, we consider a photon or electron colliding with the screen after entering through either slits of the two-slit diaphragm placed between the particle source and the screen. The physical effects observable after collision with the screen confirms the wave-like properties of the photon/electron because the resulting interference pattern on the screen strongly hints on that possibility; furthermore, the pattern of interference suggests a probabilistic range since one is not able to determine with absolute certainty as to where on the screen the particle would show up. Hence, the particulate electron now has demonstrable wave-like qualities. What is seemingly paradoxical about wave-particle duality can be resolved if we begin thinking about the photons and electrons within the framework of a field.

While the concept of fields is not new in classical mechanics, it takes on a different dimension when considered at the level of quantum mechanics, for it is now use to represent the probabilistic uncertainty in the observation of particles moving at, or close to, the speed of light. The conceptualization of the field allows us to think about the multiplicity of dimensions, and phenomenon, by which we can view the interaction of a single particle, while also formally accommodating interactions beyond a single particle.[16] Hence, as we move from thinking about the isolated interaction of a single particle with the screen to considering the collisions between particles, the particle's fundamental ontology remains the same, even if the phenomena manifested are different.

[16] Also known as quantization (quantum mechanical calculations of a single or many-body particle) beyond the first order.

At this stage, the local qualities of the particles change when an interaction takes place between them. To provide a clearer picture of what I mean, let us consider the interaction of macro-level particles, such as two billiard balls, as they collide with one another. At the point of collision, which could be elastic (the balls bouncing off each other) or inelastic (the balls sticking together), energy and momentum is always conserved. This is because, considered overall, the collision is not happening at a sufficiently high enough velocity to cause any major disturbances to the intrinsic energy level of the ball, nor would the collisions change the internal characteristics of the ball.

However, when a photonic laser beam is directed at one of the balls, the high intensity of the beam, with the ray of photons operating at the speed of light, causes the billiard ball to alter the very nature of its macro-physical structure, altering its innate properties altogether. This is because the interactions are now taking place at the level where the properties of each atomic and subatomic particle making up the ball matter, as do the individual spin, momentum, and rotational direction of each of these particles. The highly energetic beam also breaks down the atomic particles into its more fundamental constituents via decay while recombining these subatomic particles into less naturally common atomic composites (fusion). The process of interactions are mediated by the forces between these micro-entities, similar to the kind Newton had once predicted concerning two gigantic bodies operating at a considerable distance from each other (such as two planetary objects). Depending on what the particles are, and the range of action between them, these forces can be either weak or strong.

After the discovery of the electromagnetic force by Maxwell, the second force to be discovered, through work in nuclear and radioactive physics, is the weak force. This is because the earliest experiments involved protons, electrons, and the muons; they are known as the fermions, which are categories of particles that obey specific statistical distribution known as the Fermi-Dirac statistics, where the weak interacting forces are detected. These forces are basically part of a theoretical structure used to systematically organize the fundamental entities and forces of nature to achieve unitarity, and through that, symmetry. Hence, the particles that are part of the interactional processes of the force are what constitute the Standard Model. There were also later attempts to unify the electromagnetic with weak forces, bringing about the electro-weak force.[17] With the unification of these forces,

---

[17] I will not attempt a mathematical dissection of this since it would merely become too involved and convoluted. Suffice to say that the mathematics is a very specific form of group theoretical algebra that has had its earlier, and less complex, incarnation in classical mechanics. Hence, the quantum mechanical version maintains the same mathematical ontology, but with greater degree of complexity since we now have to move from thinking about the more 'global' and 'neat' interaction between

physicists also began working on experiments that could indicate the existence of specific bosonic particles known as gauge particles.

Now that I have discussed the electroweak force, the next important force to be considered is the QCD (quantum chromodynamics), a strong nuclear force. Prior to the introduction of the quark theory, the bound state of a nucleon was studied mostly in nuclear physics, but not given as much attention in particle physics. The binding energies studied in nuclear physics involve the decay of heavy nuclei in the form of protons (and neutrons) at non-relativistic level energies, while the high-energy physics of quark operates at a relativistic level.

Much of the early developments in QED concentrate on the free particle in the form of electrons and muons, although the bound state of the nucleon (where the protons and neutrons reside) would also become important due to the interactions between a proton and an electron. In the early years of the development of the quark theory, QED is used to discuss interactions between hadronic particles and leptons even if the flavor of the quarks are not conserved in the weak interactions. In other words, quarks develop lepton-like qualities in weak interactions.

---

billiard balls to a 'messier' local-level interaction between subatomic particles. However, mathematical symmetry, and thus physical symmetry, is always an epistemic priority.

While I have concluded my discussion into the known forces that are part of the Standard Model, there is still another force not featured here because much of its properties have not been finessed and also because there is a problem in unifying relativity and quantum mechanics conclusively, notwithstanding the semi-classical attempts that one finds in quantum field theories. In fact, one could say that the theories of the gravitational force have remained a product of speculation for decades, appearing both in speculative scientific retellings and also in science fictional works. In the following section, I will discuss a development in the strong nuclear force that has itself been a product of skepticism and speculation for many decades prior to its final confirmation: the quark.

## A.3 The Enigma of the Quark

Even as the quark model was increasingly better understood, there was still a problem in its inability to explain why the flavor (up, down, strange, etc) properties of the quarks can be transformed during the process of decay. This was only resolved, in a manner of speaking, with the application of a mechanism, known as the Cabibbo mechanism, to the neutral processes proposed by Glashow, Illiopoulos, and Maiani (also known as GIM) in 1970, and then extended down by Kobayashi and Maskawa (KM) through all the three generations of quarks known by 1973.[18]

---

[18] For more information, see David Griffith's *Introduction to Particle Physics*. New York: John Wiley and Sons, 1987. See chapters two and ten. Even though there are not as much updates on the properties of

These suggested mechanisms solved the paradox of non-conservation through the utilization of linear algebra. The GIM scheme would become useful for thinking about the different handedness (left or right) of the quarks, which are intrinsic not only to the constitution of the Standard Model, but also in understanding how symmetry is maintained in the process of transformation and rotation of the particle through real space. The Standard Model, after all, is constituted of chiral symmetry.

If electric force in QED is carried by the photon, in the nuclear force, it is represented by the pion exchange, which was first predicted by Yukawa as a nuclear-force counterpart to the beta-decay for the 'free' electronic and neutrino particles described above. According to Robert Oerter,

> Yukawa has been struggling with the nature of the nuclear force holding the protons together. Several years before, Heisenberg had suggested that nucleus was held together by an exchange of some particle among the protons and neutrons, rather like the electric force is carried by exchange of (virtual) photons. Yukawa realized, as Heisenberg had, that the particle being exchanged could not be an electron, for the same reason as in beta decay – the spins wouldn't add up. Finally, after hearing about Fermi's theory of beta decay and the invention of a new particle, Yukawa decided to assume the nuclear force was caused by the exchange of some particle, which came to be called pion…" (Oerter 140)

The decay of the pion, either in its negative or positive condition, provides the 'hidden' explanation and balance for the two-way transformation between a proton

---

the interactions I refer to here, one can still check out the second edition of the book, published in 2008, for the other updates.

and a neutron. This virtual particle holds for nuclear force what the 'virtual' photon does for the electron. Unlike the photon, however, the pion has mass and must hence borrow its energy from a particle that travels at the speed of light and crosses into the nucleus. What is interesting about the prediction and discovery of the pion in the photographic emulsions stemming from a study into the cosmic rays was how it became the precursor to understanding quarks and symmetry breaking, the latter important in the story of the Higgs boson. The pion began life as a mathematical construct of particular properties that are then fitted into material traces exhibiting particular properties that bear relationality to theoretical predictions, not unlike the case of the quark.

If the cosmic ray experiments of the 1930s provided the first plausible hint of other leptons beside the electron, it was only with the improvement in the sensitivity of the detection methods, and techniques that could function independently of the presence of the experimenters, that the first evidence of the pion was detected from the spray of particles coming from cosmic rays (a particle of spin 0 and about 270 times heavier than the electron mass). By the 1950s, many of these pions were produced through the machine-mediated collisions of nucleonic atoms (the nature of the strong force was still not completely understood at this point in terms of the quark model and gluons) and play an important role as the 'virtual' mediator for the electron-positron collisions. The discovery of the pion was a start to uncovering the identity of 'strange' particles that were later known as quarks. But before that, one must understand the beginnings of the composite model of point-particles (which would be later be called the quark model) through the narrative of the Eightfold way.[19]

The Eightfold way was to the subatomic particles what the Mendeleyev periodic table was to chemistry. Murray Gell-Mann first drafted a version of the Eightfold way in 1961, where a particle isospin, the abstract 'imaginary' spin that behaves differently with each force of the Standard Model, was plotted onto an x-axis. The isospin has nothing to do with space and time, but rather, with the relationships between the different particles referred to as internal symmetry. There is also the concept of strangeness on another axis, and a hexagon of particles was plotted in, giving us the Eightfold way. Strangeness is a property of a strange particle that decays either into a normal (such as a proton or anti-proton), or another strange particle (xi-minus and xi-positive).[20] Nevertheless, there were teething

---

[19] Just to note that the 'particle,' used in physics, represents the best conceptual way of describing fragments and fractions of matter of the atom or subatom; it is not meant to represent a literal embodiment of a particle. As with any historic terms in science, and an issue that would be raised in all the chapters of this dissertation, there are always misnomers.

[20] The Eightfold way went from a hexagonal to an octuplet representation, thanks to Gell-Mann's discovery of an algebraic group theory known as the Lie group that later became the mathematical foundation of symmetry in particle physics, and of the unitary groups that we see manifested in the form of U(1), SU(2) and SU(3) for representing electromagnetism, weak interaction, and strong interaction. However, the octuplet was soon upgraded to the decuplet, and Gell-Mann was able to predict the existence of yet-to-be-discovered particles, such as the Omega-minus.

problems during the earliest days of the model's formation that could only be dealt with over time.

The first three quarks to be discovered by 1964 were the up quark, down quark, and strange quark. Quite some years later, they were followed by the charm, bottom, and top quark; with the latter two quarks extensively studied at the experimental level due to the unknowability and greater complexity of their qualities. The final two known quarks were discovered at a time much closer to the present than the 1960s (the bottom/beauty quark discovered in 1977 and the top quark in 1995). Nevertheless, the quark model provides the complete explanation needed to speculate more intelligently on the observed properties of subatomic particles mentioned, especially the pion-exchange process that is integral for understanding how a proton becomes a neutron, and vice versa.

An inelastic collision process between the protons and electrons, performed in varying degrees of sophistication at different accelerators on the East and West coast of the US, is known as scaling and said to provide the earliest indication of the constitution of a proton as point-like composites of the quarks. The scaling process (sometimes referred to as Bjorken scaling) refers to how, at a high enough energy, the scattering condition of the quarks are no longer dependent on energy loss. One has to do with the fact that a free quark has never been noted, even though the idea of scaling is premised on the assumption that the quarks are not strongly bound to each other. The other has to do with the physical impossibility of having all quarks of similar spin state aligned with each other.

Therefore, Gell-Mann and his colleagues attempted to solve the problem by constructing the idea of three colors for each flavor of the quarks. The 'colors' were arbitrarily given the color of the primary colors: blue, green, and red. The plan is not to insinuate that the quark flavors come in colors, but to demonstrate the need for another property in order to explain away the paradoxes of the quark model, paradox that stems from our insistence at finding the most elegant solution to pressing ontological dilemmas.[21] The addition of the color properties helps explain why it is possible for three strange quarks with similar flavors and spin direction to exist. The color describes the force that keeps the quarks bound into a hadron, such as a proton. The theory predicts that a force, behaving like glue that holds the quarks together, is carried by a set of massless particles referred to as gluons. More of this will be discussed as I move on to an elaboration of QCD below.

Before we enter more extensively into the current-day quark model and its place within the quantum chromodynamics network, let us take a step back and consider the gauge principle that is the basis of the relativistic quantum field theory of the original classical Maxwellian electromagnetic equations. However, the

---

[21] See more of this in Sheldon Lee Glashow's 1975 Scientific America article "Quarks with Color and Flavor," pp 38-50.

Maxwellian equations have to be modified using Dirac's equations so as to ensure that the evolution of the electron for compliance with experiment. In order to do so, the equations for conserved current are utilized to construct a wavefunction of electrons in a four-vector form that takes into account three spatial dimensions and time while considering the intrinsic properties of an electron such as spin.

Without going into too many details, the gauge principle is needed for creating the mathematical foundation for demonstrating the different physical properties of the particles situated within the Standard Model, and for understanding the properties of the quarks. The Glashow-Weinberg-Salam (GWS) model for weak interaction is also used for describing the interactions of leptons with that of the hadrons. While the weak interaction properties (including the flavor symmetries) and certain aspects of the electromagnetic properties of the quarks were understood, the strong interactional properties were not properly understood until after the November revolution of 1974; after the discovery of the J/psi mesons, neutrinos, Drell-Yan process of the quark and anti-quark annihilation process, composite quark model of the hadrons sometimes referred to as the partonic model, as well as the transverse momentum of the hadronic particle so intrinsic to experimental high-energy physics and to the foundation of QCD.

According to QCD, color can be changed when a quark emits a gluon: a gluon has to carry two colors - the color of the incoming quark and the anti-color of the outgoing quark. While QED and QCD are structurally quite similar, the emission of the gluon is what differentiates QCD most strongly from the QED. Another difference is that, in QED, the photon that mediates between electrically-charged particles is not itself electrically charged, while in the QCD, the gluon itself carries color (which is equivalent to the charge in QED), meaning that gluons can interact with each other while the photons could not. Hence, we have quarks with six flavors, each with three different colors, bringing us to 18 quarks (not all are depicted in the usual Standard Model diagram) given that each quark has an antiquark, so there are 18 antiquarks as well.

Another main difference between QCD and QED related to the first difference is the existence of the *asymptotic freedom* in the latter. In QED, there is the screening effect of the cloud of particles, not unlike the screening effect that one might have encountered in high school chemistry or the discussion of electron shells in physics. However, this can be circumvented when higher energy is exerted and one is able to penetrate deeper into the cloud, leading the electric force to increase as well. The opposite is true for QCD as higher energy and shorter distances decrease the strength of the color forces (Oerter 180).

The Feynman diagram is the best way for demonstrating this feature because calculations can become messy when attempted algebraically. However, as a single diagram demonstrates, the more interactions there are, the messier their representations become. Further, the greater number of interactions within a single diagram also means that the effect of the diagram on the overall result of the QCD interaction, especially at high energies, decreases. The quark-quark interaction also becomes weaker, causing them to become increasingly likely to behave as free particles (Oerter 285-88).

Up to this point, I have provided an overview to the development of the mechanics influencing the development of the Standard Model while discussing briefly all the important components that make up the table: the leptons, quarks, forces (electromagnetic, weak, strong, but excluding the gravitational force), and the bosons. The bosons are known also as intermediate vectors and force-carriers: their existences are needed to explain why the other entities behave the way they do.

The known mediators of the three forces include one photon, eight gluons, and three weak bosons known as ±W and the Z. Of course, we also now have the Higgs boson, whose properties are still being studied. While the spin properties of these bosons matter in the physics, it is less important, for the purpose of my dissertation, to go into the details of their technical characteristics. Nevertheless, they are important for QED-related interactions and validity tests. It has been a puzzle as to why a photon would not have mass while W and Z do. The Higgs mechanism, which also predicts the possibility of a new particle called the Higgs boson, seeks to explain this.

The Higgs mechanism introduces a new kind of field, known as the Higgs field, that is over and above the known forces. The purpose of the mechanism is to provide plausible explanation for the mass of bosons without leading to infinitely huge divergences that could quickly turn ugly, mathematically-speaking. The Higgs mechanism attempts to introduce a scalar field that would enable the Higgs field to interact with all particles, bringing about the aforementioned symmetry breaking, which is necessary for attributing mass to each of the interacting particles. The Higgs field starts out the journey of fundamental inquiry that now can possibly extend beyond the Standard Model, and thus into an area that is ripe for systematic consideration of various emergent theoretical models.

## A.4 Beyond the Standard Model

There has been a definite difficulty with the Standard Model, which has caused physicists to return to the drawing board to work out the problems that the model has not been too able to resolve thus far. In the chapter of "Beyond the Standard Model" of his 2010 *Introduction to Particle Physics and the Standard Model*, Robert Mann outlines what he considers to be the three main reasons:

1. Too many adjustable parameters: too many constructions of the particles, geometrical presumptions, and mathematically adjustable possibilities. There are 18 arbitrary parameters that have not been predicted that are mathematically valid and theoretically self-consistent in that the mathematical manipulations performed to extract the physical predictions from the Standard Model are as valid as if other values have been used for these parameters. There are just too many possibilities that have not been confirmed as implausible, leaving the model open to subjectivity.

2. There are also problems with including gravity with the other three interactional processes. The quantum corrections to the gravitational processes are infinite and do not lend themselves to possible resolutions to these infinities. To do that would require the redefinition of every possible physical quantity in physics. Moreover, time has a different ontology in gravity and causality is uncertain.[22]

3. Our universe is biophilic in that it supports life but the Standard Model is unable to explain why the existence of life is so tightly connected to the parameters that have been predicted by the elementary particles, or why any adjustments to the parameters as we know it would render life impossible in our Earth (483-5).

4. The impossibility of the Grand Unified Theory where all the aforementioned forces, plus gravity, would be unified to create one simple scheme, and one coupling constant, that will provide an elegant explication as to why our universe is the way it is.

---

[22] I have recently reviewed a book that discusses the problem of timelike and spacelike in general relativity, in my review of Tim Maudlin's *Philosophy of Physics: Space and Time*. See the review "Philosophical about Space-Time" in <http://physicsworld.com/cws/article/indepth/2013/jan/24/philosophical-about-space-time>. This will help address some of the problems of general relativity as existing in a different ontology from quantum mechanics that I will not be going into detail here.

Dark matter, as much of cosmological as it is of science-fictional interest, demonstrates the failure of the Standard Model to explain the existence of a form of matter that is weakly detected and whose properties are not readily observable. Evidence for the existence of dark matter stems from the observation of what gravitational effect has on the motion of the stars in galaxies, of how the clusters of galaxies are moving, and of the still-speculative theory on the expansion of the universe which supposedly has found confirmation through the observation of a supernovae that earned three physicists a shared Nobel Prize in physics in 2011.

In the first instance, the motion of stars is detected by measuring the shifts of light within the electromagnetic spectrum (the shift is known as the Doppler shift). By comparing light from one edge of the universe to that of another, one is able to measure the rotation of the galaxies. By using the Newtonian gravitational calculus to measure the speed of the stars as one moves further away from the center for the galaxy, assuming that there is no friction caused by other matter in space, the speed of the stars at the outer-region of the galaxy should therefore decrease since the number of stars pulling inward to the center would decrease. However, it was discovered that this was not the case; instead, the speed stabilizes into a constant value. In addition, the stars are also moving too quickly, lending to the destabilization of the galaxy due to insufficient mass for holding it together under such high speed.

The same light shifts for measuring the motion of the stars are used for measuring the rotation of the galaxies. An example of this is the nearest galaxy to our Milky Way, the Andromeda galaxy. Measurements have indicated that Andromeda is moving towards the outskirt of galaxy at 100 kilometers per second, which takes place at a rather large speed. Andromeda would have left the local cluster were it not for the pull of the other members of the galaxies within the same cluster. A problem arises because our sun orbits the center of the Milky Way at a distant of 20000 light years at a speed of about 200 kilometers per second. However, the Andromeda galaxy is 1100 times further away from the center of our galaxy, with half the speed, and yet contains sufficient gravitational force to hold it at this distance to enable the Andromeda to remain at the cluster.

The dark matter, as it is predicted now, are thought to be constituted Weakly Interacting Massive Particle (WIMP) considered as part of the weak interaction but escaping accelerator detection due to their very large mass. More recently, experiments from a satellite known as Planck has semi-confirmed that neutrino, which was once considered a candidate for explaining dark matter, may not possibly exist.[23]

---

[23] See more details at < http://blogs.scientificamerican.com/observations/2013/03/22/cosmos-study-dashes-hope-for-new-neutrino/> "Cosmos Studies Dashes New Hope for Neutrino." Accessed March 29, 2013.